\documentclass{aa}  
\usepackage{multirow}
\usepackage{fontawesome}
\usepackage{placeins}
\usepackage[colorlinks=true,linkcolor=blue,citecolor=blue,urlcolor=blue]{hyperref}
\usepackage{graphicx}
\usepackage{txfonts}

\def\MS{\text{M}_{\odot}}
\def\MSt{M_{\star}}
\def\pc{\text{pc}}
\def\kpc{\text{kpc}}
\def\eV{\text{eV}}

\begin{document} 

\title{Confronting fuzzy dark matter  with the rotation curves of nearby dwarf irregular  galaxies}
 
\author{Andr\'es Ba\~nares-Hern\'andez
\inst{1,2}
\and 
Andr\'es Castillo
\inst{1,2}
\and
Jorge Martin Camalich
\inst{1,2}
\and
Giuliano Iorio
\inst{3,4,5}
}
\institute{Instituto de Astrof\'\i sica de Canarias, La Laguna, Tenerife, E-38200, Spain
\and
Departamento de Astrof\'\i sica, Universidad de La Laguna
\and
Dipartimento di Fisica e Astronomia Galileo Galilei, Universit\`a di Padova, Vicolo dell’Osservatorio 3, I–35122 Padova, Italy
\and
INFN-Padova, Via Marzolo 8, I–35131 Padova, Italy
\and
INAF-Padova, Vicolo dell’Osservatorio 5, I–35122 Padova, Italy}

   \date{\today}

 
  \abstract
   {}
{In this paper, we carry out a phenomenological investigation of the viability of fuzzy dark matter, which is composed of coherent waves of non-interacting ultralight axions with a mass of $m_a\approx10^{-22}$ eV. We did so  by confronting the predictions of the model, in particular, the formation of a solitonic core at the center of dark matter halos, with a homogeneous and robust sample
of high-resolution rotation curves from the LITTLE THINGS
in 3D catalog. This comprises
a collection of isolated, dark matter-dominated dwarf-irregular
galaxies that provides an optimal benchmark for cosmological
studies. Our aim is to find evidence of fuzzy dark matter in the observations; alternatively, we seek to set exclusion ranges for its mass.}
{We  used a statistical framework based on a $\chi^2$ analysis of the rotation curves of the LITTLE THINGS in 3D catalog using a fuzzy dark matter profile as the theoretical model. This allows us to extract relevant parameters such as the axion mass and mass of the solitonic core, as well as the mass of the dark matter halo and its concentration parameter. We fit the data using current Markov chain Monte Carlo techniques  with a rather loose set of priors, except for the implementation of a core-halo relation predicted by simulations. The results of the fits were then used to perform various diagnostics on the predictions of the model.}
 {Fuzzy dark matter provides an excellent fit to the rotation curves of the LITTLE THINGS
in 3D catalog, with axion masses determined from different galaxies clustering around $m_a\approx2\times10^{-23}$ eV. However, we find two major problems from our analysis. First, the data follow scaling relations of the properties of the core, which are not consistent with the predictions  of the soliton. This problem is particularly acute in the core radius-mass relation with a tension that (at face value) has a significance of $\gtrsim5\sigma$. The second problem is related to the strong suppression of the linear power spectrum that is predicted by fuzzy dark matter for the axion mass preferred by the data. This can be constrained very conservatively by the galaxy counts in our sample, which leads to a tension that exceeds  $5\sigma$. We estimate the effects of baryons in our analysis and discuss whether they could alleviate the tensions of the model with observations.}  
  {}


   \maketitle
%
\section{Introduction}
\label{sec:intro}

One of the most pressing open problems in contemporary physics is the fundamental nature of dark matter (DM). The prevailing hypothesis is that DM is formed by a new type of particle, or a ``dark sector'' that is not contained within the standard model of particle physics and behaves as a cold and collisionless fluid on large cosmological scales~\citep{Bertone:2010zza,Bertone:2016nfn,Chou:2022luk}. This cold DM (CDM) is a crucial ingredient of the standard cosmological model, known as $\Lambda$CDM, which has proven to be very successful in describing observations of the universe at these large scales~\citep{2020A&A...641A...6P,Alam2017,ParticleDataGroup:2022pth,Peebles:2022akh}. 
However, observations probing the formation of structure at small galactic scales might pose a challenge to $\Lambda$CDM. 

Stellar and gas kinematics indeed provide one of the most direct observational probes of the internal structure of DM halos~\citep{1978PhDT.......195B,1980ApJ...238..471R,1987ARA&A..25..425T}. While pure $N$-body CDM simulations generically predict density profiles peaking as $\rho(r)\propto1/r$ at small radii~\citep{1991ApJ...378..496D,1996ApJ...462..563N,1997ApJ...490..493N}, the rotation curves of DM-dominated dwarf galaxies point instead to the existence of flat cores at their centers~\citep{1994ApJ...427L...1F,1994Natur.370..629M,2002A&A...385..816D,2011AJ....141..193O,2015AJ....149..180O,2017MNRAS.467.2019R,2020ApJS..247...31L,
2020A&A...643A.161D}. This is known as the "core-cusp" puzzle that is among the small-scale problems related to $\Lambda$CDM (see~\cite{2017ARA&A..55..343B} and~\cite{2022NatAs...6..897S} for recent reviews). Other possible discrepancies between the predictions of the model and observations are referred to as the "missing satellites" problem that is related to the observation of many fewer low-mass galaxies in the Local Group than otherwise expected from CDM simulations~\citep{1999ApJ...522...82K,1999ApJ...524L..19M,2012AJ....144....4M}. Then there is the "too-big-to-fail" problem, whereby the central masses are too low compared to the most massive (sub)halos predicted in $\Lambda$CDM~\citep{2011MNRAS.415L..40B,2014MNRAS.439.1015K}.

These problems have prompted a flurry of activity over the past decade theorizing and analyzing phenomenologically models beyond CDM. One prominent candidate is fuzzy DM (FDM), which is composed of ultralight axions with masses in the range of $\sim 10^{-23}-10^{-21}$ eV~\citep{Sin:1992bg,Hu:2000ke,Matos:2000ss,Chavanis:2011zi,Schive:2014dra,Schive:2014hza}. This type of particle has been predicted by several theoretical models,  preserving the features of CDM at large cosmological scales and which can be produced ``naturally'' in the early universe to match the DM relic abundance~\citep{Marsh:2015xka,Hui:2016ltb,Niemeyer:2019aqm,Ferreira:2020fam,Hui:2021tkt}. On the other hand, on distances comparable to their de Broglie wavelength, which can be in the kiloparsec (kpc) scale, these particles populate the galactic halos with large occupation numbers and behave as self-gravitating DM waves. One of the consequences of this is the appearance of a pressure-like effect on macroscopic scales leading to the formation of a flat core, or ``soliton,'' at the center of galaxies with a relatively marked transition to a less dense outer region that follows a CDM-like distribution~\citep{Schive:2014dra,Schive:2014hza,Veltmaat:2018dfz,Mocz:2018ium,Nori:2020jzx,2020arXiv200704119M,Chan:2021bja}. This effect also produces a suppression of the linear power spectrum on small scales and to a decrease of the number of low-mass halos~\citep{1985MNRAS.215..575K,Schive:2015kza,2017MNRAS.465..941D,May:2021wwp,May:2022gus}. The FDM provides, then, an elegant solution to the small-scale issues of $\Lambda$CDM, which has motivated several phenomenological studies and searches in astrophysical and cosmological probes of low-scale structure (see~\cite{Ferreira:2020fam} for a review and Sect.~\ref{sec:discussion} below for a discussion). 

Other attempts to solve some of the small-scale issues of $\Lambda$CDM by modifying the fundamental properties of DM are warm DM~\citep{Sommer-Larsen:1999otf,Bode:2000gq,Rubakov:2017xzr} including degenerate fermion DM~\citep{Tremaine:1979we,Gorbunov:2008ka,Domcke:2014kla,Bar:2021jff,Chavanis:2021jds,Krut:2023drx}, self-interacting FDM~\cite{Goodman:2000tg,Chavanis:2011zi,Chavanis:2011zm,Delgado:2022vnt} and, more generally, self-interacting DM~\citep{Spergel:1999mh,Rocha:2012jg,Vogelsberger:2012ku,Tulin:2013teo,Ren:2018jpt,Correa:2022dey}, as well as phenomenological approaches (see e.g.,~\cite{2020A&A...642L..14S}). However, these problems could also be the result of baryonic physics at play in galaxy formation and evolution within $\Lambda$CDM~\citep{2022NatAs...6..897S}. For instance, the missing satellites could be due to a loss of efficiency in galaxy formation within low-mass halos as a result of the suppression of gas accretion during reionization~\citep{1992MNRAS.256P..43E,Bullock:2000wn,Sawala:2015cdf}. In addition, baryonic feedback from supernovae (SN) could provide a mechanism that is powerful enough to flatten CDM profiles at the core of dwarf galaxies~\citep{Navarro:1996bv,Pontzen:2011ty,DiCintio:2013qxa,Sawala:2015cdf,Tollet:2015gqa,Chan:2015tna,Fitts:2016usl,Read:2015sta}. The extent and reach of these mechanisms is yet unclear because it depends on modeling assumptions. In the case of SN feedback, there is a certain consensus that it is particularly efficient for galaxies with stellar masses in the range of $10^8~\MS \lesssim \MSt\lesssim10^9~\MS$~\citep{Tollet:2015gqa} and it could still play a role for masses all the way down to $\MSt\sim10^6~\MS.$~\footnote{The lack of a first-principles treatment of baryonic feedback makes the assessment of their precise effects uncertain and somewhat controversial. For instance, simulations from~\cite{Sawala:2015cdf} and~\cite{Zhu:2015jwa} do not find baryonic-feedback induced cores for dwarf galaxies while those from~\cite{Read:2015sta} find the cores for all of them (see ref.~\cite{2017ARA&A..55..343B} for a brief discussion). Note: also the systematic errors in the modeling required for extracting the rotation curves from data have been suggested to introduce considerable uncertainties, explaining the observed diversity~\citep{Roper:2022umd}.}  

The aim of this paper is to compare the predictions of FDM with the dynamics of dwarf galaxies, as observed through their HI rotation curves. While recent works in this direction (e.g., in~\cite{Bernal:2017oih,Bar:2018acw,2019MNRAS.483..289R,Bar:2021kti,2021ApJ...913...25C,Street:2022nib,Khelashvili:2022ffq,2023arXiv230404463D}) have focused on the galaxies compiled by the SPARC data base~\citep{Lelli:2016zqa}, we focus here on a specific group of nearby isolated low-mass irregular galaxies from the LITTLE THINGS survey~\citep{Oh:2015xoa}. Our selection is based on rotation curves published in~\cite{2017MNRAS.466.4159I,2016MNRAS.462.3628R}, which were obtained using state-of-the-art 3D reconstruction techniques to reanalyze the HI data cubes. Galaxies that could have a biased dynamical analysis, dubbed ``rogues,'' were flagged and excluded from the core data set. 
This catalog, referred to as LITTLE THINGS in 3D ~\citep{2017MNRAS.466.4159I}, constitutes a robust and homogeneous collection of high-resolution rotation curves of galaxies that lie on the edge of galaxy formation. It provides an excellent laboratory to investigate the small-scale issues of $\Lambda$CDM and their potential interconnections with the nature of DM.

\section{FDM Generalities}
\label{sec:FDM}

We start by considering a real scalar field, $\phi(t,\vec r),$ with a mass, $m_{a}$, satisfying a Klein-Gordon equation coupled to the gravitational field, $\Phi(t,\vec r)$.~\footnote{The theoretical framework presented in this section is obtained specifically from~\cite{Bar:2018acw,Guzman:2004wj,Moroz:1998dh,Hui:2016ltb,Ruffini:1969qy}. In our formulae, we implicitly use natural units $\hbar=c=1$.} In the non-relativistic limit, it is convenient to decompose it as $\phi(t,\vec r)=1/\sqrt{2}m_{a}\left(e^{-im_{a}t}\psi(t,\vec r)+{\rm c.c.}\right)$ in terms of a new complex scalar field $\psi(t,\vec r)$ and its complex conjugate (c.c.) that we assume to be slowly varying in space and time, $\vert\nabla\psi\vert\ll m\vert\psi\vert$ and $\vert\dot{\psi}\vert\ll m\vert\psi\vert$. The field $\psi$ verifies, then, the Schr\"odinger-Poisson (SP) system of classical wave equations,
\begin{align}
i\frac{\partial}{\partial t}\psi(t,\vec r)  &= -\frac{1}{2m_{a}}\nabla^{2}\psi(t,\vec r)+m_{a}\Phi(t,\vec r)\psi(t,\vec r), \nonumber\\
\nabla^{2}\Phi(t,\vec r)&=4\pi G \vert\psi(t,\vec r)\vert^{2}.\label{eq:SP}
\end{align}
Space and time dependence of the field factorize in the quasi-stationary solutions of these equations,

\begin{align}
\label{eq:ansatz}
    \psi(\vec r,t)=\left(\frac{m_{a}}{\sqrt{4\pi G}}\right)\exp(-i\gamma m_{a} t)\chi(\vec r).
\end{align}
We have chosen the normalization such that the physical density of the field is  
\begin{align}
\label{eq:sc_dens}
    \rho(\vec r) = \frac{m^{2}_{a}m_P^2}{4\pi}\chi^{2}
(\vec r),
\end{align}
and have introduced the Planck mass $m_P=G^{-1/2}\approx10^{19}$ GeV for convenience. 
Assuming spherical symmetry, $\chi$ obeys a simpler version of the SP equations, 
\begin{align}
\partial_x^2[x\chi(x)]&=2x(\Phi(x)-\gamma)\chi(x),\nonumber\\
\partial_x^2[x\Phi(x)]&=x\chi^2(x),\label{eq:SP2}
\end{align}
where we have rescaled the radial extension as $x=m_a r$ ($x$ is dimensionless, while $r$ is the usual radial coordinate). The solutions to these equations that are regular at $x=0$ and approaching 0 at $x\to\infty$ are obtained from solving an eigenvalue problem for the parameter $\gamma$, which encodes the energy per unit mass. In particular, by integrating the energy density that can be derived from Eqs.~\eqref{eq:SP} over a virialized density distribution, it is easy to show that $\gamma = 3 E / M,$ where $E, M$ are the energy and mass of the distribution respectively~\citep{Bar:2018acw}. We also note  that the expressions in Eq.~\eqref{eq:SP2} are invariant with respect to a scale transformation, \begin{align}
\label{eq:scale}\Big(\,\chi(x),\,\Phi(x),\,\gamma\Big)\to\lambda^2\,\Big(\,\chi(\lambda x),\,\Phi(\lambda x),\,\gamma\Big),
\end{align}
and we can set $\chi(0)=1$ without loss of generality. 

\begin{figure}[t]
    \centering   \includegraphics[scale=0.57]{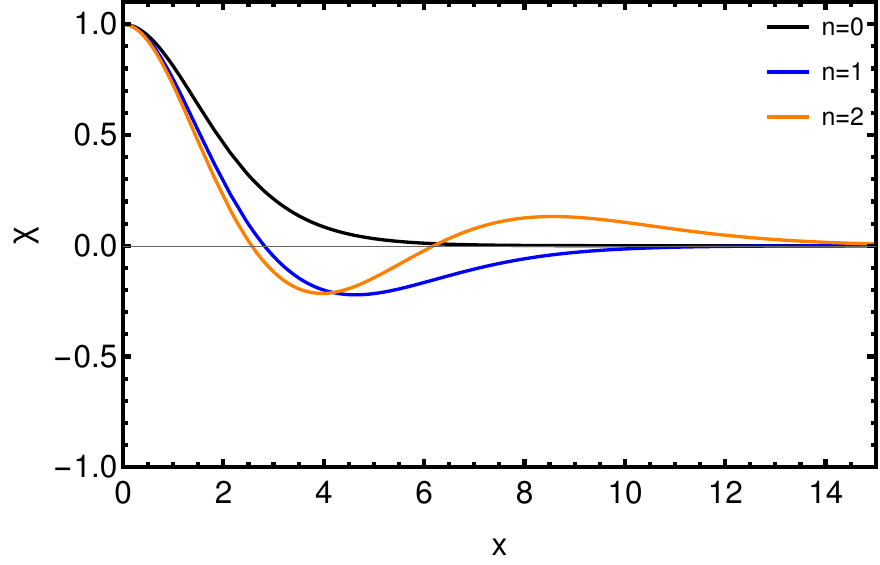}
    \caption{Solutions to Eq.~\eqref{eq:SP2} for the ground state and first two excitations, where the scale invariance was used to set $\chi(0)=1$.}
    \label{fig:soliton}
\end{figure}

The gravitationally bound eigenstates require $\gamma<0$ and they are characterized by their number of zeros, $n$, with the ground-state solution corresponding to $n=0$ and $\gamma\approx-0.69$. Eq.~\eqref{eq:SP2} needs to be determined numerically, but these expressions admit an approximate analytical solution around $x=0$. In particular, using~\cite{Moroz:1998dh},
we have:\ \begin{align}
\chi(r)=1-\frac{1}{3}(\gamma-\Phi(0))x^2+\mathcal O(x^4).
\end{align}
Hence, the resulting density distributions are flat at the center or ``cored.'' That of the ground state solution is referred to as a ``soliton'' in the literature. For completeness, we show in Fig.~\ref{fig:soliton} the solution  to Eqs.~\eqref{eq:SP2} $\chi(x)$ for the soliton and the two first excited states.

The FDM corresponds, then, to a realization of the solutions to the SP equations whose mass density at the center is dominated by the soliton. This is determined by the axion mass, $m_a$, and a rescaling parameter $\lambda$ that can be fixed using a physical property of the halo, such as the central density, via Eq.~\eqref{eq:sc_dens}, the size, or the mass of the core. Interestingly, this implies that we obtain ``soliton scaling relations'' between different observable properties of the cores for a given mass of the axion. In particular, we obtain:
\begin{align}
M_c &\simeq  5.4 \times 10^9\,\bigg(\frac{m_a}{10^{-23}\eV} \bigg)^{-2} \bigg(\frac{r_c}{\kpc} \bigg)^{-1}\MS, \label{eq:rcmc}\\
\rho_c&\simeq1.9\,\bigg(\frac{m_a}{10^{-23} \eV} \bigg)^{-2} \bigg(\frac{r_c}{\kpc} \bigg)^{-4}~~\MS\,\pc^{-3}, \label{eq:rhocrc}
\end{align}
where $r_c$ is the core radius defined as the radius where the density drops by a factor of 2, $M_c$ is the core mass, which is the mass enclosed within $r_c$, and $\rho_c$ is the central density. Therefore, for a fixed $m_a$ the soliton is heavier and denser the smaller the corresponding core. Lastly, the soliton solution can be approximated by an analytical profile that is obtained by fitting to results of simulations as~\cite{Schive:2014hza}:
\begin{align}
\label{eq:an_sol}
    \rho_{\text{sol}}(r) \approx \frac{\rho_c}{\big[1 + 9.1 \times 10^{-2} (r/r_c)^2 \big]^8}.
\end{align} 

The FDM can, in principle, be composed of different eigenstates of the SP equations but they will all eventually decay down to the solitonic ground state by expelling scalar field to infinity~\citep{Seidel:1993zk}. This process would relax the FDM distribution relatively quickly inside of a certain radius within the halo~\citep{Guzman:2006yc,Schwabe:2016rze,Hui:2016ltb,Li:2020ryg}, while the outer region is expected to form a virialized halo. This configuration of the FDM has been confirmed by simulations~\citep{Schive:2014dra,Schive:2014hza,Veltmaat:2018dfz,Mocz:2018ium,Nori:2020jzx,Chan:2021bja}. The outer halo follows the universal Navarro-Frenk-White (NFW) density profile characteristic of CDM~\citep{1997ApJ...490..493N}:
\begin{align}
\label{eq:densityNFW}
    \rho_{\rm NFW}(r) = \frac{\rho_{s}}{\frac{r}{r_{s}}\left(1+\frac{r}{r_{s}}\right)^{2}},
\end{align}
where $r_{s}$ and $\rho_{s}$ are the scale radius and density, respectively. Equation~\eqref{eq:densityNFW} parametrizes the mass distribution of a CDM halo formed by the gravitational collapse of an initial overdensity that is spherically symmetric and homogeneous. Following the convention of \cite{1998ApJ...495...80B}, the mass of the halo, $M_h$, is:
\begin{align}
\label{eq:bryan}
M_h &\equiv \frac{4}{3} \pi r_{\rm vir}^3 \zeta(z) \rho_{m0},\\
\zeta(z)&
\equiv \frac{ 18 \pi^2 + 82(\Omega_m(z) - 1) - 39(\Omega_m(z) - 1)^2}{\Omega_m(z)},
\end{align}
where $\rho_{m0}$ denotes the cosmological background matter density ($\rho_{m0} = \Omega_{m0}\rho_{c0}$, with $\rho_{c0} = 3 H_0^2 /8 G$), and $r_{\rm vir}$ is defined as the radius where the average density falls to a critical threshold $\zeta(z) \rho_{m0}$, marking the boundary of the halo. Throughout this paper, we set our cosmological parameters to the Planck 2015 cosmology results in \cite{ade2016planck}, with $h = 0.678, \Omega_{m0} = 0.308$ (all our calculations are at $z = 0$).~\footnote{Some of the authors that we cite and compare with used somewhat different cosmologies and related conventions. While this is a technical consideration one should be aware of, such modifications have little effect on the results, allowing for adequate comparisons to suit our purposes.} An additional characteristic property of the halo is the concentration parameter which is defined as $c=r_{\rm vir}/r_s$ and can be related to $\rho_s$ in Eq.~\eqref{eq:densityNFW} by using
\begin{align}
\label{eq:Relrhos_c}
\rho_s=\frac{\xi(z)\rho_{m0}}{3}\frac{c^3}{\log(1+c)-\frac{c}{1+c}}.
\end{align}

Another important result of the simulations has been finding a relation that links the mass of the soliton to the mass of the host halo. This ``core-halo relation'' was originally found by~\cite{Schive:2014dra,Schive:2014hza} and can be written as
\begin{equation}
\label{eq:Schive_2014}
M_c \approx 3.1\times 10^9 (1+z)^{1/2} \bigg(\frac{m_a}{10^{-23} \text{ eV}}\bigg)^{-1}
\bigg(\frac{M_h}{10^{12} \text{ M}_{\odot}}\bigg)^{1/3} \ \text{M}_{\odot},
\end{equation}
which depends on the redshift, $z$.
At the moment there is no comprehensive understanding of the physical mechanism underlying the core-halo relation, although some ideas have been put forward in~\cite{Schive:2014hza,Hui:2016ltb,Bar:2018acw,Eggemeier:2019jsu}. 
In fact, subsequent simulations have found that there is a significant scatter that was studied in detail in \cite{Chan:2021bja} by using a large sample of cosmological and soliton-merger simulations. This resulted in a more general relation that at $z = 0$ can be parametrized as:
\begin{equation}
\label{eq:ferreira}
M_c = \beta\,\bigg(\frac{m_a}{8 \times 10^{-23} \text{ eV}} \bigg)^{-3/2}
+ \bigg( \frac{M_h}{\gamma} \bigg)^{\alpha}\bigg(\frac{m_a}{8 \times 10^{-23} \text{ eV}}\bigg)^{3(\alpha - 1)/2} \; \text{M}_{\odot},
\end{equation}
where $\beta = 8.00^{+0.52}_{-6.00} \times 10^6 \text{ M}_{\odot}, \;
\log_{10} ( \gamma / \text{ M}_{\odot} ) = -5.73 ^{+2.38}_{-8.38} \; $ and 
$\alpha = 0.515^{+0.130}_{-0.189}$ are the best-fit parameters obtained in \cite{Chan:2021bja} with their respective errors determining the scatter. 

\begin{table*}[t]
\caption{Priors used to fit the DM and baryon contributions to the rotation curve data from the LTs catalog.} 
    \centering
    \setlength{\tabcolsep}{0.7em}
\renewcommand{\arraystretch}{1.5}
  \setlength{\arrayrulewidth}{.30mm}
    \begin{tabular}{c|c c c}
     \hline\hline
        \textbf{Parameter} & \textbf{Units} & \textbf{Prior Type} & \textbf{Range}  \\ \hline 
        $m_{a}$ & eV & flat &  $[10^{-25}, \; 10^{-19}]$
        \\
        $M_c$ & $\text{M}_{\odot}$  & flat &  Range spanned by Eq.~\eqref{eq:ferreira} \\

        $M_{\text{sol}}$ &  $\MS$ & flat & $[10^{4.5}, \; M_h]$ \\
        
        $v_{\text{vir}}$ &  km s$^{-1}$ & flat & $[5, \; 130]$ \\
        $c$ & none & flat & $[1, \; 100]$ \\
       $\MSt$ & $\text{M}_{\odot}$ & log-normal & 
       \begin{tabular}{@{}c@{}}$\mu = M_{\star0} ,$ 
       \\  $  \sigma = \frac{1}{2}\log_{10}\frac{ M_{\star0} + \delta M_{\star+}}{M_{\star0} - \delta M_{\star-}}$ dex
       \end{tabular}
       \\
       
       $(\MSt + M_{\text{gas}})/M_h$ &  none & flat & $[0, \; 0.2]$ \\
       $r_t$ &  kpc & flat & $ \geq r_c$ (if $r_t < r_{\text{max}}$) \\

       $i$ & degrees & normal & 
       \begin{tabular}{@{}c@{}} $\mu = i_0,$ 
       \\  $  \sigma = (\delta i_+ + \delta i_-)/2 $
       \end{tabular}
       \\
        \hline\hline
    \end{tabular}
\tablefoot{The variables used directly in the MCMCs are $\log_{10}m_a, \; \log_{10} M_c, \; \log_{10} c, \; \log_{10} M_h, \; \log_{10} \MSt,$ and $\log_{10} (\sin i)$ in the case of inclination rogues, with all other variables being derived from these. $M_{\rm sol}$ corresponds to the total mass of the soliton and it is related to $M_c$ via $M_c\simeq 0.24\,M_{\rm sol}$. $r_{\rm max}$ denotes the maximum observable radius from the rotation curve data. In the case of stellar mass, $M_{\star0}$ and $\delta M_{\star\pm}$ denote the central values with errors reported in \cite{2017MNRAS.467.2019R} from population synthesis modeling, while  $i_0, \delta i_{\pm}$ correspond to the same quantities for the case of the inclination that was obtained through their analysis of rotation curve data.}
\label{tab:priors}
\end{table*}  

Another feature of FDM halos is that the characteristic flatness in their density profiles is not limited to the core region where the soliton dominates, but is also reflected in their NFW-like tails at greater radii. This manifests in the form of a smaller concentration parameter compared to the typical CDM halo. This leads to a modified concentration - mass relation characterized by a suppression on lower mass scales, which, following \cite{2022MNRAS.515.5646D}, can be parametrized as:
\begin{equation}
c^{\text{FDM}}(M_h) = \bigg[1 + \gamma_1 \bigg(\frac{M_0}{M_h}\bigg)\bigg]^{-\gamma_2} c^{\text{CDM}}(M_h),
\label{eq:cm}
\end{equation}
where $\gamma_1 = 15$,  $\gamma_2 = 0.3$ and we have the characteristic mass scale $M_{0}=1.6\times10^{10}\left(m_{a}/10^{-22}\text{ eV}\right)^{-4/3} \MS$, while $c^{\text{CDM}}(M_h)$ denotes the concentration parameter in CDM. As we show later in this work, the CDM relation is modeled using the relation from \cite{2014MNRAS.441.3359D} as:
\begin{equation}
\log_{10} c^{\text{CDM}} = 1.025 - 0.097 \log_{10} \Big(M_h / [10^{12} h^{-1} \MS] \Big),
\label{eq:dutton}
\end{equation}
with a scatter of $0.16$ dex.~\footnote{This was not the scatter originally reported in \cite{2014MNRAS.441.3359D} ($0.11$ dex), but we instead decided to use the more conservative one suggested by \cite{diemer2015universal}. We also note that the CDM relation used in \cite{2022MNRAS.515.5646D} was somewhat different from the one used here. However, we checked that, when applied to Eq. \eqref{eq:cm}, the resulting FDM relation was very similar.}

As a point of clarification, we note that 
the concentration parameter is defined in Eqs.~\eqref{eq:densityNFW} and \eqref{eq:Relrhos_c} with respect to the NFW component of the density profile. This is irrespective of whether the point $r = r_s$, which can be generally defined as the point where the logarithmic slope equals $-2$, is located in the NFW-like region or not, meaning that the two definitions are not necessarily equivalent.
Nevertheless, as we discuss in Sect.~\ref{sec:fits} for the halos that we model in this work, we find that the median values of $r_s$ lie in the NFW-like regions and the two definitions are consistent with each other in our case. 

A final property that is characteristic of FDM and is particularly useful in the context of small-scale structure issues of $\Lambda$CDM is the suppression of the primordial power spectrum that inhibits the formation of halos below a certain mass scale~\citep{1985MNRAS.215..575K,Hu:2000ke}. Moreover, the subhalo population is expected to be further reduced by dynamical effects such as tidal disruption~\citep{Hui:2016ltb}. This offers an alternative ``exotic'' solution to the missing satellites problem and, by extension, to the associated too big to fail problem. One observable statistic that is relevant to test this suppression of formation of structures at small scales is the halo mass function (HMF) for low halo masses of $M_h\lesssim10^{10}~\MS$. This can be derived by a simple parametrization derived from numerical simulations~\citep{Schive:2015kza,May:2021wwp,May:2022gus}:
\begin{align}
\label{eq:HMF-FDM}
    \left.\frac{dn}{dM_h}\right\vert_{\rm FDM}=\left(1+\left(\frac{M_{h}}{M_0}\right)^{-\alpha}\right)^{-\beta} \left.\frac{dn}{dM_h}\right\vert_{\rm CDM},
\end{align}
where, as previously for the concentration - mass relation, $M_{0}=1.6\times10^{10}\left(m_{a}/10^{-22}\text{ eV}\right)^{-4/3} \text{M}_{\odot}$, $\alpha=1.1$ and $\beta=2.2$. The characteristic mass $M_{0}$ defines the regime below which the HMF starts dropping substantially. 

\begin{table*}[t]
\caption{Median and 68\% confidence level (CL) of the  posterior distributions of the parameters $m_a$, $M_c$, $c$, $M_h$, and $\MSt$, and the reduced chi-square of the maximum posterior fit to the LTs catalog, including some of the rogues.}
\centering
\setlength{\tabcolsep}{0.7em}
\renewcommand{\arraystretch}{1.5}
  \setlength{\arrayrulewidth}{.30mm}
\centering
  \begin{tabular}{lcccccc|ccc}
  \hline\hline
   Galaxy & $\MSt$   & $m_a$  & $M_c$ & $M_h$  & \multirow{2}{*}{$c$} &  \multirow{2}{*}{$\chi_{\nu}^2$} & $r_c$  & $r_t$ & $10^3\times\rho_c$\\
   &[10$^7$ $\MS$]&[$10^{-23}$ eV]&[$10^{8}$ $\MS$]&[$10^{10}$ $\MS$] &  &  & $\kpc$  & $\kpc$ & $\MS/\pc^3$\\ \hline
  NGC 2366&$6.5^{+1.7}_{-1.3}$&$1.26^{+0.13}_{-0.13}$&$11.3^{+1.7}_{-1.4}$&$2.5^{+4.6}_{-1.3}$&$5^{+15}_{-4}$& $1.29$ & $3.00^{+0.24}_{-0.21}$ &
  $6.33^{+0.00}_{-0.77}$ & $15.0^{+1.6}_{-1.5}$\\
  DDO 168 &$5.6^{+1.4}_{-1.1}$&$1.26^{+0.27}_{-0.26}$&$12.3^{+4.7}_{-2.9}$&$3.2^{+6.3}_{-1.8}$&$5^{+15}_{-4}$& 0.56 & $2.76^{+0.44}_{-0.32}$& $4.69^\dagger$ & $20.9^{+2.9}_{-2.7}$\\
   DDO 52 &$5.4^{+1.4}_{-1.2}$&$1.72^{+0.43}_{-0.31}$&$7.5^{+2.3}_{-1.9}$&$2.2^{+3.3}_{-1.3}$&$7^{+12}_{-5}$& $0.13$ & $2.45^{+0.36}_{-0.35}$& 
   $5.4^{+0.00}_{-1.9}$&$18.2^{+4.0}_{-2.9}$\\
    DDO 87 &$3.37^{+0.88}_{-0.69}$&$1.77^{+0.37}_{-0.32}$&$7.0^{+2.1}_{-1.5}$&$1.9^{+3.1}_{-1.1}$&$6^{+13}_{-4}$& $0.27$ & $2.47^{+0.37}_{-0.31}$& $5.2^{+0.00}_{-1.2}$
    & $16.5^{+3.1}_{-2.6}$\\
       DDO 126 &$1.58^{+0.41}_{-0.33}$&$2.22^{+0.53}_{-0.53}$&$4.9^{+2.2}_{-1.2}$&$1.2^{+2.2}_{-0.6}$&$5^{+14}_{-3}$& $0.49$ & $2.25^{+0.41}_{-0.29}$& 
       $3.3^\dagger$& $15.4^{+2.4}_{-2.0}$\\
WLM &$1.92^{+0.56}_{-0.42}$&$3.77^{+0.49}_{-0.43}$&$2.61^{+0.46}_{-0.40}$&$0.9^{+1.7}_{-0.5}$&$7^{+12}_{-5}$& $0.68$ & $1.46^{+0.12}_{-0.11}$ & $2.87^{+0.00}_{-0.49}$&$30.2^{+2.9}_{-2.5}$\\       
  DDO 154 &$1.00^{+0.28}_{-0.22}$&$1.83^{+0.13}_{-0.09}$&$6.94^{+0.47}_{-0.61}$&$1.7^{+2.8}_{-0.8}$&$7^{+21}_{-5}$& $0.94$ & $2.33^{+0.09}_{-0.12}$ & 
  $5.2^{+1.1}_{-1.1}$&$19.6^{+1.1}_{-1.1}$\\
    UGC 8508 &$0.78^{+0.21}_{-0.17}$&$6.9^{+2.8}_{-2.6}$&$1.4^{+1.3}_{-0.5}$&$0.7^{+1.8}_{-0.5}$&$6^{+15}_{-4}$& $0.083$ & $0.83^{+0.27}_{-0.16}$& 
    $1.39^\dagger$& $87^{+21}_{-17}$\\
  CvnIdwA &$0.41^{+0.11}_{-0.09}$&$8.9^{+5.3}_{-5.0}$&$0.7^{+1.5}_{-0.3}$&$0.22^{+0.67}_{-0.15}$&$5^{+12}_{-3}$& $0.43$ & $1.02^{+0.61}_{-0.26}$& 
  $1.57^\dagger$& $22.8^{+7.2}_{-5.4}$\\
  DDO 210 &$0.068^{+0.019}_{-0.015}$&$9^{+55}_{-8  }$&$1^{+22}_{-1}$&$0.3^{+8.6}_{-0.3}$&$3.3^{+5.8}_{-1.8}$& $0.71$ & $0.9^{+1.8}_{-0.6}$& $0.44^\dagger$ &
  $45^{+32}_{-19}$\\
   &&&\emph{- rogues -}&&& &&&\\
 DDO 50 &$10.4^{+2.6}_{-2.1}$&$5.4^{+1.0}_{-0.8}$&$1.72^{+1.28}_{-0.57}$&$1.3^{+1.3}_{-0.57}$& $4.1^{+3.6}_{-2.0}$ &$0.53$ & $1.09^{+0.14}_{-0.13}$
 & $2.66^{+0.52}_{-0.52}$&  $48^{+15}_{-12}$
 \\   
 DDO 133   &$3.18^{+0.85}_{-0.68}$&$3.39^{+0.58}_{-0.50}$&$3.47^{+0.84}_{-0.68}$&$1.4^{+2.8}_{-0.9}$&$7^{+13}_{-5}$&$0.33$& $1.36^{+0.15}_{-0.14}$ & 
 $2.8^{+0.00}_{-0.51}$& $49.9 ^{+9.0}_{-7.3}$\\
NGC 6822$^{*}$ &$1.80^{+0.27}_{-0.24}$&$1.86^{+0.11}_{-0.10}$&$7.45^{+0.57}_{-0.58}$&$6.7^{+3.6}_{-3.0}$&$10.9^{+4.8}_{-2.2}$& $1.84$ & $2.11^{+0.08}_{-0.08}$ & $3.19^{+0.28}_{-0.33}$ & $28.5^{+1.3}_{-1.3}$\\
  \hline
\hline
\end{tabular}
\tablefoot{ We also include the derived core radius, transition radius and central density of the distribution. We flag the galaxy NGC~6822 with * because it was not initially selected as a rogue galaxy (see text for details). The transition radii with an upper uncertainty equal to 0 indicates that less than 84\% of the posterior distribution lies below the maximum observable radius $r_{\rm max}$, while for those not showing an uncertainty band and marked by $\dagger$, at least 84\% of the posterior distribution lies above $r_{\rm max}$.}
\label{tab:ResultsLTs}
\end{table*}

\section{Fitting FDM to the LITTLE THINGS in 3D}
\label{sec:fits}
\subsection{LITTLE THINGS in 3D}
\label{sec:LTs3D}

Local Irregulars That Trace Luminosity Extremes, The HI Nearby Galaxy Survey (LITTLE THINGS) is a Very Large Array HI survey which has produced high-resolution rotation curves for 26 nearby dwarf irregular galaxies  within 11 Mpc of the local volume~\citep{2015AJ....149..180O, 2012AJ....144..134H}. Most of these curves are able to resolve the inner part (within $\sim 1$~kpc of the center) of the galaxies, making this survey an optimal laboratory to investigate the core-cusp problem. We considered a representative and robust subset 
of these galaxies, with stellar masses in the range of $10^5\,\MS\lesssim \MSt\lesssim 10^{8}\,\MS$,  obtained in~\cite{2017MNRAS.466.4159I,2017MNRAS.467.2019R} by reanalyzing the data cubes with 3D reconstruction techniques. We will refer to this as the LITTLE THINGS in 3D (LTs) sample. The data set consists of circular velocities after the observed velocities have been corrected  for the pressure support of the random gas motions.~\footnote{The correction takes into account that both the velocity dispersion and the velocity rotation are important to trace the gravitational potential. This effect is appreciable when the dispersion velocities are comparable with the maximum of the velocities in the rotation curves, which is the typical behavior in many LTs dwarf galaxies. The correction is equivalent to the asymmetric drift correction for the stars.} A fraction of these galaxies termed ``rogues'' were flagged and excluded from the core LTs data set because they were nearly face-on (five inclination rogues), their measured distance was not well-determined (one distance rogue) or because they presented indications of being far from equilibrium (two disequilibrium rogues). Our fiducial cosmological analysis will include 11 of the remaining galaxies (first corresponding lines in Table 2 of~\cite{2017MNRAS.467.2019R}). For these galaxies, we will consider the measured rotation velocities in the whole measured range, which are quite regular even if some of them manifest the effect of HI holes, which could indicate recent (fast expanding) or old (quiescent) stellar activity. The disequilibrium rogues present irregular rotation curves while the distance rogue, DDO~101, could have a stellar mass quite above $\MSt\sim10^8\,\MS$, depending on the assumed distance, see~\cite{2016MNRAS.462.3628R}. Therefore we did not include any of these galaxies in our analysis. Finally, we did not include the inclination rogues in our fiducial analysis either, but we do show the results obtained for DDO~133 and DDO~50 for illustration purposes later in this work.\footnote{The galaxies DDO~47 and IC~1613 in the LTs sample present, in addition, significant irregularities in the rotation curves correlated with fast expanding HI bubbles, while DDO~53 is almost face on.} We also note that some galaxies in our sample have significant errors in the inclination values. However, this is self-consistently accounted for in the rotation curve errors~\citep{2017MNRAS.466.4159I}, so that it is unnecessary (and potentially inconsistent) to use the inclination angle as a fitting parameter for LTs.

\subsection{Fitting strategy}
\label{sec:FitStrategy}

The FDM density model is a piecewise function of the soliton and NFW profiles matched at a transition radius $r_{t}$, namely, $\rho_{\rm sol}(r_{t})=\rho_{\rm NFW}(r_{t})$. In this model, the soliton describes the inner part of the galaxy, while a NFW halo does the outer part:
\begin{align}
\label{eq:densityFDM}
    \rho_{\rm FDM}(r)=\begin{cases}
    \hspace{0.1cm}\rho_{\rm sol}(r) \hspace{0.9cm} r\leq r_{t},\\
    \hspace{0.1cm}\rho_{\rm NFW}(r) \hspace{0.65cm} r> r_{t}.
\end{cases}    
\end{align}
We use the numerical ground-state solution to the SP Eqs.~\eqref{eq:SP2} for $\rho_{\rm sol}(r)$ and the NFW density profile in Eq.~\eqref{eq:densityNFW} for $\rho_{\rm NFW}(r)$. We note that the numerical solution for $\rho_{\rm sol}(r)$ that we use is very similar to the phenomenological function in Eq.~\eqref{eq:an_sol} obtained in~\cite{Schive:2014dra,Schive:2014hza} from simulations and which is the basis for most of the analyses of FDM.  

Overall, the FDM density profile has four free parameters that we chose to be: \textbf{(1)} the axion mass $m_a$, \textbf{(2)} the solitonic core's mass $M_c$, \textbf{(3)} the concentration parameter $c$ and, \textbf{(4)} the halo mass, $M_h$, defined as in Eq.~\eqref{eq:bryan}. Other parameters such as the core radius, $r_c$, the transition radius, $r_t$, the central density, $\rho_c$, or the virial radius $r_{\rm vir}$ can be obtained from these four parameters. In particular, the transition radius, $r_{t}$, can be determined by searching
the first point in $r,$ where $\rho_{\rm sol}(r)=\rho_{\rm NFW}(r)$ starting  from $r\to\infty$. 

There are two major approximations when using the FDM density profile in Eq.~\eqref{eq:densityFDM}. Firstly, simulations have shown that the central soliton can still be affected significantly by interference with residual excited states of the SP equations. This produces $\mathcal O(1)$ oscillations in the central density, which is an effect not captured by the simple solitonic distribution~\citep{Veltmaat:2018dfz,Li:2020ryg}. This also includes contributions from non spherically symmetric eigenstates of the SP equations, as described in~\cite{Vicens:2018kdk,Li:2020ryg}, for instance.
Secondly, the SP expressions in Eq.~\eqref{eq:SP} describe configurations of the FDM field neglecting the contribution of the baryons to the gravitational potential. The baryonic contributions, including baryonic SN feedback, have been recently taken into account in simulations~\citep{2018MNRAS.478.2686C,Mocz:2019pyf,Mocz:2019uyd,Veltmaat:2019hou} with the general result that, for the same given $m_a$, a soliton of a given mass is still formed but more compressed and denser than in its pure FDM configuration. This effect should be especially relevant for galaxies whose central masses are dominated by the baryons and it can be estimated semi-analytically as discussed in~\cite{Bar:2018acw,Bar:2019bqz} and followed in Sect.~\ref{sec:discussion}.   

Moreover, the structure of the boundary between the soliton and the halo at $r_t$ can be modeled differently. This can have a non-trivial effect~\citep{Bernal:2017oih,Vicens:2018kdk} especially if it introduces further constraints between the four parameters of the basic model.  

The theoretical radial profile of the total circular velocity squared is given by the sum of the various mass contributions to the gravitational potential,
\begin{align}
\label{eq:modelpredict}
    V_{\rm th}^{2}(r;\,\vec\theta_{\rm FDM};\MSt)= V_{\rm FDM}^{2}(r;\,\vec\theta_{\rm FDM})+V_{\star}^{2}(r;\,\MSt)+V_{\rm gas}^{2}(r).
\end{align}
In this equation $V^2_{\rm FDM}(r;\,\vec\theta_{\rm FDM})=GM_{\rm FDM}(r;\,\vec\theta_{\rm FDM})/r$, where $M_{\rm FDM}(r;\,\vec\theta_{\rm FDM})$ is the enclosed mass within radius $r$ obtained from the density distribution in Eq.~\eqref{eq:densityFDM} and $\vec\theta_{\rm FDM}=(m_{a},M_{c},c,M_h)$ is the four-vector of FDM fitted parameters. To describe the baryonic components in Eq.~\eqref{eq:modelpredict}, we follow the reference~\cite{2017MNRAS.467.2019R} and model the distributions of gas and stars as exponentially thin discs~\citep{Binney2008},
\begin{align}
\label{eq:LH}
  V^2_{\star/{\rm gas}}=\frac{2 GM_{\star/{\rm gas}}}{R_{\star/{\rm gas}}}\mathcal{I}(y), 
\end{align}
with $M_{\star/{\rm gas}}$ and $R_{*/{\rm gas}}$ being the mass and the exponential scale length  of the disc, respectively, and with  $I(y)=y^{2}\left[I_{0}(y)K_{0}(y)-I_{1}(y)K_{1}(y)\right]$, where $I_{0,1}(y)$ and $K_{0,1}(y)$ are Bessel functions that depend on the dimensionless radius parameter $y =r/(2R_{\star/{\rm gas}})$.  We fixed the values of $R_{\star}$, $R_{\rm gas}$, and $M_{\rm gas}$ to the values reported in~\cite{2017MNRAS.467.2019R} and we added the stellar mass $\MSt$ as a new parameter in the fit.

\begin{figure*}[hbt!]
\centering
\includegraphics[width=0.45\textwidth]{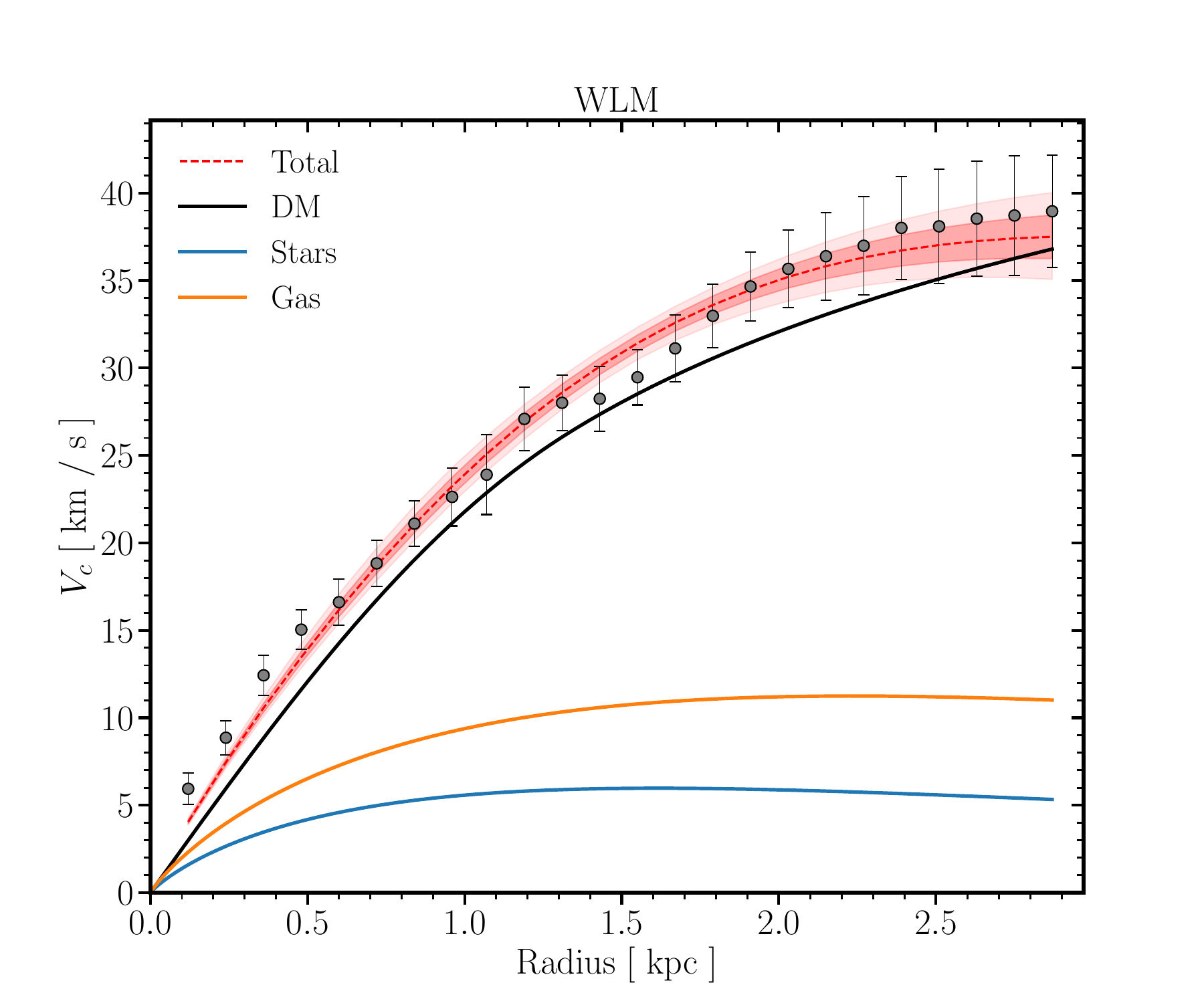} 
\includegraphics[width=0.45\textwidth]{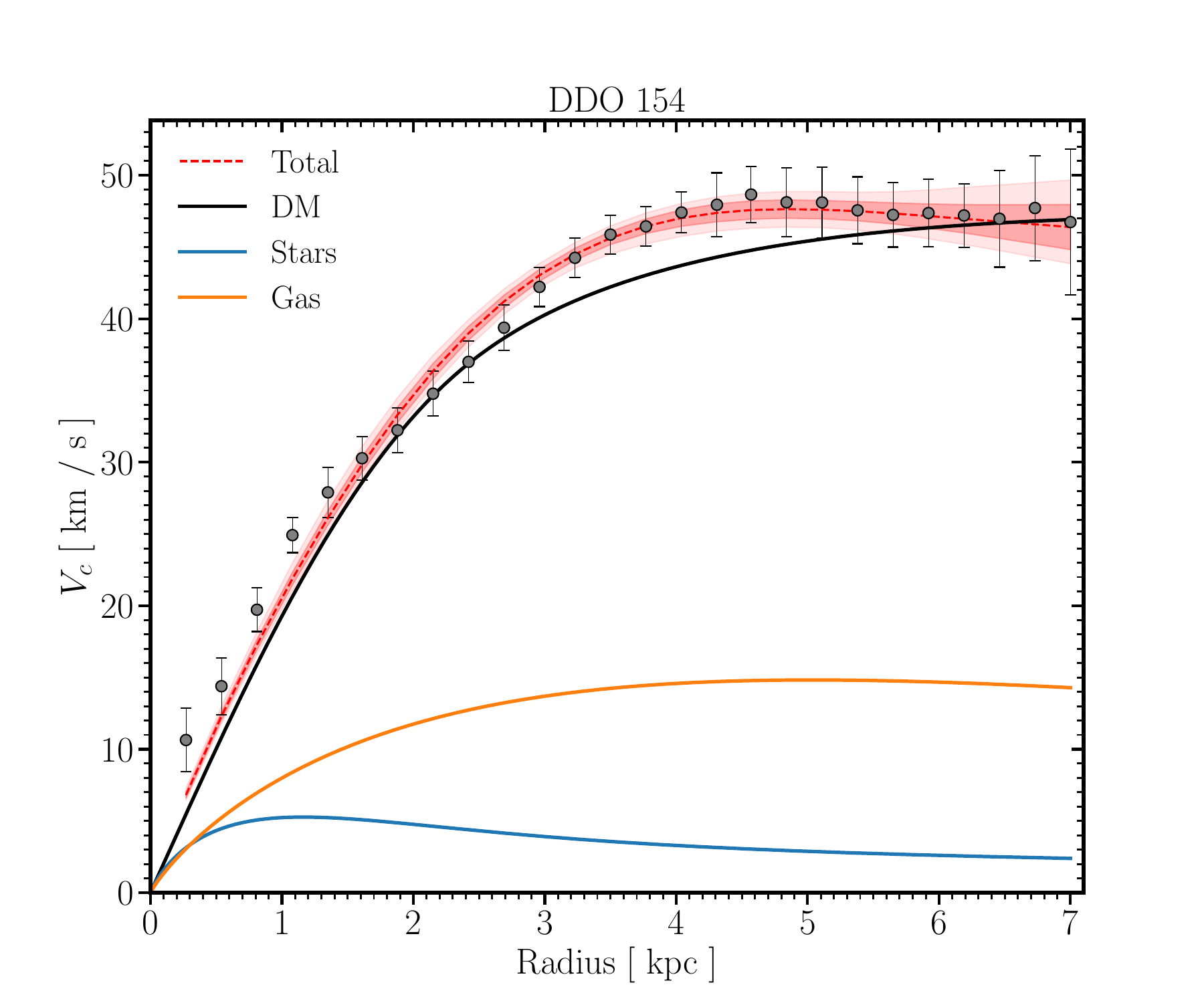}
\caption{Rotation curves in FDM for two representative galaxies in the LTs catalog, WLM, and DDO~154. We plot the DM, gas, and stellar components obtained from the fits (maximum posterior values) and compared to the data. The red dashed lines are the median of the posterior distribution for the total velocity and the colored band is the corresponding uncertainty at $68\%$ and $95\%$ CL.}
  \label{fig:my_rc}
\end{figure*}

\begin{figure*}[hbt!]
\centering
\includegraphics[width=0.45\textwidth]{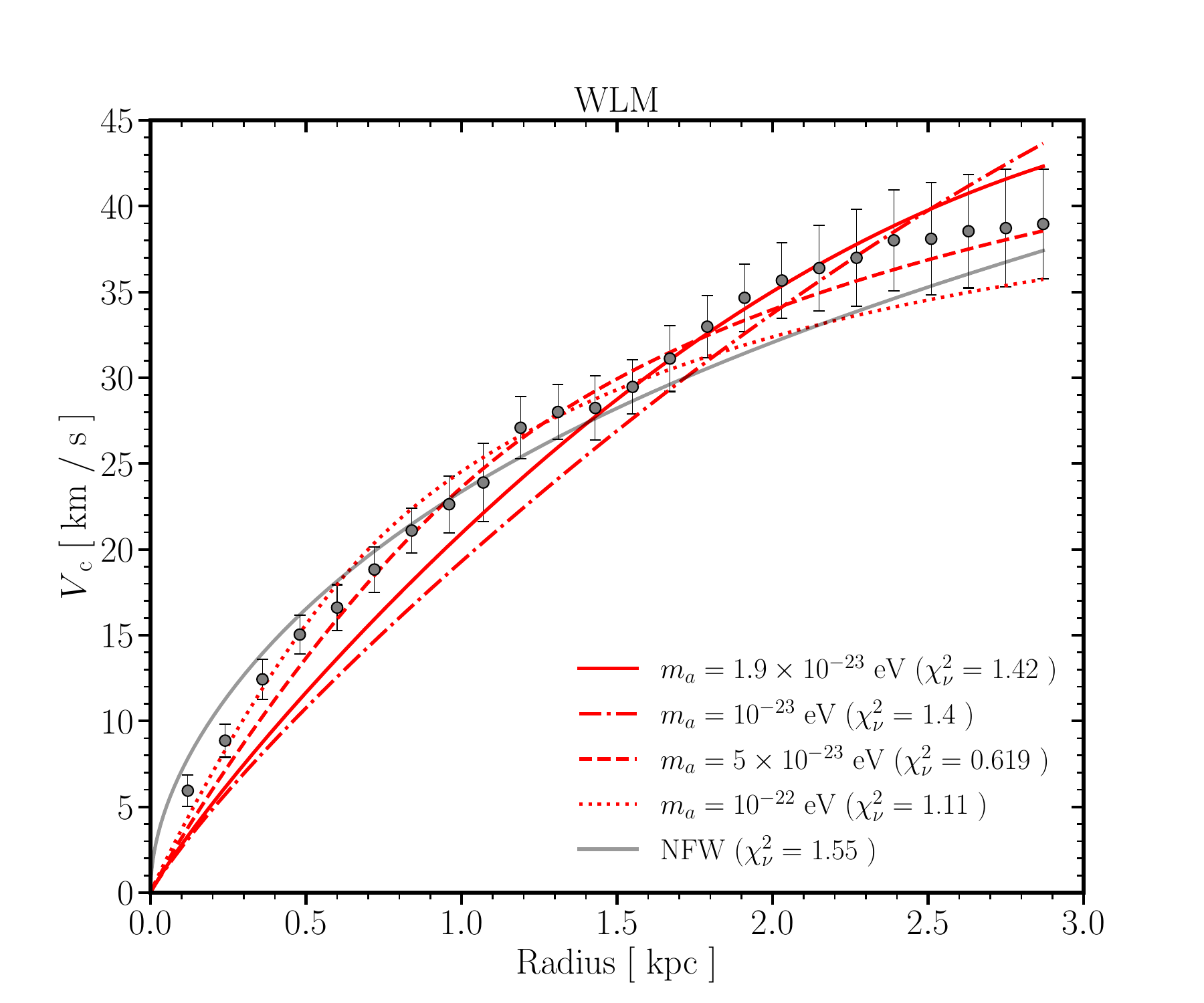} 
\includegraphics[width=0.45\textwidth]{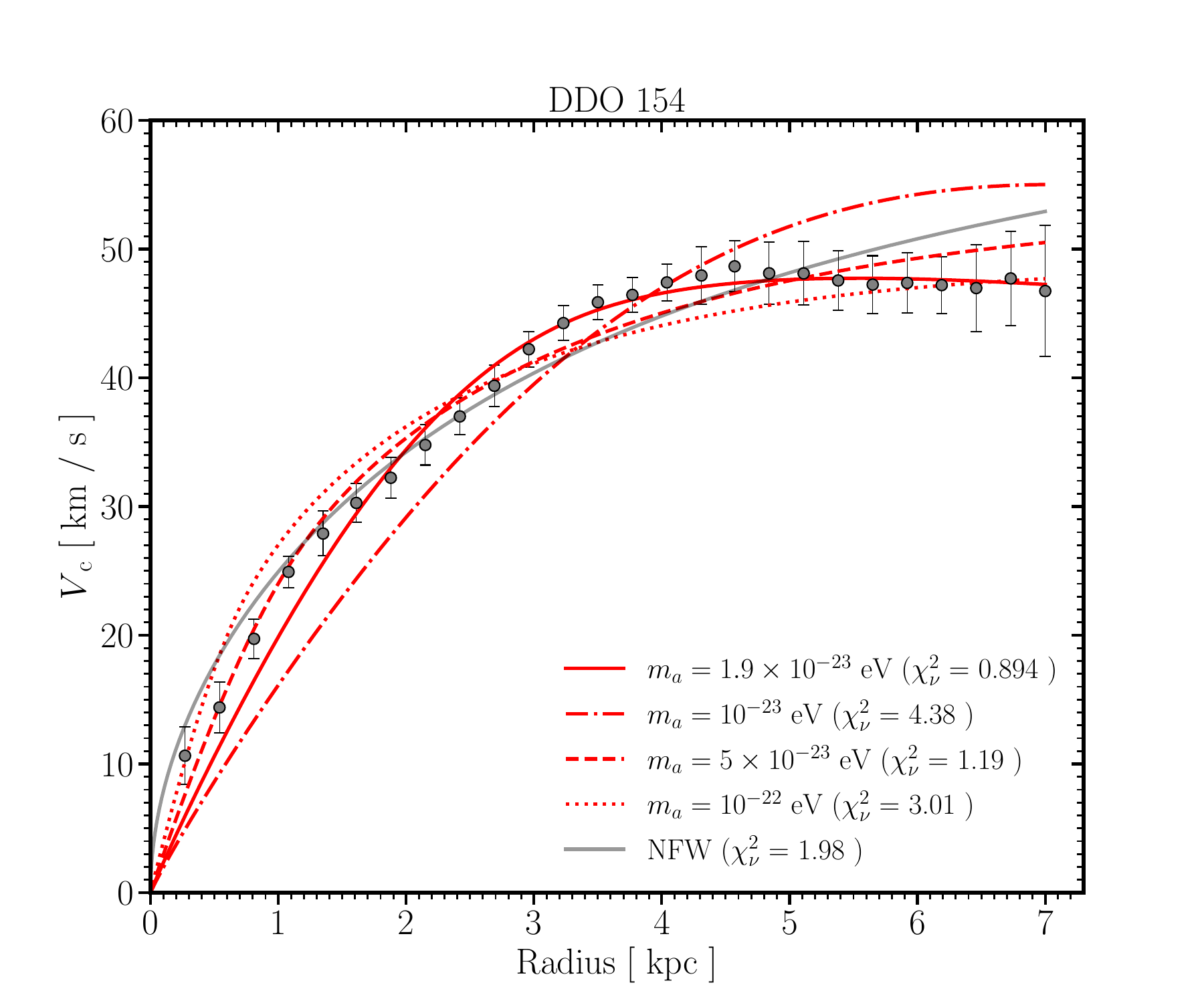}
\caption{Rotation curves in FDM for two representative galaxies in the LTs catalog: WLM and DDO~154. We plot the total circular velocity and reduced chi-squares resulting from the maximum posterior fits to the data when fixing the axion mass at different values. We compare the results also to the fit to a NFW profile.}
  \label{fig:my_rc2}
\end{figure*}

The theoretical velocity is compared to the observed rotation curves using a log-likelihood
\begin{align}
\label{eq:loglike}
    \ln \mathcal{L}(\vec\theta_{\rm FDM},\MSt)=-\frac{1}{2}\sum_{i}\left(\frac{V_{\rm obs}(r_{i})-V_{{\rm th}}(r_i;\, \vec\theta_{\rm FDM};\,\MSt)}{\delta V_{\rm obs}(r_{i})}\right)^2,
\end{align}
(up to an additive constant), where $V_{\rm obs}(r_{i})$ and $\delta V_{\rm obs}(r_{i})$ are the velocity and corresponding uncertainty of the $i$-th data point in a galaxy’s rotation curve (the summation is over the radial data bins). We also assumed uncorrelated Gaussian errors for the data points distributions. Finally, for the inclination rogues, DDO~50 and DDO~133, we also fit the inclination as an additional parameter (see Appendix~\ref{app:info_fits:irogues} for details).

To  effectively scan the parameter space of the FDM model, we used the \texttt{emcee} package, an affine-invariant ensemble sampler for Markov chain Monte Carlo (MCMC) \citep{Foreman_Mackey_2013}, based on the algorithm proposed by \cite{Goodman2010}. 
To deal with possible multi-modal posteriors and enhance sampling efficiency, we implemented a combination of MCMC moves, which significantly reduces autocorrelation times.~\footnote{See~\href{https://emcee.readthedocs.io/en/stable/tutorials/moves/}{\texttt{emcee.readthedocs.io}}, 
where this is explained further using the same combination of moves.} We also initialized MCMCs under different local optima to ensure that the posterior distributions converged to the global one. We ran the MCMCs with 48 walkers for a total of $20 \, 000$ steps, discarding the first $10 \, 000$ as a conservative burn-in period. We found these to produce consistent results under different runs, as well as meeting the criterion of exceeding $100$ autocorrelation times, with acceptance fractions in the $\sim 0.2 - 0.4$ range, in agreement with the criteria established in \cite{Foreman_Mackey_2013}.

The priors used for
the initial analysis in this study (where the FDM mass is allowed to vary for each galaxy) are summarized in Table~\ref{tab:priors}.  The prior on $\MSt$ is designed to yield results consistent with stellar population synthesis models, as reported in \cite{2016MNRAS.462.3628R}, using the same central values and defining the errors given as the 68 \% CL bounds under a log-normal distribution. A similar type of prior was used for the angle $i$ for inclination rogues, using a normal distribution instead of a log-normal one. 
Priors are generally designed to be very loose, covering ranges well beyond expected values and have, for the most part, little effect on the fits. The less trivial priors that are able to meaningfully constrain the parameter space of the FDM distribution, besides the priors on $\MSt$, are the core-halo relation from Eq.~\eqref{eq:ferreira} and demanding $r_t \ge r_c,$ (when $r_t < r_{\rm max},$ where $r_{\text{max}}$ is the maximum radius where rotation curves are measured), which are both expected from simulations. Fits with $r_t \ge r_{\text{max}}$ would yield identical soliton-dominated fits (for a given $m_a$ and $M_c$). For the core-halo relation we set $z=0$ using Eq.~\eqref{eq:ferreira}, where we have assumed that the uncertainty on the formation history of the halo, see~\cite{Burkert:2020laq}, is within the intrinsic scatter comprised by this formula. 

\begin{figure*}[!htb]
    \centering
   \includegraphics[scale=0.27]{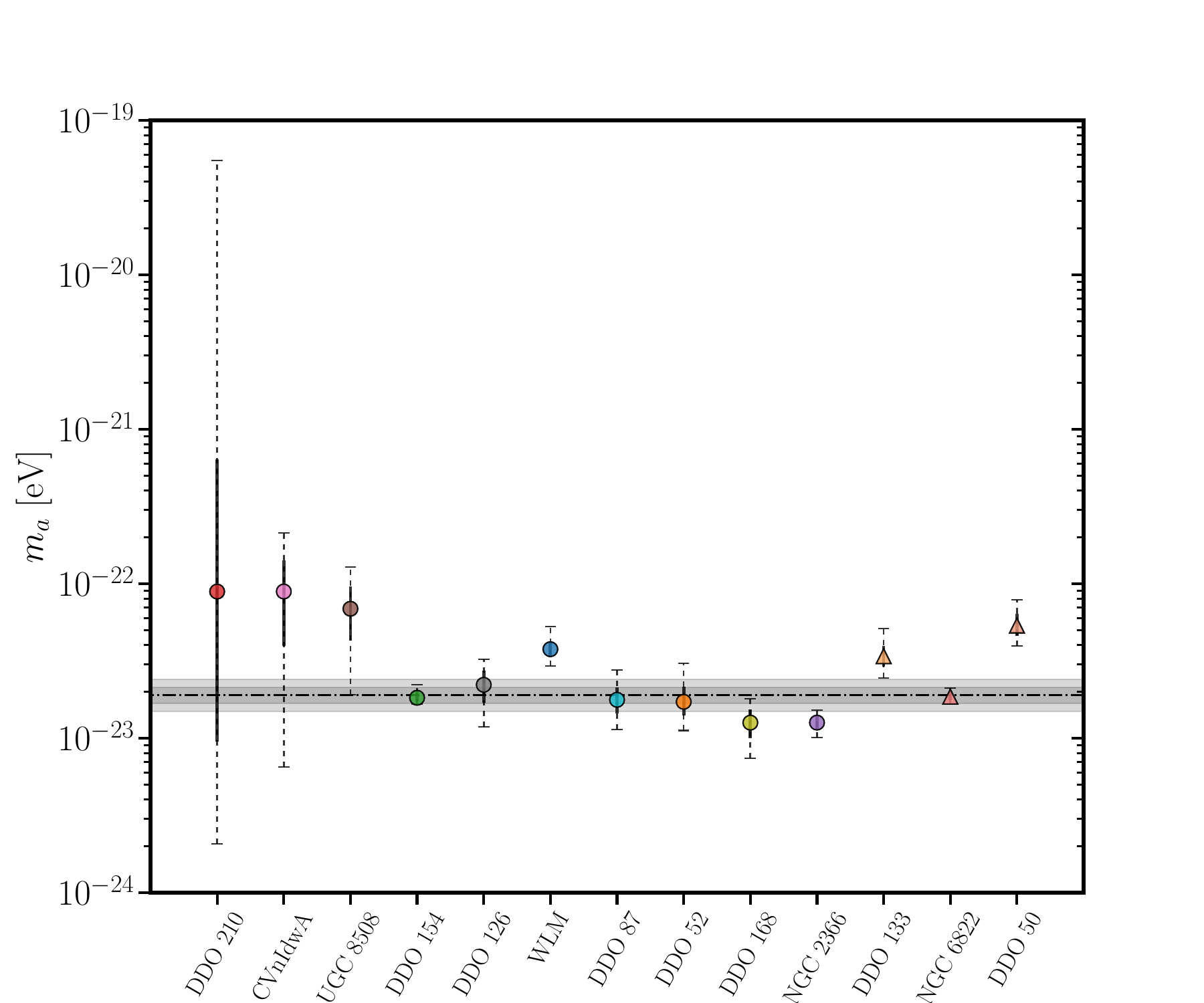}  
\hspace{1.cm}\includegraphics[scale=0.27]{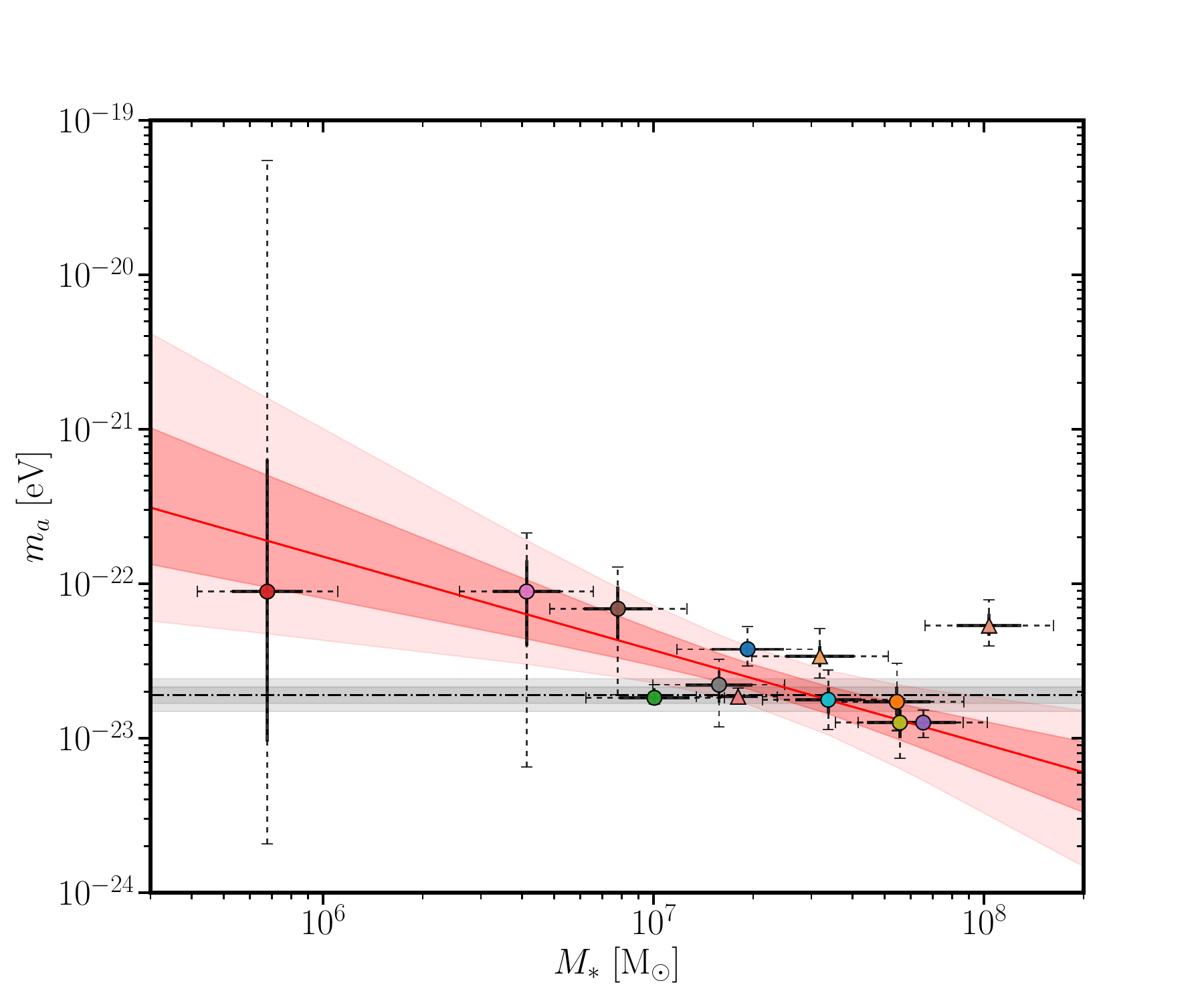}
    \caption{Axion masses obtained from the posterior distributions of the fits to the rotation curves of the LTs galaxies where  we show medians and $68\%$ and $95\%$ CL errors. The 10 galaxies entering our fiducial analysis are represented by circles while the rogues are represented by triangles. The dot-dashed line and bands represent the mean, $68\%$ and $95\%$ CL regions of a weighted average to these determinations (see Eq.~\eqref{eq:benchmarkmass} and associated discussion). \textit{Left:} Galaxies are ordered by the central value of the stellar mass reported in \cite{2017MNRAS.467.2019R}, which where found to be in good agreement with our posterior distributions. Stellar masses are increasing from left to right.
    \textit{Right:} Empirical fit to a power law between $m_a$ and $\MSt$ showing evidence for a correlation between the two as extracted from the galaxies in our sample. The red solid line denotes the maximum posterior fit with the $68\%$ and $95\%$ CL regions.
    \label{fig:AxionMass}}
\end{figure*}

\section{Results}
\label{sec:results}
\subsection{Fits to rotation curves}
\label{sec:res:RCs}

The results of the fits to the rotation curves of the 11 LTs galaxies and the 2 selected inclination rogues (DDO~50 and DDO~133) are shown in Table~\ref{tab:ResultsLTs}. For each galaxy we list the median and 68\% CL region of the posterior distributions of the five fitted parameters, $\MSt$ and $\vec\theta_{\rm FDM}$, the reduced chi-square and a subset of derived quantities characterizing the FDM distribution, including the core radius, $r_c$, the transition radius, $r_t$, and the central density, $\rho_c$. In Fig.~\ref{fig:my_rc}, we show the results of the fits for two representative galaxies in our sample, WLM and DDO~154. Besides the data, we show the median, the 68\% and 95\% CL of the posterior distribution of the total theoretical rotation curve and the individual contributions from the FDM and baryonic density distributions for the central values of the parameters. Further details, including the corner plots and mean autocorrelation times of the MCMC for the fits of these two galaxies are shown for reference in the Appendix~\ref{app:info_fits}. The full set of rotation curves, corner plots, and mean autocorrelation times obtained from the fits in Table~\ref{tab:ResultsLTs} are publicly available under the category \texttt{Variable Axion Mass} in the \href{https://github.com/acastillodm/FuzzyDM}{\texttt{FuzzyDM}\faGithub} GitHub repository. 

As shown by the reduced chi-square values obtained in the fits, the performance of the FDM model describing the rotation curve data is, in general, excellent. In two cases, DDO~52 and UGC~8508, the value of the chi-square is too low suggesting there is overfitting or overestimated errors in the data.  Our fits favor large solitons and with transition radii to the halo typically overlapping with $r_{\rm max}$. This reduces the power of the data to constrain the halo parameters $c$ and $M_h$, explaining their large uncertainties. The direct relation between the scale density of the NFW distribution and $c$ in Eq.~\eqref{eq:Relrhos_c} imposes an upper bound on this parameter because, given a soliton describing the core of a galaxy, the density at $r\leq r_t$ cannot be arbitrarily large. In case of the halo mass an increased constraining power is achieved by imposing the core-halo relation in the priors. On the other hand, the parameters of the soliton, $m_a$ and $M_c$, are quite well determined, with relative uncertainties even reaching $\mathcal O(10\%)$ in some cases. The sensitivity to the axion mass of these data is illustrated in Fig.~\ref{fig:my_rc2} for the rotation curves of WLM and DDO~154. The fit is also compared to the to a pure NFW profile that is included for reference. The description of the data can be quite insufficent, and even worse than for NFW for some choices of $m_a$. In these cases, the fit does not converge to the one of NFW (that would result from suppressing the soliton by effectively running $r_t$ to zero) because the soliton cannot be arbitrarily light due to the soliton-halo relation imposed in the priors. 

As illustrated in Fig.~\ref{fig:my_rc}, the rotation curves are dominated by the DM contribution in the full radial domain, although the baryonic contribution can also be significant in the central regions of DDO~210, NGC~2366, DDO~168, and NGC~6822. In fact, NGC~6822 is the only galaxy for which the FDM presents a relatively poor fit. The main tension arises from the two innermost points (at $\sim70$ pc and $\sim140$ pc), where the velocities can be extracted with a relatively high precision. For the nominal stellar mass of NGC~6822, $\MSt=(7.6\pm1.9)\times10^{7}\,\MS$ (and fixed scale radius $R_\star$), the mass distribution at the center of the galaxy is dominated by stars and the FDM model cannot fit the rotation curve without significantly reducing the stellar content down to $\MSt\simeq1.8\times10^7~\MS$. In this context, it is important to note that there is independent evidence for the baryonic dominance of the mass at the core of NGC~6822 based on stellar kinematics~\citep{2014MNRAS.439.1015K}. The difficulty of our model to account satisfactorily for these data does not pose, a priori, a serious problem to FDM because the central densities of the soliton can be significantly affected by interference effects, as discussed above. Apart from this, a fast-expanding HI bubble was identified in this galaxy for radii below 2.5 kpc \citep{2017MNRAS.467.2019R}. For these reasons, we removed NGC~6822 from our core data sample, adding it to the list of rogue galaxies in Table~\ref{tab:ResultsLTs} and for the purposes of the subsequent analyses. In the next subsections, we present different consistency and diagnostic tests on the behavior of the parameters obtained in the fitting procedure.

\begin{figure*}[t]
\centering
\includegraphics[width=9.4cm]{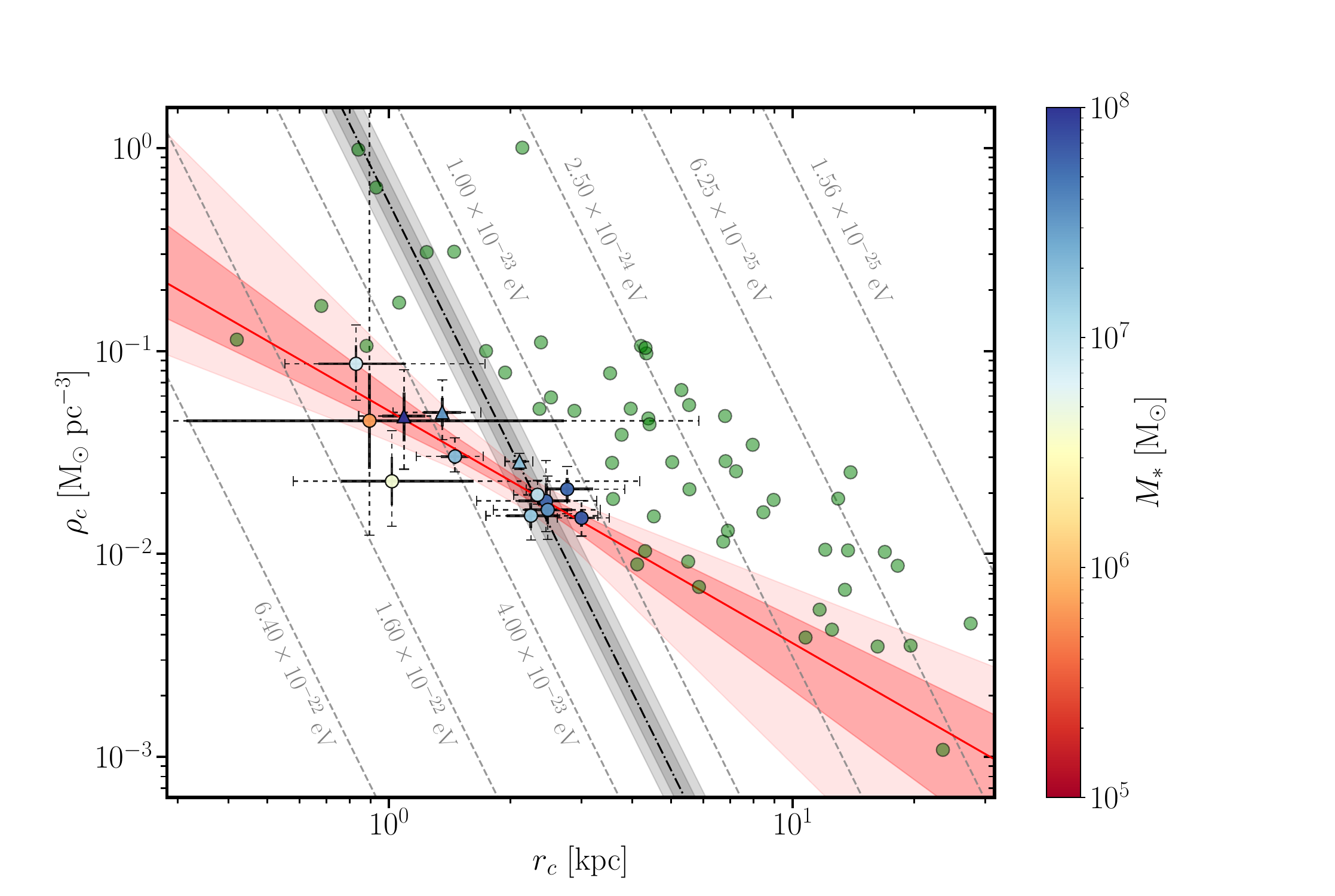} 
\hspace{-0.5cm}
\includegraphics[width=9cm]{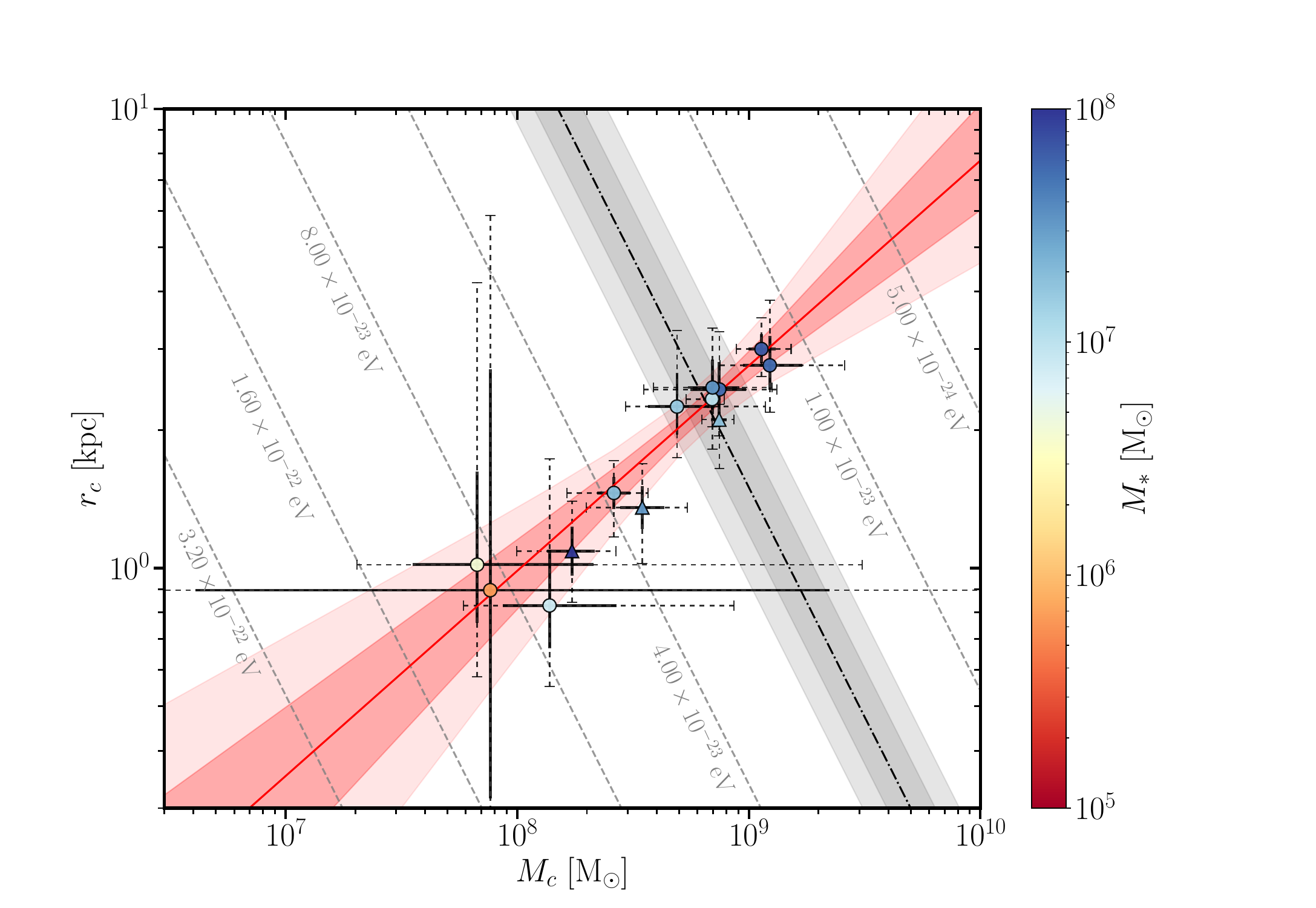}
\caption{Soliton scaling relations compared to the results of the fits to LTs, where each point is colored according to the value of $\MSt$ of the corresponding galaxy. Dot-dashed lines and gray bands represent the mean and the 68\% and 95\% CL regions for the correlation expected from FDM for the average axion mass in Eq.~\eqref{eq:benchmarkmass}. The solid line and surrounding bands represent the maximum posterior and 68\% and 95\% CL regions of the fit to a power law of the results extracted from the individual galaxies. \textit{Left: } Results in the $r_c-\rho_c$ plane compared to Eq.~\eqref{eq:rhocrc} and where we also show the results obtained independently for a different sample of galaxies by using a Burkert profile~\citep{Rodrigues:2017vto,Deng:2018jjz}. \textit{Right:} Results in the $r_c-M_c$ plane compared to Eq.~\eqref{eq:rhocrc}.
\label{fig:solitonScaling}}
\end{figure*}

\subsection{The axion mass and the soliton scaling relations}
\label{sec:res:soliton}

In our initial analysis, we fit the fundamental axion mass individually for each galaxy, allowing for a broad range of values, from $10^{-25}$ eV  to $10^{-19}$ eV. As shown in Table~\ref{tab:ResultsLTs} and in the left panel of Fig.~\ref{fig:AxionMass}, the obtained values of $m_a$ lie within the range of $\sim10^{-23}$ eV $-$ $10^{-22}$ eV. In particular, the values for the most constraining galaxies are different, at most, by a factor of $2-3$, which constitutes a remarkable consistency check for FDM with the LTs data sample. Moreover, the values of $m_a$ extracted from the selected rogues are also within this range of masses.

We can obtain the preferred value of $m_a$ by taking a weighted average of $\log_{10} m_a$ for the ten galaxies in our sample (excluding the rogues), which is the parameter actually determined in the fits and whose posterior distributions are approximately symmetric around the mean. We then obtain $m_a=1.90 \pm 0.08\times 10^{-23}$ eV, corresponding to an uncertainty of 0.019 dex, and a reduced chi-square $\chi^2_\nu=7.68$. This relatively poor value stems from the small uncertainties derived for $m_a$ in some of the galaxies (see Fig.~\ref{fig:my_rc}) and it could be due to unknown systematic uncertainties in the data or the model. Under this premise, there are several ways to account for them in the analysis~\citep{ParticleDataGroup:2022pth,2010arXiv1008.4686H}. We follow here the prescription followed by the particle data group for averages of experimental results in this case, which consists in increasing the final uncertainty by a scale factor $S=(\chi^2_\nu)^{1/2}$~\citep{ParticleDataGroup:2022pth}. With this method we obtain a more conservative error:
\begin{equation}
\label{eq:benchmarkmass}
\text{\it Optimal axion mass:}\hspace{0.8cm} m_a = 1.90^{+0.24}_{-0.21}\times 10^{-23}\text{ eV},
\end{equation}
corresponding to an uncertainty of 0.052 dex. For completeness we have also obtained an alternative determination of this optimal mass by performing a global simultaneous fit to the parameters involving the rotation curves of all ten galaxies in the sample and using the same $m_a$. This procedure is more robust as it accounts for possible correlations and features in the fit and posterior distributions. Nonetheless, we obtain the same central value although with a relatively smaller uncertainty. The full set of results obtained from the fits using Eq.~\eqref{eq:benchmarkmass} are publicly available under the category \texttt{Benchmark analysis} in the \href{https://github.com/acastillodm/FuzzyDM}{\texttt{FuzzyDM}\faGithub} GitHub repository. 

\begin{figure*}
    \centering
\includegraphics[width=8cm]{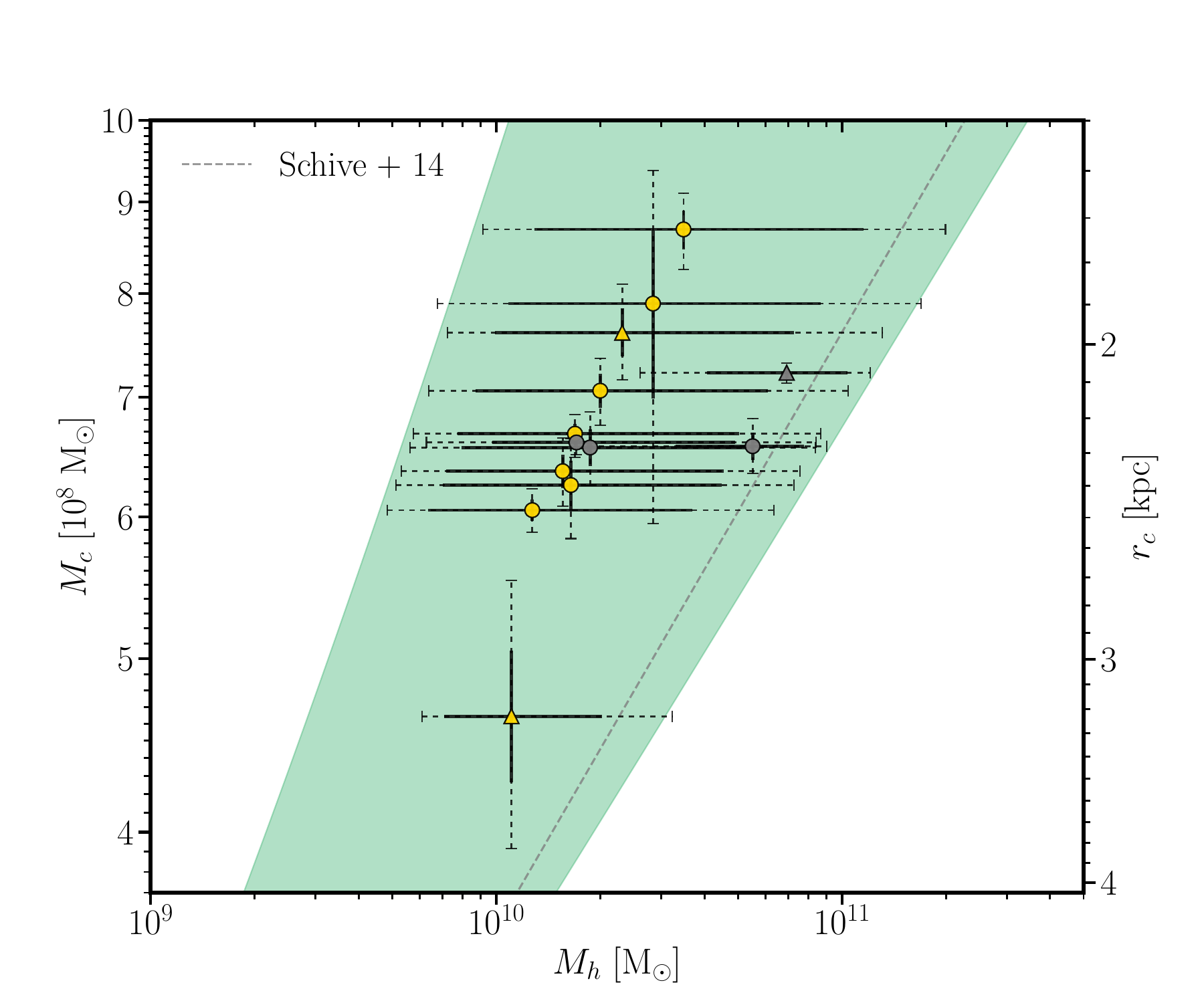}\hspace{1cm}
\includegraphics[width=8cm]{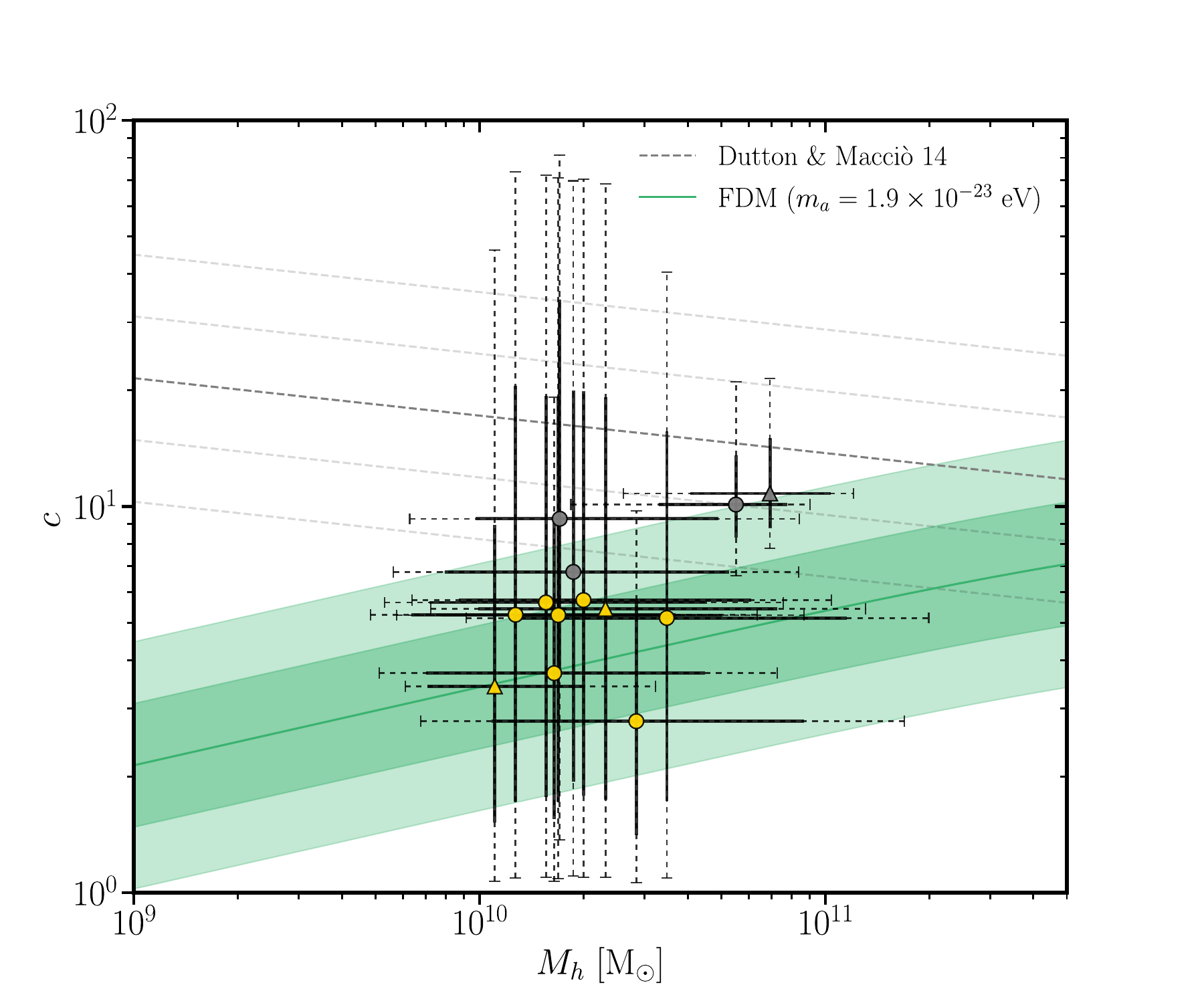} 
\includegraphics[width=8cm]{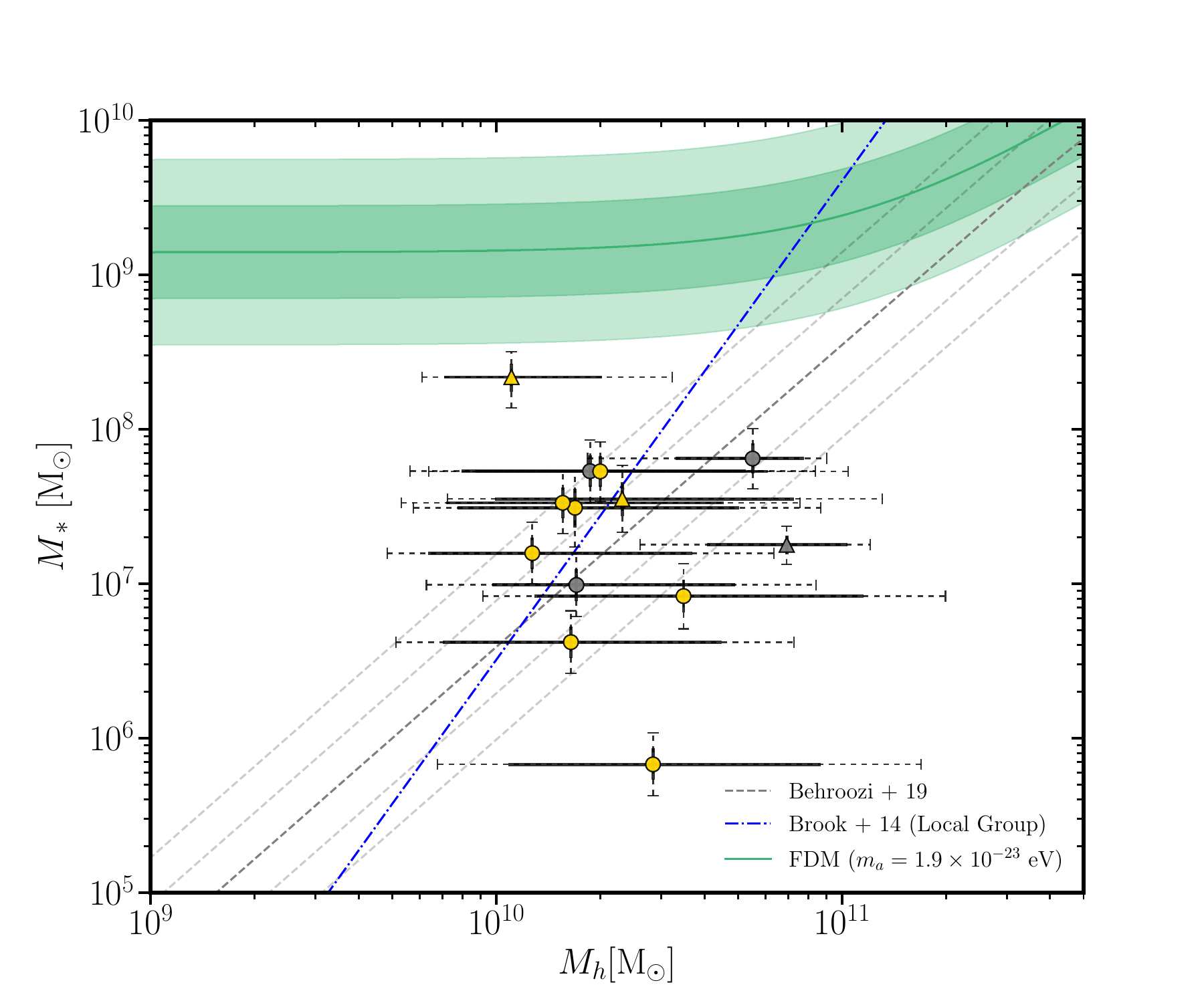}\hspace{1cm}
\includegraphics[width=8cm]{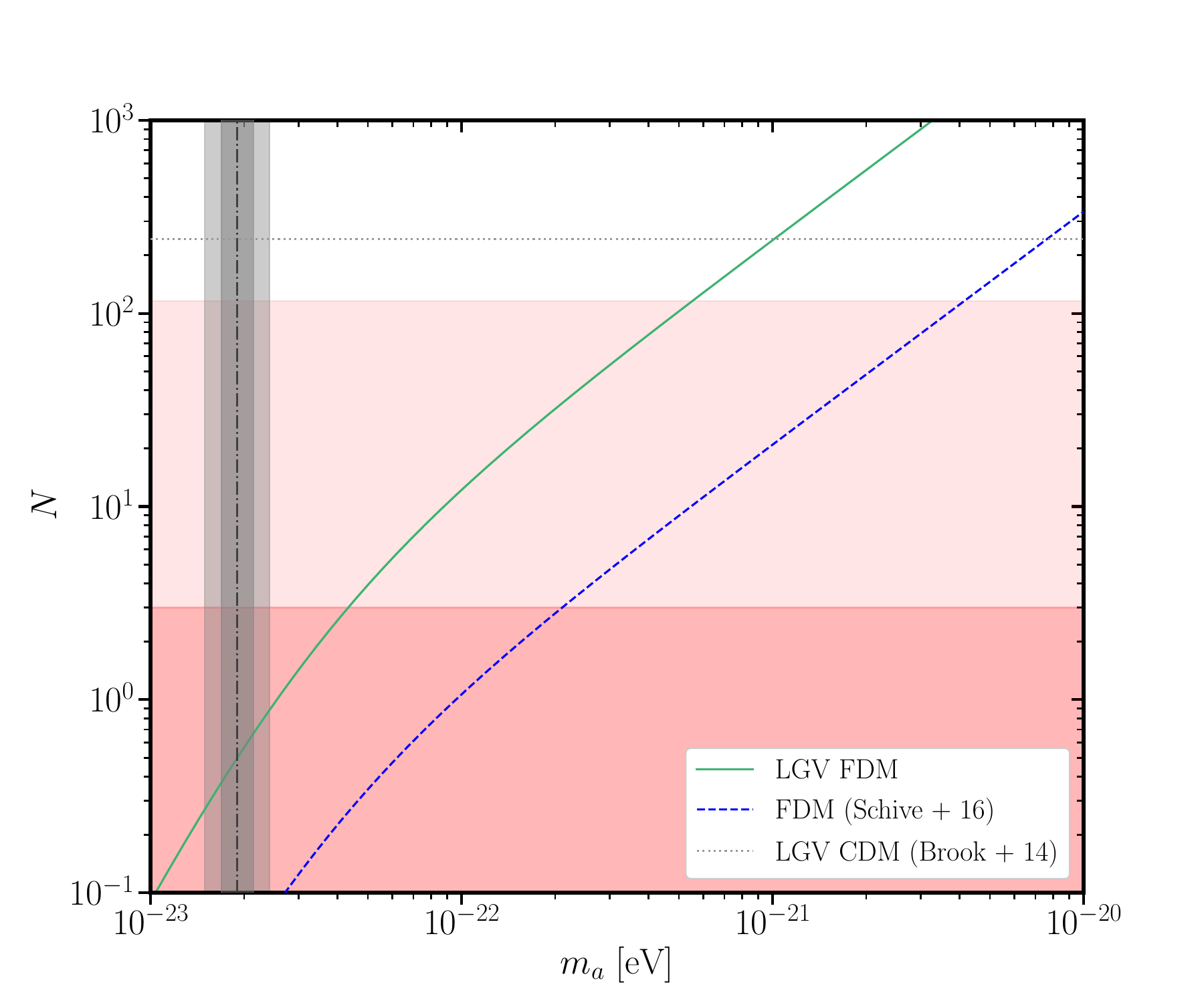}
 \caption{Halo properties obtained from the posteriors of the FDM fits to the rotation curves of the LTs galaxies for a fixed axion mass $m_a=1.9 \times 10^{-23}$ eV. Gray-colored points are galaxies for which both median and maximum posterior values of $r_t$ are within the observed range, while gold-colored are soliton-dominated by virtue of not satisfying that condition. Uncertainties in the data points describe the credible intervals at $68\%$ (solid) and $95\%$ (dashed) CL. Uncertainties in the theoretical bands refer to the 
 1 and 2$\sigma$ scatter in the predictions. \textit{Top-left:} $M_c-M_h$ relation compared to the region allowed by Eq.~\eqref{eq:ferreira} and the prediction in Eq.~\eqref{eq:Schive_2014} from~\cite{Schive:2014hza} (dotted line). 
 \textit{Top-right:} $c-M_h$ relation compared to Eq.~\eqref{eq:cm} and to the prediction in CDM,  Eq.~\eqref{eq:dutton} from \cite{2014MNRAS.441.3359D} with $\sigma = 0.16$ dex (gray dashed line). 
 \textit{Bottom-left:} Stellar-to-halo mass relation from abundance matching compared to the FDM prediction using Eq.~\eqref{eq:HMF-FDM}, and the CDM predictions from \cite{2019MNRAS.488.3143B} (dashed gray line) and \cite{2014ApJ...784L..14B} (dot-dashed line) with a reference $\sigma = 0.30$ dex scatter in both cases.
 \textit{Bottom-right:} Expected number of halos with $M_h\lesssim10^{11} \: \MS$ in a Local Group volume of radius 1.8 Mpc predicted in FDM as a function of $m_a$, using Eq.~\eqref{eq:HMF-FDM} and calibrated with the CDM prediction of the HMF from~\cite{2014ApJ...784L..14B}. This is compared to the HMF from~\cite{Schive:2015kza} (dashed line). For the CDM prediction, we take 
 $10^9 \lesssim M_h \lesssim 10^{11} \: \MS .$ The dot-dashed line and gray bands represent the mean and the 68\% and 95\% CL regions for the axion mass in Eq.~\eqref{eq:benchmarkmass}.
 \label{fig:halo}}
\end{figure*}

On the other hand, the values of $m_a$ determined from the different galaxies do not seem to be randomly scattered around the optimal value but rather following a negative correlation with the stellar mass. Specifically, the lower the $\MSt$ value, the higher these $m_a$ values tend to be. This correlation is visible in the left panel of Fig.~\ref{fig:AxionMass}, where we have ordered the values of $m_a$ according to increasing values of the stellar mass of the galaxies. Since $m_a$ is a fundamental constant in the model, this suggests that there are astrophysical processes influencing the observed cores beyond the structure predicted by a FDM soliton with the optimal mass in Eq.~\eqref{eq:benchmarkmass}. We can try to quantify the statistical significance of the correlation by introducing an empirical power law between the two quantities, $m_a=\alpha_\star\,\MSt^{\,\beta_\star}$. We then performed a linear fit of $\alpha^\prime_\star=\log_{10}\alpha_\star$ and $\beta_\star$ to the 10 $(\log_{10}\MSt,\,\log_{10}m_a)$ data points, excluding the three rogue galaxies we have studied. For this, we use MCMCs and also include a possible source of unknown systematic uncertainties in the data points as described in Sect.~8 of~\cite{2010arXiv1008.4686H}. The results of this fit are shown in the right panel of Fig.~\ref{fig:AxionMass}. We find that $\beta_\star=-0.62^{+0.21}_{-0.25}$, corresponding to evidence at a $\sim3.3\sigma$  level (with respect to $\beta_\star=0$) of a correlation between the values of $m_a$ determined from the rotation curves of the galaxies and the relevant $\MSt$. We describe the implementation of these fits  and the interpretation of the CL regions in Appendix~\ref{app:info_fits}.  As shown in Fig.~\ref{fig:AxionMass} the two rogue galaxies NGC~6882 and DDO~133 seem to also follow the correlation, while DDO~50 is an outlier. We discuss the possible implications of this correlation in more detail in Sect.~\ref{sec:discussion}.

Another perspective on the consistency of the inferred axion masses can be obtained  by testing the soliton scaling relations Eqs.~\eqref{eq:rcmc} and ~\eqref{eq:rhocrc}. To quantify the agreement of the model with observations, we fit our results of $\rho_c$ and $M_c$ to power laws in $r_c$ with exponents $\beta_{\rho}$ and $\beta_{M}$, respectively. These fits are done again with MCMCs, including possible systematic uncertainties from~\cite{2010arXiv1008.4686H}, and they are described in Appendix~\ref{app:info_fits}. In the left panel of Fig.~\ref{fig:solitonScaling}, we present our results in the $r_c-\rho_c$ plane. The data show that the central densities fall with the size of the core, although at a pace much slower than predicted by FDM. Indeed, our fit to the data yields $\beta_\rho=-1.18^{+0.22}_{-0.30}$, disagreeing with the prediction of FDM, $\beta_\rho=-4$, with a significance larger than $3\sigma$. It is remarkable that our result is compatible with $\beta_\rho\approx-1.3$, obtained by~\cite{Deng:2018jjz} using a different sample of galaxies analyzed with a Burkert profile~\cite{Rodrigues:2017vto}. Our results also agree within 1$\sigma$ with $\beta_\rho\approx-1$, which is an empirical result derived from the fact that the so-called ``column density,'' $\Sigma_0=\rho_c\times r_c$~\citep{Donato:2009ab}, is nearly constant for a large sample of different galaxies (see also~\cite{Kormendy:2014ova}, \cite{Burkert:2015vla}, \cite{2017MNRAS.465.4703K}, and ~\cite{2019MNRAS.490.5451D}). In fact, we obtain $\Sigma_0=51^{+12}_{-9}\,\MS\,\text{pc}^{-2}$ which undershoots slightly the empirical value $\Sigma_0=75^{+55}_{-45}\,\MS$ pc$^{-2}$~\citep{Burkert:2020laq}.    

In the right panel of Fig.~\ref{fig:solitonScaling}, we show our results in the $r_c-M_c$ plane. In this case the data manifest a correlation which is almost orthogonal to the FDM prediction: The cores are heavier, the bigger (not the smaller) they are. Quantitatively we find $\beta_M=0.45^{+0.10}_{-0.08}$ which (at face value) disagrees with the FDM value $\beta_M=-1$ with a significance larger than $5\sigma$. In addition, and as shown by the color code used in the data points of    Fig.~\ref{fig:solitonScaling}, $r_c$, and $M_c$ are also correlated with $\MSt$, as the heavier cores contain also more stars. This suggests again that there is a non-trivial interplay between the structure of the DM cores at the center of the dwarf galaxies in LTs and baryonic physics.  

\begin{figure*}[h]
    \centering \includegraphics[scale=0.25]{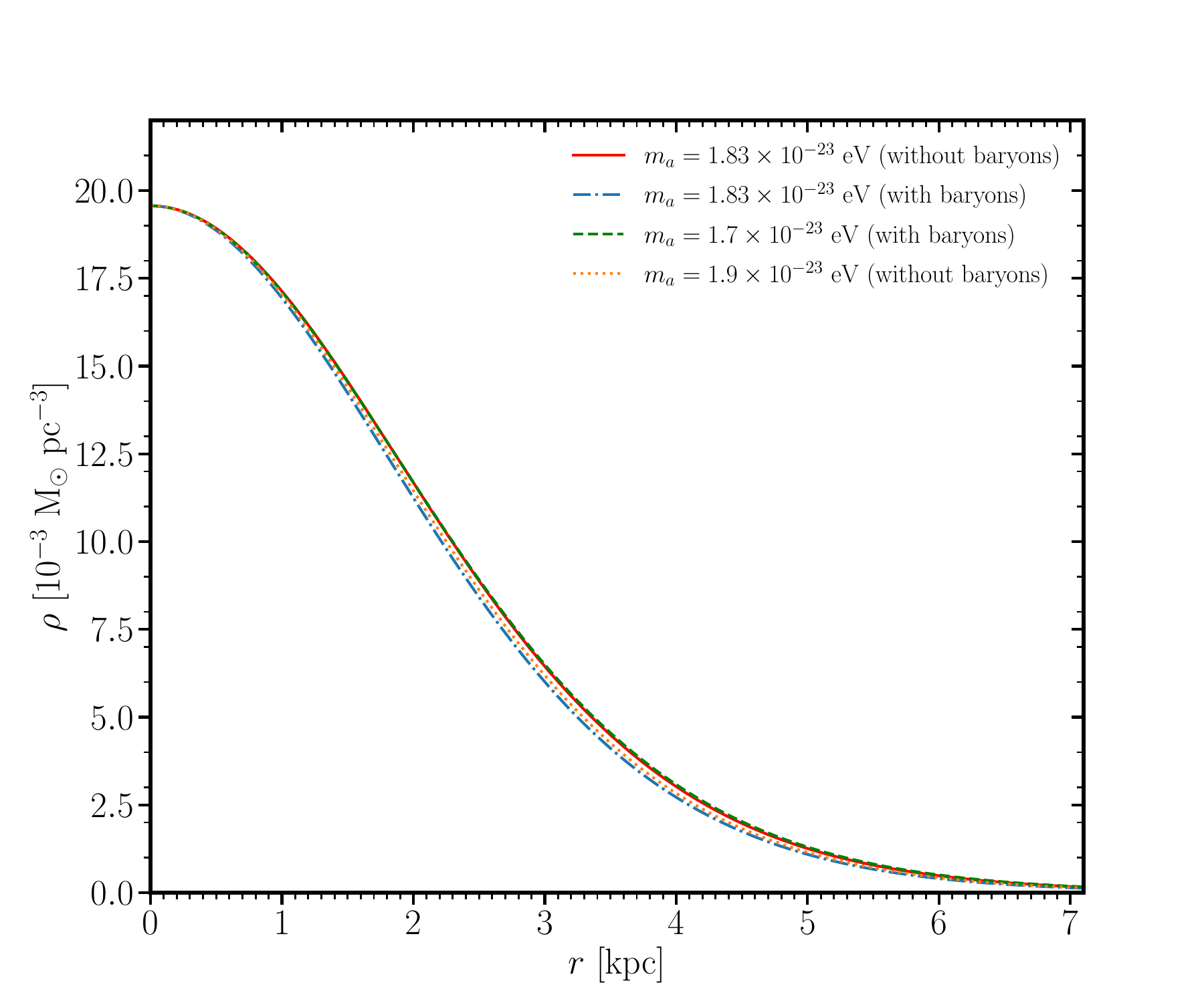}  
\hspace{1.cm}\includegraphics[scale=0.25]{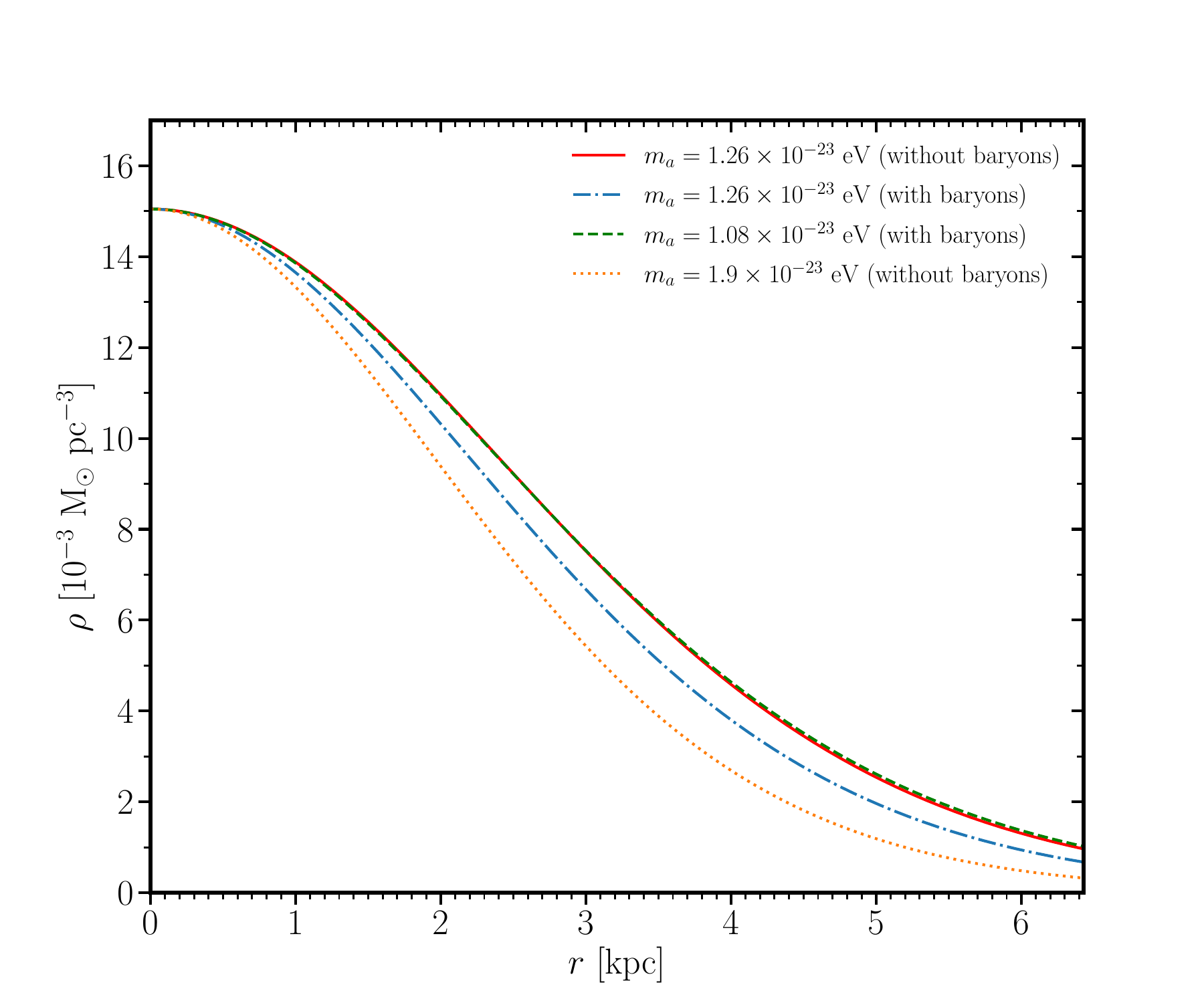}
    \caption{Soliton density profiles for DDO 154 (left) and NGC 2366 (right) including the baryonic contributions to the gravitational potential. The red solid line denotes the original soliton matching the median values of the FDM mass and central density (with the latter being fixed in all cases). The blue dot-dashed line corresponds to the same FDM mass with the inclusion of baryons, while the green dashed line to the mass needed to replicate the same density profile as the original soliton. Lastly, the orange dotted line corresponds to a soliton without baryons fixed at the value of the global optimal mass from Eq.~\eqref{eq:benchmarkmass}.
    \label{fig:densities}}
\end{figure*}

\subsection{Properties of the halo}
\label{sec:res:halo}

In addition to the properties of the solitons we can test the predictions of FDM regarding the halos they inhabit. To do this we set $m_a$ to the optimal value found in Eq.~\eqref{eq:benchmarkmass}, $m_a\approx1.90\times10^{-23}$~eV, disregarding the difficulties of the model to explain the data discussed in the previous section. First, we show on the top-left panel of Fig.~\ref{fig:halo} our results in the $M_h-M_c$ plane compared to the core-halo relation predicted from simulations in Eqs.~\eqref{eq:Schive_2014} and~\eqref{eq:ferreira}. This relation is imposed as a prior in the fits and it provides the most important constraint on the halo masses for most of the galaxies. This is why the uncertainties of $M_h$ span almost the entire allowed range for a given $M_c$. Next, we show on the top-right panel our results in the $M_h-c$ plane comparing them to the predictions of CDM and FDM. The uncertainties on the concentration parameter $c$ obtained from the fits are large although, as discussed in Sect.~\ref{sec:res:RCs}, this is constrained from above by the central densities reached by the soliton. This reduces systematically the values of $c$ obtained from the data as compared to the predictions of CDM and agreeing with the expectations of FDM.  

The most important test that can be performed on the properties of the FDM halos with our results is related to the abundance of dwarf galaxies. As discussed in Sect.~\ref{sec:FDM}, FDM predicts a suppression of structure formation at small scales that translates into a low-mass cutoff in the HMF. One classical method to test observationally the HMF is via abundance matching methods with the observed stellar mass function (SMF)~\citep{1988ApJ...327..507F,2004MNRAS.353..189V,2010ApJ...710..903M,2013ApJ...770...57B}. Given the non-monotonic behavior of the HMF in Eq.~\eqref{eq:HMF-FDM}, we follow a method inspired by~\cite{2004MNRAS.353..189V}, whereby we find a $\MSt-M_h$ stellar-to-halo mass function by matching the cumulatives of the SMF and HMF,
\begin{align}
\label{eq:abundance_matching}
\int^\infty_{\MSt} \frac{dn}{dM}\Big|_{\rm SMF} dM=\int^\infty_{M_h} \frac{dn}{dM}\Big|_{\rm HMF} dM. 
\end{align}
In the bottom-left panel of Fig.~\ref{fig:halo} we show the results of abundance matching using the stellar-to-halo mass function reported in~\cite{2019MNRAS.488.3143B} for CDM and corrected for FDM by Eq.~\eqref{eq:HMF-FDM} and the optimal axion mass in Eq.~\eqref{eq:benchmarkmass}. We also add the prediction for CDM reported in~\cite{2014ApJ...784L..14B}, which implements a HMF specific for the environment of the Local Group. The FDM prediction for the stellar-to-halo mass function features a  flattening starting at values as high as $M_h\approx5\times10^{11}\,\MS$, which results from the strong suppression of structure formation in FDM for the axion mass value favored by the fits to the rotation curves of LTs. In particular, we observe that according to this prescription of abundance matching, galaxies with stellar masses $\MSt\lesssim10^{8.5}$ should be extremely rare, which is clearly in plain contradiction with the posterior distributions of the very same fits in the $\MSt-M_h$ plane. On the contrary, the data seem to be more consistent with the stellar-to-halo mass function predicted by CDM~\citep{2019MNRAS.488.3143B, 2014ApJ...784L..14B}.
 
A more transparent way to illustrate this tension between FDM and observations is by directly comparing the expected number of halos that are predicted to exist  in a specified local volume to the corresponding number of observed galaxies. 
Or, inversely, we can  translate the abundance of observed galaxies to a lower bound on the FDM mass that is necessary to make the existence of such galaxies statistically plausible in the model. We take as a reference the simulations of galaxy formation in the Local Group within the $\Lambda$CDM that are reported in~\cite{2014ApJ...784L..14B} using 1.8 Mpc for the radius of the Local Group Volume (LGV), which is the one we use. In the bottom-right panel of Fig.~\ref{fig:halo}, we show the
number of FDM halos with  $M_h\lesssim10^{11}\MS$ that are expected to be contained in this volume as a function of $m_a$. For this we use the HMF in Eq.~\eqref{eq:HMF-FDM} where we implement the Sheth-Tormen prediction for the HMF of CDM~\citep{1999MNRAS.308..119S} and we normalize it to reproduce the results of \cite{2014ApJ...784L..14B} for the LGV (see Appendix~\ref{app:abundance_hmf}). For reference, we also show the abundance that is obtained using the HMF originally reported in \cite{Schive:2015kza}.

The predicted abundances are to be compared with the three galaxies contained in the LGV that we have found in our analysis. More generally, we should also compare these predictions with the properties of 116 such galaxies that can be extracted from the most recent census~\cite{2021ApJ...913...53P}.~\footnote{This bound can be applied using galaxies beyond those in our sample (for which we have extracted $M_h$) because we are conservatively integrating from $10^{11} \, \MS$ down to negligible masses.} Using the former, we find a very conservative lower bound of $m_a \gtrsim 4.3 \times 10^{-23}$ eV. Nonetheless, this alone implies that our benchmark value for $m_a$ in Eq.~\eqref{eq:benchmarkmass} is in tension with observations with a significance greater than $5\sigma$. Using the 116 galaxies from  \cite{2021ApJ...913...53P}, we can obtain a much more severe lower limit on the mass, namely,  $m_a \gtrsim 5.5 \times 10^{-22}$ eV. 

Finally, we can also study the population of dwarf galaxies in the Milky-Way's halo using the subhalo mass function for FDM recently reported in \cite{du2018structure, Du:2016zcv}. Using this, and  focusing on the 59 galaxies from  \cite{2021ApJ...913...53P} that lie within $400$ kpc of the Milky Way (following~\cite{2019PhRvL.123e1103M} and which is roughly twice its virial radius $r_{\rm vir}^{\rm MW}\approx200$ kpc~\citep{2016ARA&A..54..529B}), we obtain the stronger lower bound $m_a \gtrsim 7.3 \times 10^{-22}$ eV.

\subsection{Baryonic effects}
\label{subsec:baryons}

Our analysis so far has ignored the effects of baryons and we may consider if a more realistic approach including them could address the difficulties of FDM to explain observational data that were discussed above. Intuitively, one might expect that in DM-dominated galaxies such as those in the LTs sample these effects should not lead to major changes in the analysis. However, as discussed in Sect.~\ref{sec:intro}, the consequences of baryonic feedback from SN can be significant for CDM specifically for dwarf galaxies in the mass range we are investigating. In order to  realistically account for these effects, we would require information outside the SP equation related to the baryonic components and including the modeling of star formation and evolution. Interestingly, FDM simulations of structure formation with baryonic feedback have recently been performed by~\cite{Mocz:2019pyf,Mocz:2019uyd,Veltmaat:2019hou}, finding that the inclusion of baryons in FDM halos does not impede their thermalization and formation of quasi-stationary (modified) solitons at their centers. This motivates the study of this solution even in the presence of baryonic effects.

A simple and minimal approach is to solve for the modified soliton solution under the assumption of a static background baryonic potential, which can be obtained in a very similar way to that of the original soliton. The resulting modified soliton solution has been studied in \cite{Bar:2018acw} (and subsequently in \cite{Bar:2019pnz}), which was found to accurately match simulations incorporating gravitational interactions with baryons in \cite{2018MNRAS.478.2686C}. In order to proceed, we closely follow the approach of~\cite{Bar:2018acw}. The methodology is very similar to that of the ordinary soliton in Eq.~\eqref{eq:SP}, except that in this case we introduce the baryonic density, $\rho_b$, solving for the total (baryonic + DM) potential $\Phi_{\text{tot}}$ and solve for a given central density $\rho_c.$ The resulting modified SP equation is given by
\begin{align}
i\frac{\partial}{\partial t}\psi(t,\vec r)  &= -\frac{1}{2m_{a}}\nabla^{2}\psi(t,\vec r)
+m_{a}\Phi_{\text{tot}}(t,\vec r)\psi(t,\vec r), \nonumber\\
\nabla^{2}\Phi_{\text{tot}}(t,\vec r)&=
4\pi G \big(\vert\psi(t,\vec r)\vert^{2} + \rho_{b}(\vec r)\big).\label{eq:SPB}
\end{align}
We can then apply the same quasi-stationary ansatz as in Eq.~\eqref{eq:ansatz}, along with the boundary condition $\chi'(0) = 0$, and proceed accordingly with the solution to the eigenvalue problem to obtain the modified soliton solution; that is, we solve for the unique, constantly decreasing solution with $\chi(r \to \infty) = 0.$ We note that $\rho_b$ is taken to be static in time and spherically symmetric. Nonetheless, the validity of the spherical approximation was corroborated for DM-dominated galaxies in \cite{Bar:2019pnz}, where axially-symmetric baryonic distributions were implemented in the SP equations. The spherically averaged baryonic density is computed from the baryonic mass profile 
\begin{align}
M_b(r) = \frac{r}{G}\big(V_*^2(r) + V_{\text{gas}}^2(r) \big),
\end{align}
using Eq.~\eqref{eq:LH} so that
\begin{align}
\rho_b(r) = \frac{1}{4\pi r^2} \frac{\text{d} M_b}{\text{d} r},
\end{align}
where we compute the derivative analytically.

\begin{table}[t]
\caption{Systematic correction in FDM mass central value as a result of introducing the modified soliton due to a background baryonic potential.}
\centering
\setlength{\tabcolsep}{0.7em}
\renewcommand{\arraystretch}{1.5}
  \setlength{\arrayrulewidth}{.30mm}
\centering
  \begin{tabular}{lcc}
  \hline\hline
   Galaxy & $m_a$  & $\Delta m_{a}$  \\
  &[$10^{-23}$ eV]&[$10^{-23}$ eV]\\ \hline
  NGC 2366 &  $1.26^{+0.13}_{-0.13}$ & $-0.18$\\
  DDO 168 &  $1.26^{+0.27}_{-0.26}$ & $-0.17$\\
  DDO 52 & $1.72^{+0.43}_{-0.31}$  & $-0.16$ \\
  DDO 87 &  $1.77^{+0.37}_{-0.32}$ & $-0.16$ \\
  DDO 126 &  $2.22^{+0.53}_{-0.53}$ & $-0.23$\\
  WLM &  $3.77^{+0.49}_{-0.43}$ & $-0.33$\\
  DDO 154 &  $1.83^{+0.13}_{-0.09}$ & $-0.13$ \\
  UGC 8508 &  $6.9^{+2.8}_{-2.6}$ & $-0.3$ \\
  CvnIdwA &  $8.9^{+5.3}_{-5.0}$ & $-0.9$ \\
  DDO 210 &  $9^{+55}_{-8  }$ &$ -0.3$ \\
  \rule{0pt}{4ex}
  $ $ & \emph{- rogues -} & $ $\\
  \rule{0pt}{4ex}
  DDO 50 &  $5.4^{+1.0}_{-0.8}$ & $-0.9$ \\
  DDO 133 &  $3.39^{+0.58}_{-0.50}$ & $-0.29$\\
  DDO 6822$^{*}$ &  $1.86^{+0.11}_{-0.10}$ & $-0.17$\\
  \hline
\hline
\end{tabular}
\tablefoot{  $m_a$ is the median value of the original FDM mass found in Table~\ref{tab:ResultsLTs} with corresponding errors, while $\Delta m_a$ is the correction in $m_a$ needed to reproduce the original soliton without baryons, that is, the difference between the new FDM mass and the original one.}
\label{tab:ResultsBaryons}
\end{table}

Our initial approach to assess the self-consistency of the model with baryons is to solve for the modified soliton for each galaxy fixing the axion mass and central density to the corresponding central values in Table~\ref{tab:ResultsLTs}. This leads to solitons more compact than the original ones, in agreement with~\cite{Bar:2018acw} and the results found in simulations~\citep{Mocz:2019pyf,Veltmaat:2019hou}. However, the differences between these solitons and their DM-only counterparts are less pronounced than in these simulations mainly because the relative proportion of baryons compared to DM  is significantly lower in the LTs galaxies. This is illustrated in Fig.~\ref{fig:densities}, where we show the effect of the gas and stars in the density distribution of the soliton for DDO~154 and NGC~2366. 

We confirm the findings of~\cite{Bar:2018acw} that solitons formed including the baryonic effects closely follow the same universal distribution as the DM-only soliton, for instance, the one approximated by Eq.~\eqref{eq:an_sol}. This does not imply necessarily that the two solitons (DM-only and with baryons) are the same but rather that their difference is captured by a modification of the relation between $\rho_c$ and $r_c$  for a given $m_a$ (or of any other scaling relations such as $r_c$ and $M_c$); or, conversely, that $\rho_c$ and $r_c$ parametrize the same density profile regardless of whether baryons are included or not but where each case corresponds to a different value of $m_a$. In particular, if baryons are included for a given empirical density profile, a lower $m_a$ should be obtained to compensate for the deepening of the gravitational potential. 

\begin{figure}[t]
    \centering
   \includegraphics[scale=0.3]{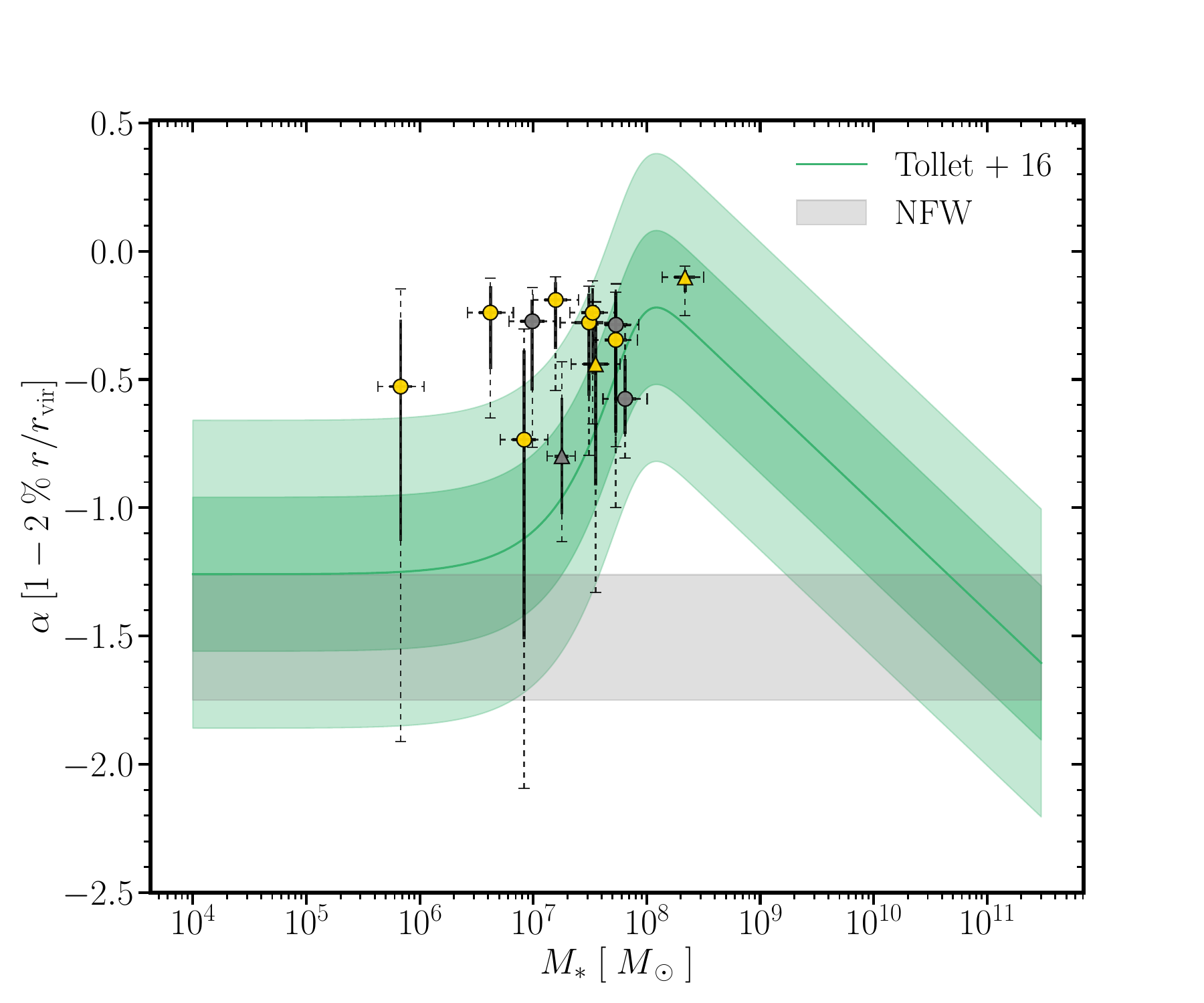}  
    \caption{Inner slope vs stellar mass for simulations running with the optimal FDM mass of $1.9 \times 10^{-23}$ eV. In green appears the prediction of baryonic feedback simulations from \cite{DiCintio:2013qxa,Tollet:2015gqa}. We use a 1$\sigma$ scatter of 0.3, matching closely with the results of \cite{DiCintio:2013qxa,Tollet:2015gqa}, while also adding an additional 2$\sigma$ scatter band. The gray band represents a conservative scatter derived analytically for NFW profiles.
    \label{fig:inslope_mst}}
\end{figure}

Therefore, we can implement the inclusion of baryons in our analysis of the rotation curves of LTs by including a systematic negative shift in the FDM mass needed to replicate the original profile favored by the fits. In Table~\ref{tab:ResultsBaryons} we list the corrections to the value of $m_a$ (with respect to the results listed in Table~\ref{tab:ResultsLTs}) that is required to fit the rotation curves including the effects of the baryons. We find that the correction needed is typically within the original $68 \%$ CL errors, with the only exception of DDO~154 and NGC~2366 (both of which appear in Fig.~\ref{fig:densities}) and NGC~6822 in case of the rogues. Based on these results, we conclude that the effect of including the background baryonic potential introduces a systematic drop of $\lesssim 15 \%$ in the axion masses which should be included as a correction in the determination of $m_a$ in Eq.~\eqref{eq:benchmarkmass}. We note that this effect does not explain the empirical dependence of $m_a$ with $\MSt$ that we found in our sample and does not help understanding the discrepancies of the data with the soliton scaling relations discussed in Sect.~\ref{sec:res:soliton}. Furthermore, this effect of the baryons only exacerbates the tension found in the previous Sect.~\ref{sec:res:halo} in relation to the excessive suppression of the HMF at small scales. 

In connection to the baryonic effects, we also investigate the inner-slope parameter of the DM distribution of the LTs galaxies as a function of their mass. This quantity characterizes the flatness of the central DM distribution and it is a figure of merit for many studies of baryonic feedback within $\Lambda$CDM. We follow~\cite{DiCintio:2013qxa,Tollet:2015gqa} and define this parameter as the logarithmic slope of the density $\rho(r) \propto r^{\alpha}$ fitted to the data between $1 - 2 \%$ of the virial radius. In Fig.~\ref{fig:inslope_mst} we show the values of $\alpha$ for the LTs sample as a function of $\MSt$ and compared to the prediction from simulations in $\Lambda$CDM~\citep{DiCintio:2013qxa,Tollet:2015gqa}.  The $\MSt$ range covered by the LTs galaxies overlaps with the region where baryonic feedback starts becoming relevant, $\MSt\gtrsim10^7\,\MS$. The most massive galaxies in the sample present cores which are consistent with those expected from baryonic feedback while the less massive ones, with $\MSt\approx10^7\,\MS$, present cores generally more extended (or flatter) than predicted by these processes. We note, however, that we are comparing results from two different DM models. 

\section{Comparison with previous works}
\label{sec:discussion}

Various works have recently investigated the solitonic profiles using galactic rotation curves. These papers use data from the 175 galaxies compiled by the SPARC database~\citep{Bar:2018acw,2019MNRAS.483..289R,Bar:2021kti,2021ApJ...913...25C,Street:2022nib,Khelashvili:2022ffq,2023arXiv230404463D}, while~\cite{Bernal:2017oih} also studied a set of low-surface brightness (LSB) galaxies. In terms of methodology, our analysis presents some similarities with some of these works:

 In their work,~\cite{Bernal:2017oih} focuses on $\chi^2$ fits to the selected 18 LSB galaxies and 6 from the SPARC data set. A large scatter of preferred axion masses is obtained although a good global fit to the LSB set is obtained for $m_a\approx0.5\times10^{-23}$ eV. The soliton or soliton-halo scaling relations are not investigated nor the cosmological implications of the model for this mass. 

In their work,~\cite{Bar:2018acw} focuses on the physical interpretation and phenomenological implications of the core-halo relation. In particular, the appearance of a ``double-bump'' structure in the  rotation curves is predicted and then used to set an exclusion of FDM in the range of $m_a\sim10^{-22}-10^{-21}$ eV range. This structure is not a generic feature appearing in our analyses due to the broader range of core-halo relations that we use, as described in detail in Appendix~\ref{app:bar18}.   
In their work, ~\cite{Bar:2021kti} derived, for each $m_a$, upper bounds on the mass of the soliton using a conservative method and the 175 galaxies in SPARC. However, this reference does not perform a full-fledged fit to the data and uses exclusively the core-halo relation as a exclusion criterion for FDM in the range $m_a\sim10^{-23}-10^{-20}$ eV.
In their work, ~\cite{Khelashvili:2022ffq} performs a Bayesian analysis of the 175 galaxies in SPARC and studies the statistical comparison between FDM and different CDM models. This yields that FDM is preferred by the data although no single value of $m_a$ gives a good fit to all galaxies. Moreover, the core-halo and soliton scaling relations are tested. 

FDM has been studied in other different cosmological and astrophysical contexts and the derived limits on the axion mass  have been recently compiled in ~\cite{Ferreira:2020fam}. 
In the following paragraphs we include some of the cosmological bounds available in the literature. 

The small-scale suppression given by FDM would impact the angular scale of the CMB acoustic peaks and anisotropies, as proposed in~\cite{Hlozek:2014lca, Hlozek:2017zzf}. Using data from Planck 2015 and WiggleZ, these studies derive the lower bound \break $m_{a}\gtrsim 10^{-24}$ eV. 

The Lyman-$\alpha$ forest is a powerful tracer of the linear matter power spectrum for sub-Mpc scales, where the FDM suppression arises ~\citep{Niemeyer:2019aqm}. With high-resolution observations of the spectra on the Lyman-$\alpha$ forest,~\cite{Nori:2018pka,Kobayashi:2017jcf,Armengaud:2017nkf,Rogers:2020ltq} have determined lower bounds in the range \break $m_{a}>(2.0-20)\times10^{-21}$ eV. 

At high redshift, the $21$ cm absorption signal of neutral hydrogen at $z\sim15-20,$ measured by EDGES~\citep{Bowman:2018yin} is also sensitive to a suppression of the power spectrum  at small scales
~\citep{Lidz:2018fqo} as it would delay galaxy formation~\citep{Schneider:2018xba}. This would lead to the lower bound $m_{a}>8\times10^{-21}$ eV~\citep{Schneider:2018xba}.

The shear correlation of galaxies can be used to study the impact of the FDM scenario. With the Dark Energy Survey Year 1 (DES-Y1) data and the Planck cosmic microwave background anisotropies information,~\cite{2022MNRAS.515.5646D} has reported a search of the effects caused by FDM in the cosmic shear. This sets $m_{a}>10^{-23}$ eV at $95\%$ CL. 

The abundances of lensed ultra-faint galaxies at high redshift provide strong constraints on the suppression of the HMF of FDM. With the measurements from Hubble Frontier Field (HFF) of the number density of $z\approx6$ galaxies,~\cite{Menci:2017nsr} found the lower limit $m_{a}\geq8\times10^{-22}$ at 3$\sigma$ CL.

We now include results from astrophysical searches and bounds.
For instance, in the regime of zero self-coupling, the non-detection of black-hole superradiance leads to two distinctive ranges of exclusion of axion masses~\citep{Arvanitaki:2010sy,Baryakhtar:2017ngi,Brito:2017zvb}: Supermassive black holes exclude $7\times10^{-20}~\text{ eV} < m_{a}  < 10^{-16}$~\text{ eV} at the 95$\%$ CL, while stellar mass black holes yield an exclusion region $7 \times 10^{-14}~\text{ eV} < m_{a} < 2 \times 10^{-11}$ eV at the 95$\%$ CL. With superradiance analyses, the Event Horizon Telescope measurement of the spin and mass of M87$^{*}$ also disfavors the range $2.9\times10^{-21}\text{ eV}< m_{a}< 4.6\times10^{-21}$ eV~\citep{Davoudiasl:2019nlo}.

The FDM's wave behavior would heat up stellar objects in galaxies~\citep{2019ApJ...871...28B}. In case of the Milky Way this effect can be search for through the velocity dispersion of stars in the solar system vicinity, leading to the lower bound  $m_{a}>6\times10^{-23}$ eV~\citep{2019MNRAS.485.2861C}. A recent analysis~\citep{2023MNRAS.518.4045C} for the Milky Way disk heating leads to the bound $m_{a} > 0.4\times10^{-22}$ eV. By accounting for different sources of uncertainties,~\cite{2023MNRAS.518.4045C} finds that the range of $ m_{a}\sim 0.5-0.7 \times 10^{-22}$ eV is favored by the observed Milky-Way disc kinematics.

FDM affects dynamical friction and the effects can be searched for with the analysis of orbits of globular clusters in galaxies~\citep{Hui:2016ltb,Bar:2021jff}. In fact, FDM can help solving the so-called Fornax timing puzzle~\cite{1976ApJ...203..345T} as long as $m_{a}>10^{-21}$ eV~\citep{Lancaster:2019mde}.  

Stellar streams are excellent tracers of the gravitational potential~\citep{Carballo-Bello:2014oka} and can be used to test predictions of the sub-halo mass function in different DM models. The fluctuations in the stellar streams  lead to imposing the lower limit $m_{a}>2\times10^{-21}$ eV~\citep{Schutz:2020jox}.

Orbital motions of stars in the galactic center can also probe the matter distribution through its gravitational potential. Using mock catalogs of astrometric and spectroscopic observations of the star S2 orbiting near Sagittarius A$^{*}$, ~\cite{2023A&A...670L...4D} obtained an upper limit of $m_{a}<10^{-21}$ eV at a 95$\%$ CL. 

The suppression of the cosmic growth in FDM reduces the number of low-mass halos, which impinges on the dwarf-spheroidal galaxies' (dSph) typical distance scales. From the observed population of satellites in the Milky Way the mass of the FDM is required to be $m_{a} >2.9 \times 10^{-21}$ eV~\citep{Nadler:2019zrb}. 

By applying a Jeans analysis for Milky Way-satellite dSphs, it is possible to study the density profile predicted by FDM.  By considering a soliton core and kinematic data of the classical dSphs,~\cite{2017MNRAS.468.1338C} obtained the value for $m_{a}=1.79^{+0.35}_{-0.33}\times10^{-22}$ eV (at 2$\sigma$). Comparing to simulated data on the line-of-sight velocity dispersion ($\sigma_{\rm LOS}$) of Fornax and Sculptor dSphs,~\cite{Gonzalez-Morales:2016yaf} found a tighter upper bound $m_{a} < 4 \times 10^{-23}$ eV at 97.5$\%$ CL.   

The existence of the sub-halo to host Eridanus II low-surface brightness dwarf galaxy and the survival of its star cluster can also constrain the FDM mass~\citep{2019ApJ...871...28B,2016ApJ...824L..31B}. As shown in~\cite{2019PhRvL.123e1103M}, the sub-halo mass function in the Milky Way  and the existence of Eridanus II implies $m_{a}\gtrsim8\times10^{-22}$ eV. Gravitational heating can also be applied in the case of Eridanus II and its star cluster, which will be destroyed by gravitational heating from the fluctuations
of soliton peak density for $m_a\lesssim10^{-19}$ eV with the possible exception of the range $10^{-21} \text{ eV} \leq m_{a} \leq 10^{-19} \text{ eV}$ which is still allowed by observations~\citep{2019PhRvL.123e1103M}. We note, however, that this lower limit was recently disputed by~\cite{2021PhRvD.103j3019C}. 

Sizes and stellar kinematics
of ultra-faint dwarf (UFD) galaxies can be used to estimate the effects of wave interference produced by FDM. In particular, in order for the FDM's gravitational heating to be weak enough to explain the observed velocity dispersions of $2.5-3$ km$/$s and sizes of $24-40$ pc in Segue 1 and 2~(\cite{2011ApJ...733...46S,2013ApJ...770...16K}),~\cite{Dalal:2022rmp} derived a lower bound of $m_{a} > 3 \times 10^{-19}$ eV at 99$\%$ CL.
On the other hand, a Jeans analysis of 18 galactic UFD galaxies finds a preference for higher values of FDM mass in~\cite{2021ApJ...912L...3H}, being the strongest one for Segue 1 with $ m_{a}= 1.1^{+8.3}_{-0.7} \times 10^{-19}$  eV, which is consistent with the results obtained directly from the MUSE-Faint survey~\citep{2021A&A...651A..80Z}.  Finally, ~\cite{2016MNRAS.460.4397C} found that FDM particle with a mass of $m_{a} \sim 3.7 - 5.6 \times 10^{-22}$ eV is in agreement with the data of half-light mass of Draco II and Triangulum II; although~\cite{2020ApJ...893...21S} later found an upper limit $m_{a}\lesssim10^{-21}$ eV using a kinematic analysis of the Milky Way's satellites.

\begin{figure}[t]
\begin{center}
\hspace*{-0.9cm}
\includegraphics[scale=0.375]{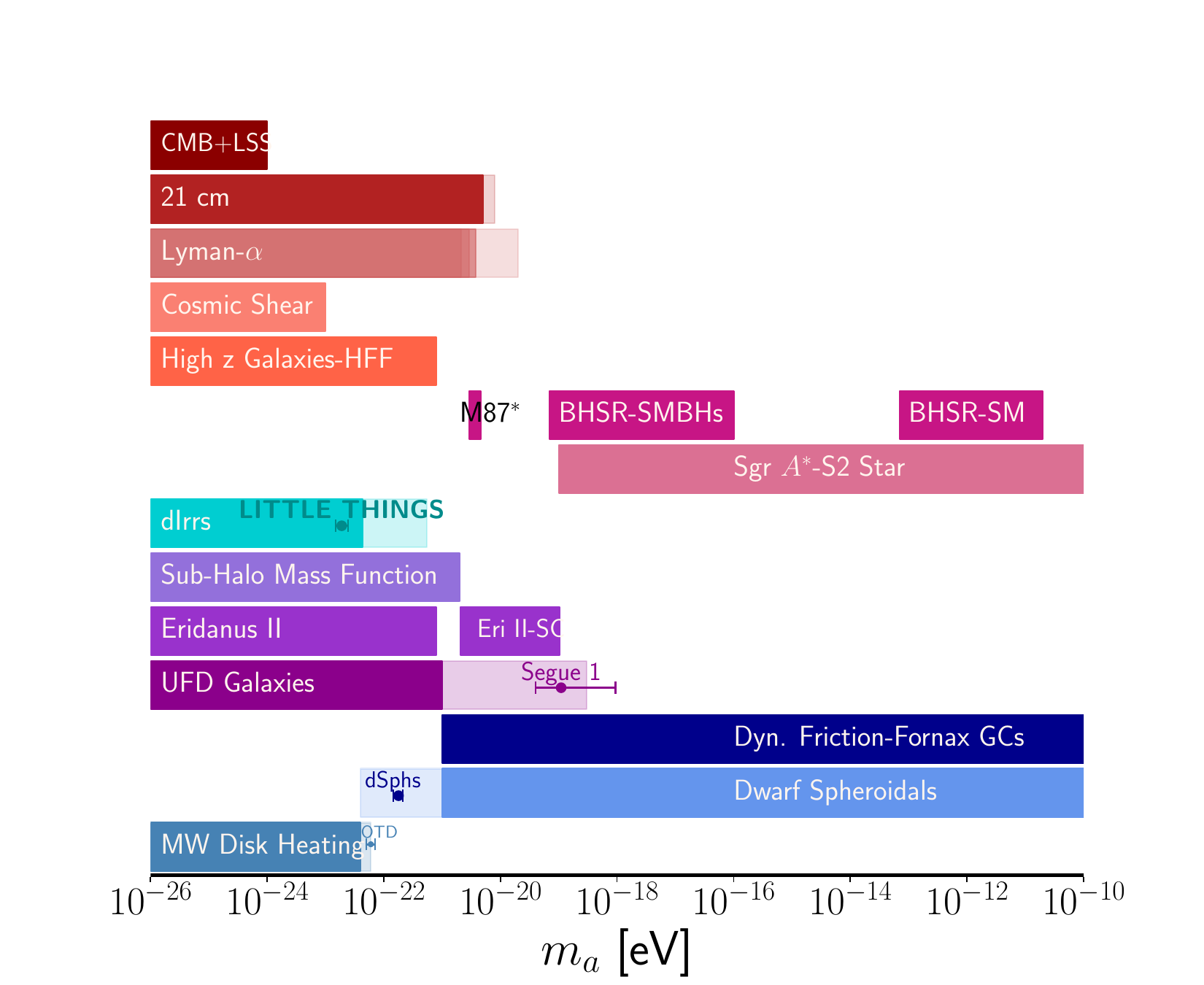}
\caption{Bounds from cosmology and astrophysics on the axion mass. The constraint from HMF in the local group with the halos of nearby dwarf irregulars (dIrrs)  from the LTs catalog is displayed with dark turquoise bars. On the dIrrs bar, we also show the optimal mass $m_{a}=1.9^{+0.5}_{-0.4}\times10^{-23}$ eV (uncertainties at 2$\sigma$ CL). Our constraints are compared to other cosmological and astrophysical limits (see main text for details).
}
\label{fig:constraints}
\end{center}
\end{figure}

In  Fig.~\ref{fig:constraints} we show a selection of the existing constraints on the axion mass along with the ones derived in this work from the rotation curves of the isolated dwarf irregular galaxies in the LTs sample. In particular, the $2\sigma$ interval stemming directly from the average of the fits to the rotation curves, Eq.~\eqref{eq:benchmarkmass}, and the lower bounds derived from the comparison of the HMF of FDM with dwarf-galaxy counts in the LGV.  Nonetheless, before closing this section, we would like to point out that many of the analyses leading to these constraints depend on astrophysical modeling assumptions and are, thus, subject to significant uncertainties. We refer, for instance, to~\cite{Ferreira:2020fam} and, in particular, to Sect.~4.3 of ~\cite{Chiang:2022rlx} for a recent critical reappraisal of some of these constraints.

\section{Conclusions}
\label{sec:Conclusions}

In this work we have investigated and tested the predictions of FDM at galactic scales using a homogeneous and robust sample of high-resolution rotation curves from the LITTLE THINGS in 3D (or LTs for short in our paper) catalog. This comprises a collection of isolated, dark matter-dominated dwarf-irregular galaxies that provides an optimal benchmark for cosmological studies. The methodological basis of our study is a statistical framework based on the $\chi^2$ analysis of the rotation curves using a soliton+NFW density profile as a theoretical model. This depends on four parameters: two for the soliton that we chose to be the axion mass, $m_a$, and the core mass, $M_c$, and two for the NFW halo that we choose to be the concentration parameter, $c,$ and the halo mass $M_h$. We fit the data using current MCMC techniques and a rather loose set of priors, except for the core-halo relation linking $M_c$ and $M_h$ for which we used a broad set of predictions stemming from the FDM simulations compiled in~\cite{Chan:2021bja}. From the results of the fits, we were able to perform various diagnostics on the predictions of FDM that allow us to draw the main conclusions of our work:
\begin{itemize}
\item The soliton+NFW model provides an excellent fit to the rotation curves of the LTs sample with the inferred axion masses clustering around a relatively narrow range of values $m_a\sim(1-5)\times10^{-23}$ eV.  Some of the galaxies lead to very stringent constraints on $m_a$ (see Fig.~\ref{fig:AxionMass}) so that combining them yields a relatively poor fit. If we conservatively attribute this to possible systematic effects or presence of outliers we obtain  $m_a=1.90^{+0.24}_{-0.21}\times 10^{-23}$ eV. 
\item However, we find that the individual determinations of $m_a$ are not scattered randomly around the average, but follow instead a mild correlation with the stellar mass of the galaxy with a significance of $\sim3.3\sigma$. Moreover, displaying the results of our fits in the $r_c-\rho_c$ and $r_c-M_c$ planes, we find scaling relations in the data that are at odds with the predictions of FDM. In case of the $r_c-M_c$ relation, the data show a scaling that is almost orthogonal to the one predicted by FDM with a significance that (at face value) is $\gtrsim5\sigma$. From this, we conclude that the cores we find in the LTs catalog do not have the properties of the standard FDM solitons.
\item In parallel, we investigated various cosmological implications related to the FDM halos for an axion mass $m_a \approx2\times 10^{-23}$ eV. The most important result is that the mere existence of the very same galaxies we are studying would be excluded by the strong suppression in the linear power spectrum that is predicted for FDM with this mass. Combining the HMF of FDM obtained in simulations with the one of CDM obtained in simulations of the Local Group, we derive a conservative bound of $m_a\gtrsim4.3\times10^{-23}$ eV. This represents a tension with the average axion mass determined from the rotation curves that (at face value)  again has a significance of $\gtrsim5\sigma$. Including all galaxies from a recent census yields a much more severe lower bound $m_a\gtrsim5.5\times10^{-22}$ eV.
\item Finally, we investigated the effects of baryons in our analysis. In particular, we estimated the contribution of the mass distribution of stars and gas to the structure of the soliton and found that this can be parametrized by a galaxy-dependent negative correction to the axion mass. However, these corrections are typically too small to alleviate any of the problems of FDM with data that we found. We also briefly discuss the possible role of baryonic feedback and pointed out that our galaxy sample is in a region of stellar masses that overlaps with the one where these effects can be significant in CDM.  
\end{itemize}

In summary, our analysis poses a serious challenge to FDM as a contending hypothesis for solving the core-cusp problem or any of the other small-scale issues related to $\Lambda$CDM. The tension that we have found between the favored FDM mass and the abundance of halos is yet another example of the problems of the model with cosmology, which is reminiscent of the so-called ``catch-22 problem'' found in warm DM~\citep{2012MNRAS.424.1105M} and renewed recently in the context of FDM~\citep{Mocz:2023adf}.
This arises when one wishes to replicate the core-like structures in dwarf galaxies by invoking a particle mass that would imply a suppression of small-scale structure that is much too strong to be consistent with observations. Baryonic physics within FDM could help relieving some of these tensions, although it seems unlikely in light of our findings and the results from pioneering FDM simulations with realistic baryonic feedback. An alternative approach would be extending the dark sector with, for instance, self-interactions, as recently explored in~\cite{Mocz:2023adf} as a possible mechanism to enhance small-scale structure formation, or by demanding that only a fraction of DM is FDM, as in, for instance,~\cite{Kobayashi:2017jcf}. Future work is expected to shed more light on these topics. On the observational side, a  reanalysis of HI data from other isolated dwarf irregular galaxies, using the same methods as in the LTs catalog, would increase the statistical power of cosmological analyses, such as the one presented in this paper. 

\begin{acknowledgements}
A.B.-H., A.C. and J.M.C. would like to thank Diego Blas, Jorge S\'anchez Almeida and Ignacio Trujillo Cabrera for introducing us to the small-scale issues of $\Lambda$CDM and for continuous guidance and encouragement. In addition, we thank Giuseppina Battaglia, Kfir Blum, Chris Brook and Arianna Di Cintio for guidance at different stages of the project. Finally, we would also like to thank Filippo Fraternali, Jorge Garc\'ia-Farieta, Th\'eo Gayoux, Se-Heon Oh, Marc Huertas-Company, Javier Olivares and Hsi-Yu Schive for useful discussions and/or for sharing data and information with us. A.B.-H. A.C. and J.M.C. acknowledge support from the grant PGC2018-102016-A-I00 and J.M.C. is also funded by the ``Ram\'on y Cajal'' program RYC-2016-20672. G.I. acknowledges financial support from the European Research Council for the ERC Consolidator grant DEMOBLACK, under contract no. 770017.
\end{acknowledgements}

%
%

\FloatBarrier
\bibliographystyle{aa} 
\bibliography{ULDM} 
\begin{appendix}

\section{Further details on the fits}
\label{app:info_fits}

In this appendix, we provide additional information about the fits and MCMCs performed in this paper. 
Various plots describing the relevant information are available in the \href{https://github.com/acastillodm/FuzzyDM}{\texttt{FuzzyDM}\faGithub} GitHub repository. There are three main directories. {\tt Variable mass} refers to the analysis  discussed in Sect.~\ref{sec:res:RCs}, where the axion mass $m_a$ is fitted independently for each galaxy. The {\tt Benchmark Analysis} contains the information regarding the fits where the axion mass in Eq.~\eqref{eq:benchmarkmass} was used for all the galaxies, as discussed in Sect.~\ref{sec:res:soliton}. Finally, {\tt NFW} contains the details regarding the fits to a NFW profile discussed in App.~\ref{app:info_fits:NFW}. The information in each class includes rotation curves, the corner plots and 1D posterior distributions of the parameters, and the MCMC auto-correlation times. In case of the fits to FDM we also include the density profiles for the DM component.

\subsection{Results for the FDM profile}
\label{app:info_fits:FDM}

\begin{figure}[h!]
\centering
\includegraphics[width=0.45\textwidth]{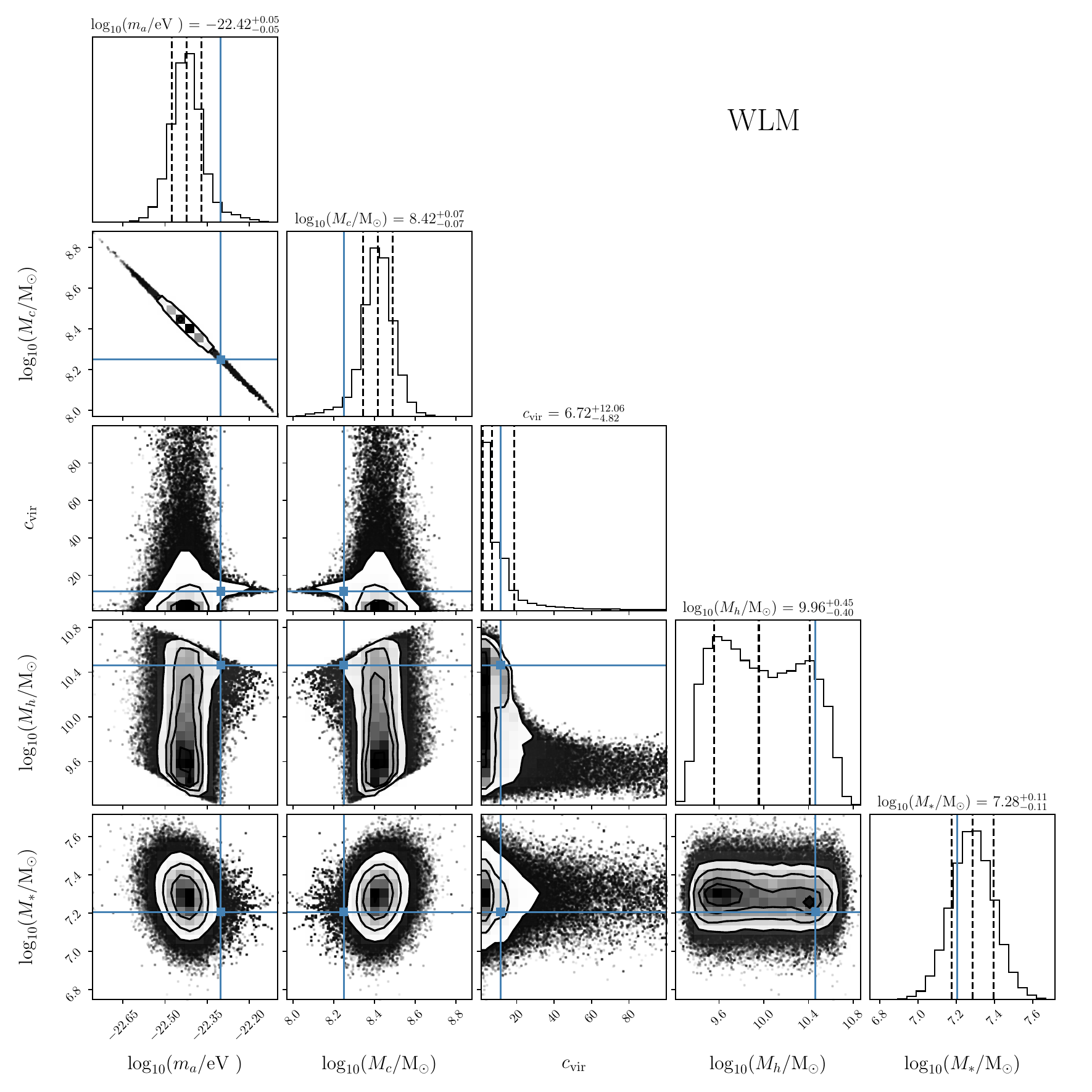}
\includegraphics[width=0.45\textwidth]{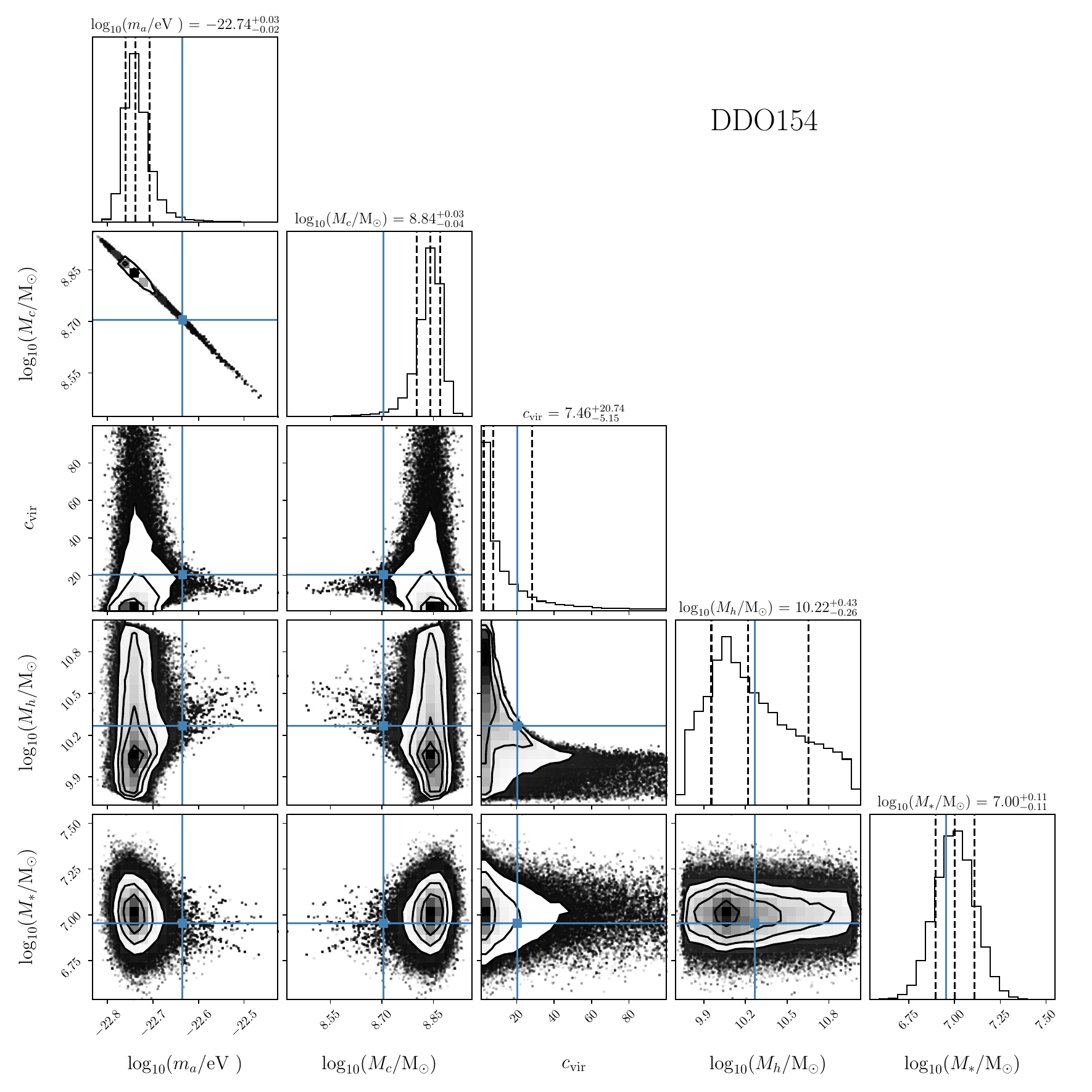} 
\caption{Corner and 1D posterior marginalized distributions for the fitting
parameters $\vec \theta_{\rm FDM}$ and $\MSt$ in the FDM plus baryons model for the two benchmark galaxies: WLM (\textit{top panel}) and DDO 154 (\textit{bottom panel}). Posteriors in the 1D plots show the 16$\%$, 50$\%,$ and 84$\%$ CL ranges. The priors are shown in Table \ref{tab:priors} and the blue lines and dots indicate the maximum posterior values of the parameters from the MCMC samples. 
Contours in the 2D plots display the $\sim 39.3 \%$, 67.5$\%,$ and 86.4$\%$ CL regions, corresponding to the 1, 1.5, and 2$\sigma$ regions of a 2D normal distribution. Cases where an additional inner contour is present denote an $\sim 11.8 \%$ CL or 0.5$\sigma$ region.} 
\label{fig:cornerplotsFDM}
\end{figure}

\begin{figure*}[t]
\centering
\begin{tabular}{lll}
\hspace{-0.15cm}\includegraphics[width=0.34\textwidth]{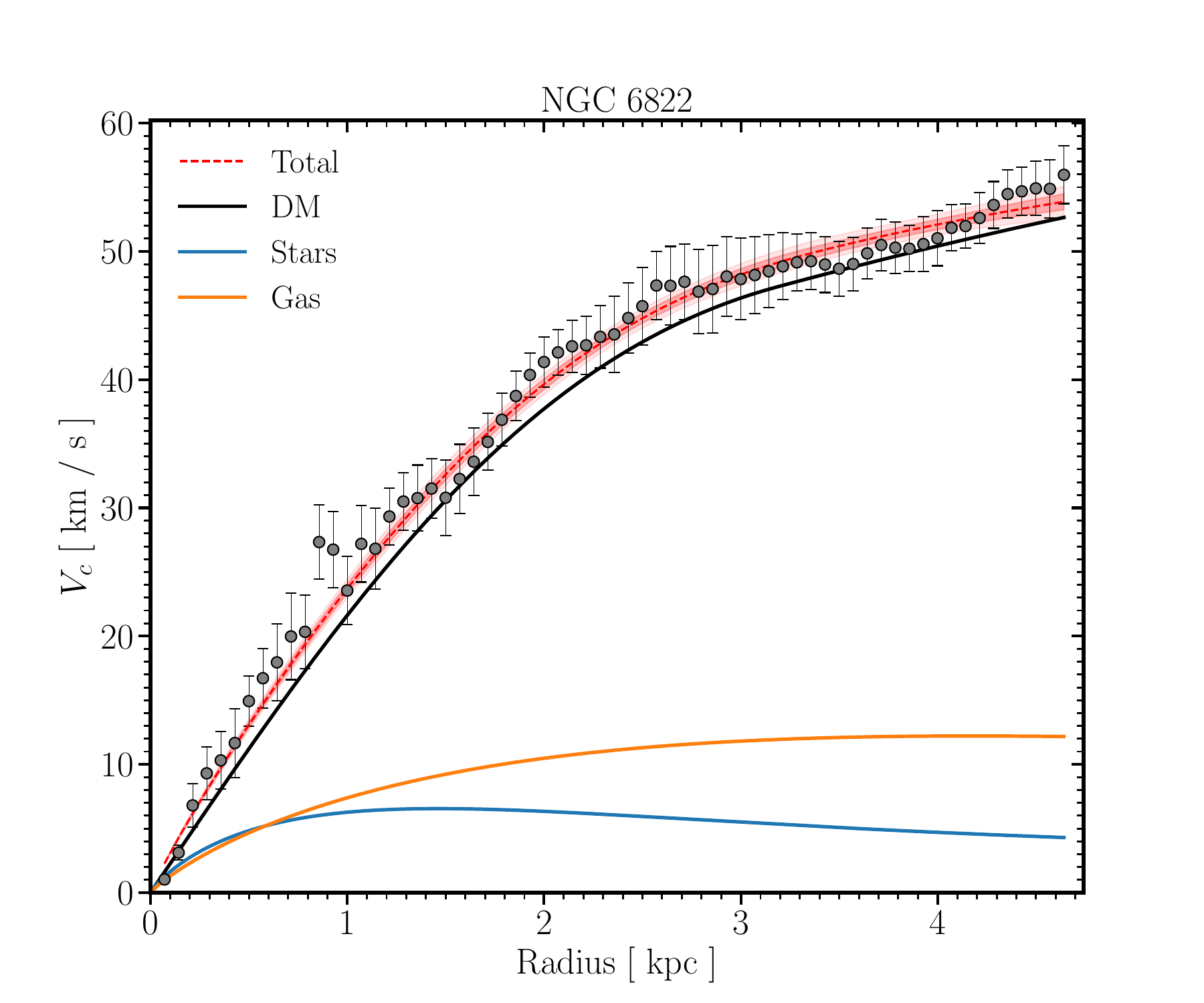} 
&\hspace{-0.35cm}\includegraphics[width=0.34\textwidth]{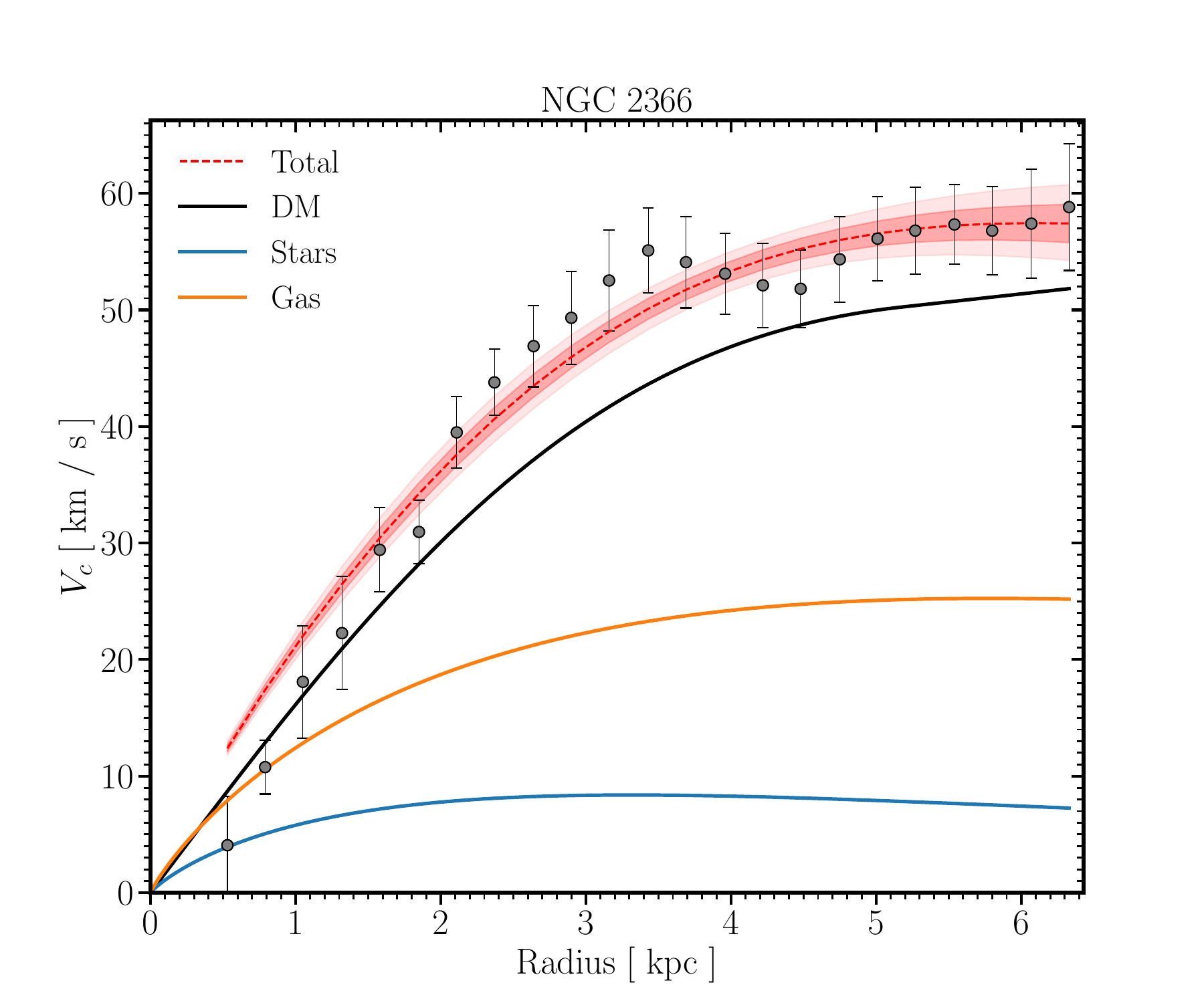}
&\hspace{-0.35cm}\includegraphics[width=0.34\textwidth]{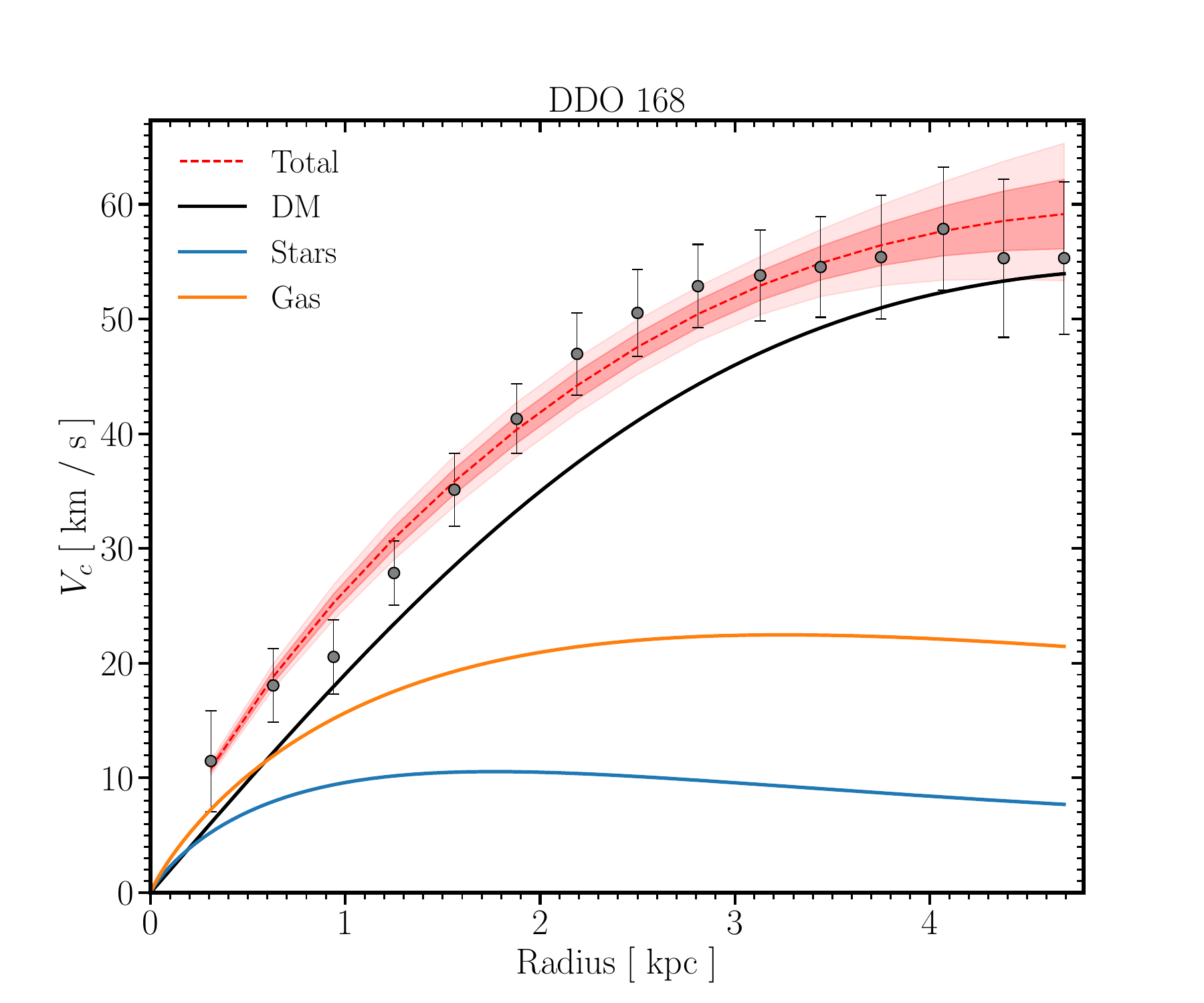}\\
\hspace{-0.15cm}\includegraphics[width=0.34\textwidth]{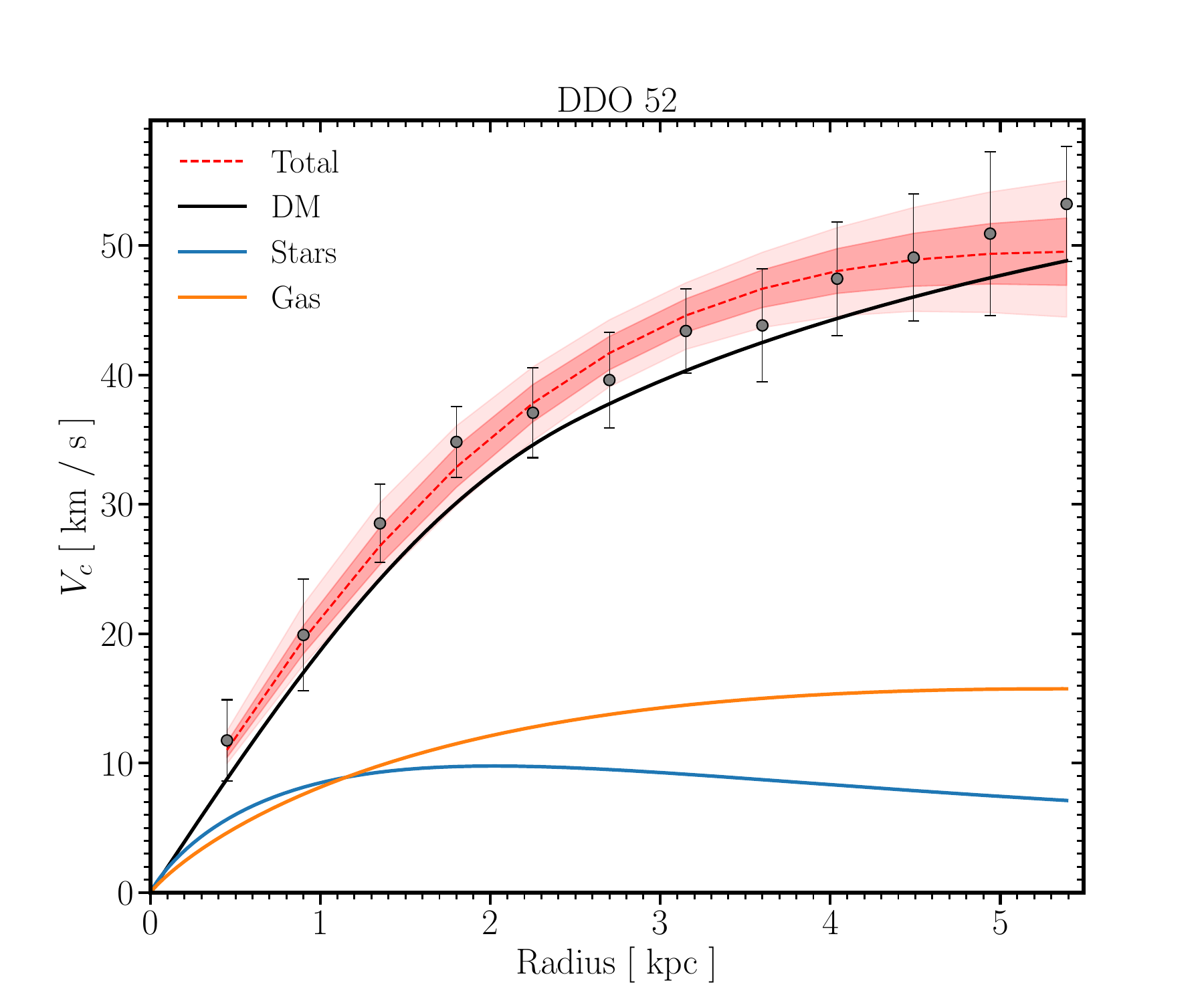} 
&\hspace{-0.35cm}\includegraphics[width=0.34\textwidth]{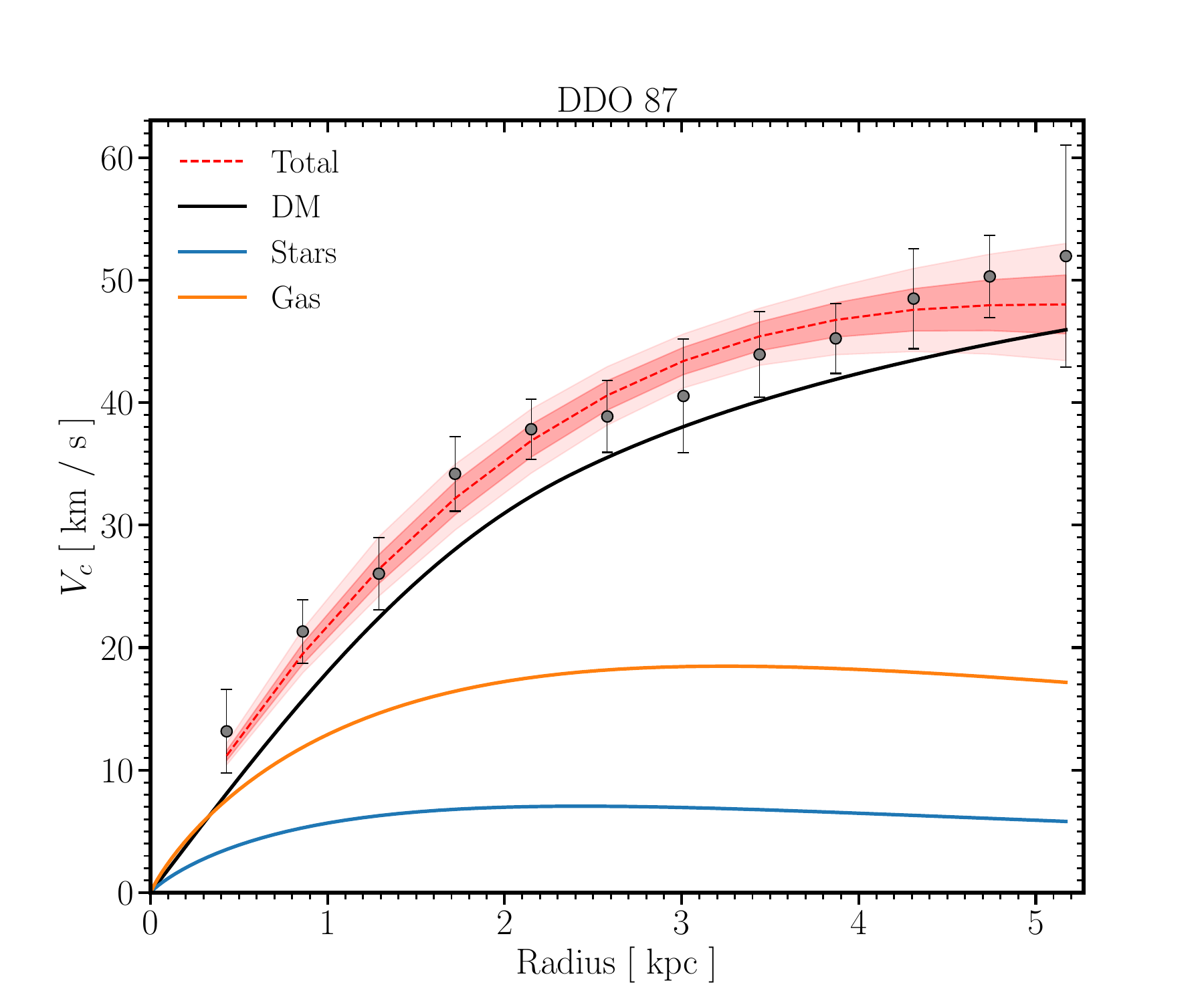}
&\hspace{-0.35cm}\includegraphics[width=0.34\textwidth]{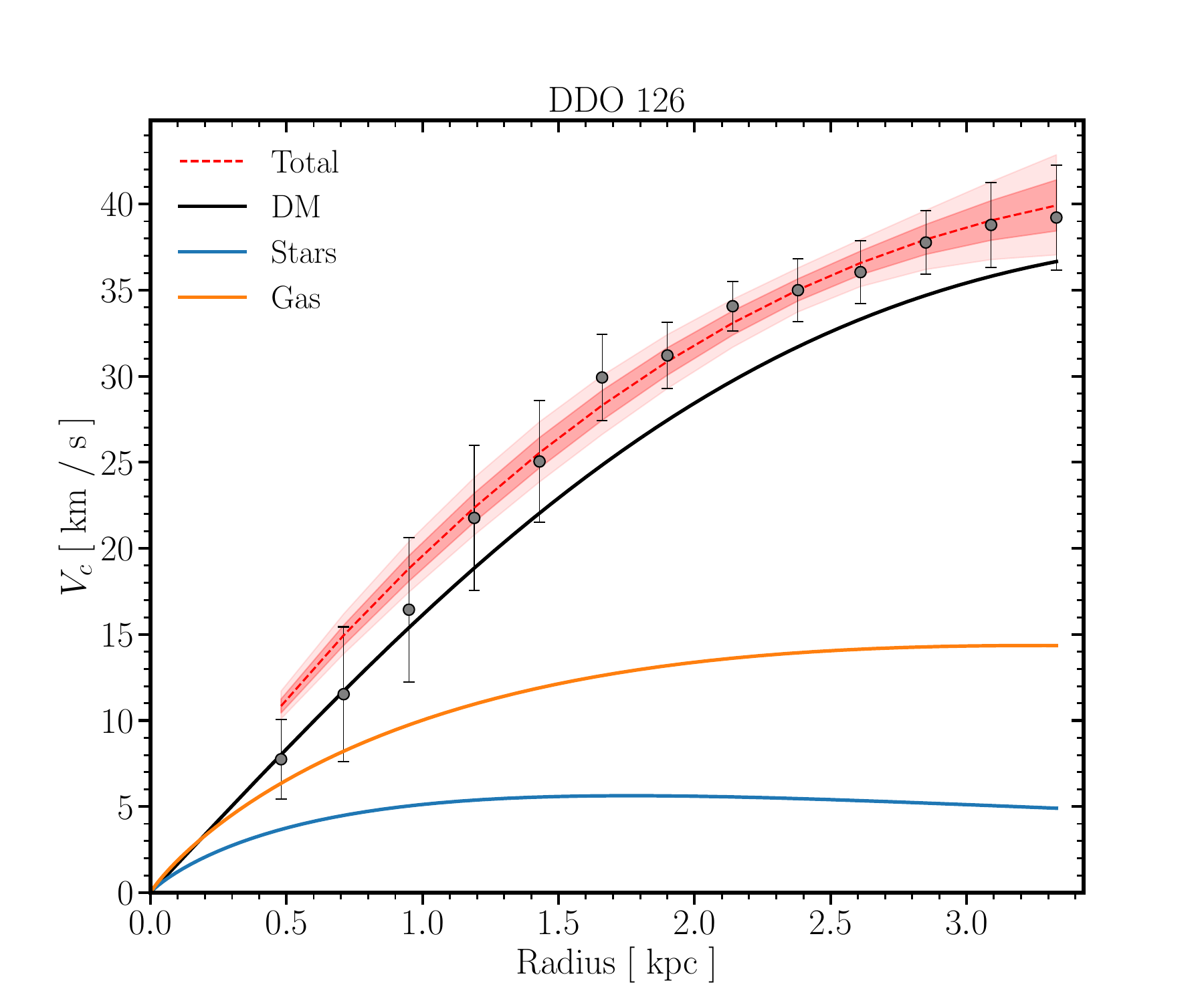}\\
\hspace{-0.15cm}\includegraphics[width=0.34\textwidth]{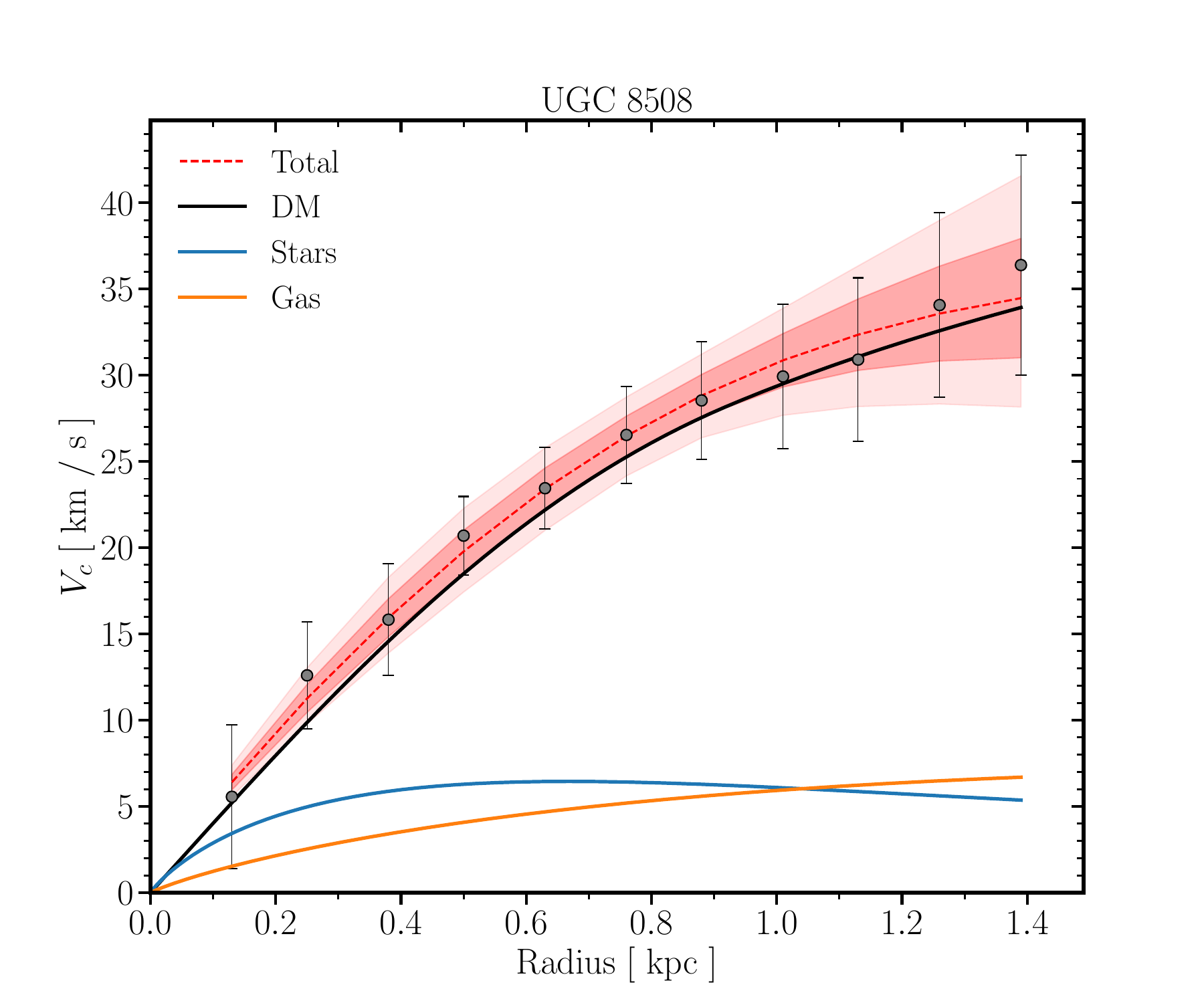} 
&\hspace{-0.35cm}\includegraphics[width=0.34\textwidth]{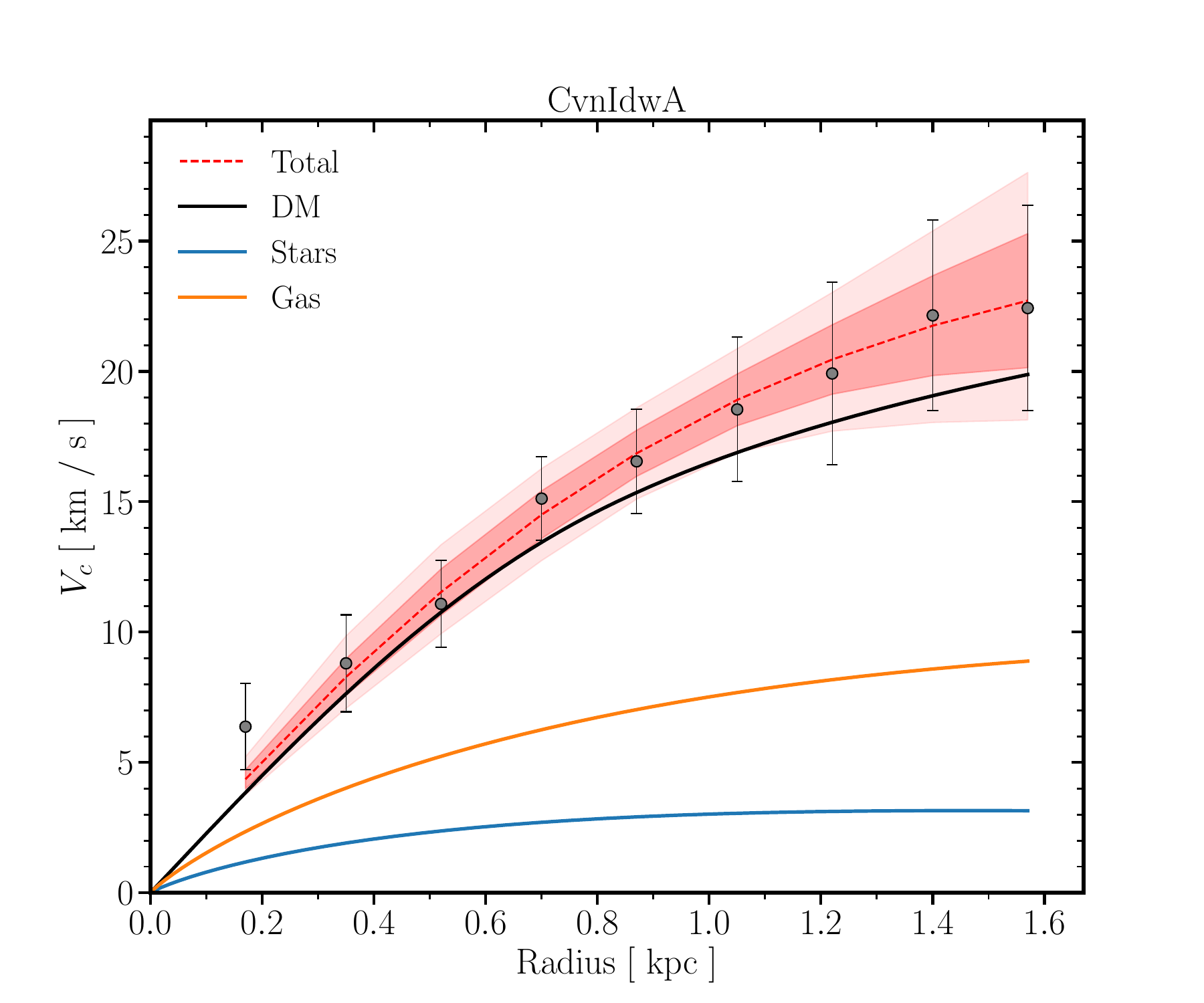}
&\hspace{-0.35cm}\includegraphics[width=0.34\textwidth]{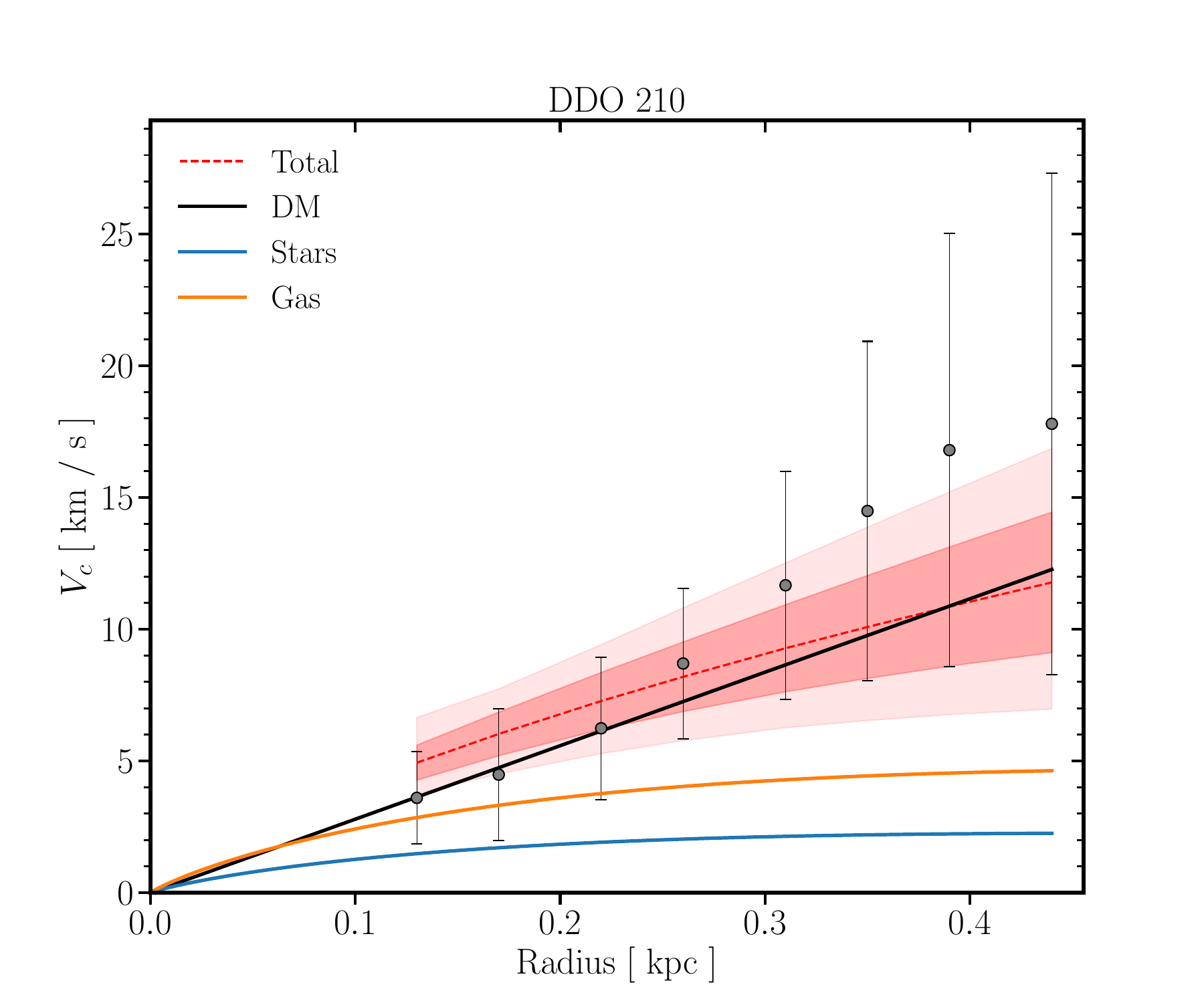}
\end{tabular}
\caption{Rotation curves in FDM for the core galaxies in the LTs catalog including NGC~6822 and excluding WLM and DDO~154, shown in Fig.~\ref{fig:my_rc}. We plot the DM, gas, and stellar components compared to data. The red dashed lines are the median of the posterior distribution for the total velocity and the colored band is the corresponding uncertainty at $68\%$ and $95\%$ CL. High-resolution figures can be found in the~\href{https://github.com/acastillodm/FuzzyDM}{\texttt{FuzzyDM}\faGithub} GitHub repository. }
\label{fig:rcs_total}
\end{figure*}

In Fig.~\ref{fig:cornerplotsFDM}, we show the corner plots and the 1D distributions of the marginalized posteriors of the parameters for the fits to WLM and DDO~154, which are the two galaxies used as a reference in the main text and in Figs.~\ref{fig:my_rc} and~\ref{fig:my_rc2}. The posteriors in the $m_{a}-M_c$ plane have elliptical shapes (with a negative correlation for both galaxies) that allow us to  precisely determine both parameters, with relative uncertainties of $\lesssim 20\%$ (see Table~\ref{tab:ResultsLTs}). Consequently, other derived core properties such as its radius, $r_c$, or central density, $\rho_c$, are also precisely determined from the fits. On the other hand, the halo  parameters (concentration factor and virial mass) are not well constrained due to the lack of data at large radii and the freedom that gives the continuity condition over transition radius, $r_{t}$. 
However, $r_{t}$ itself is determined with good precision, with relative errors also in the ballpark of $\lesssim20\%$. The rotation curves corresponding to the fits displayed in Table~\ref{tab:ResultsLTs} for the other galaxies in the core LTs catalog, besides WLM and DDO~154 and including NGC~6822, are shown in Fig.~\ref{fig:rcs_total}. 

\begin{figure*}[h!]
\centering
\begin{tabular}{cc}
\includegraphics[width=0.34\textwidth]{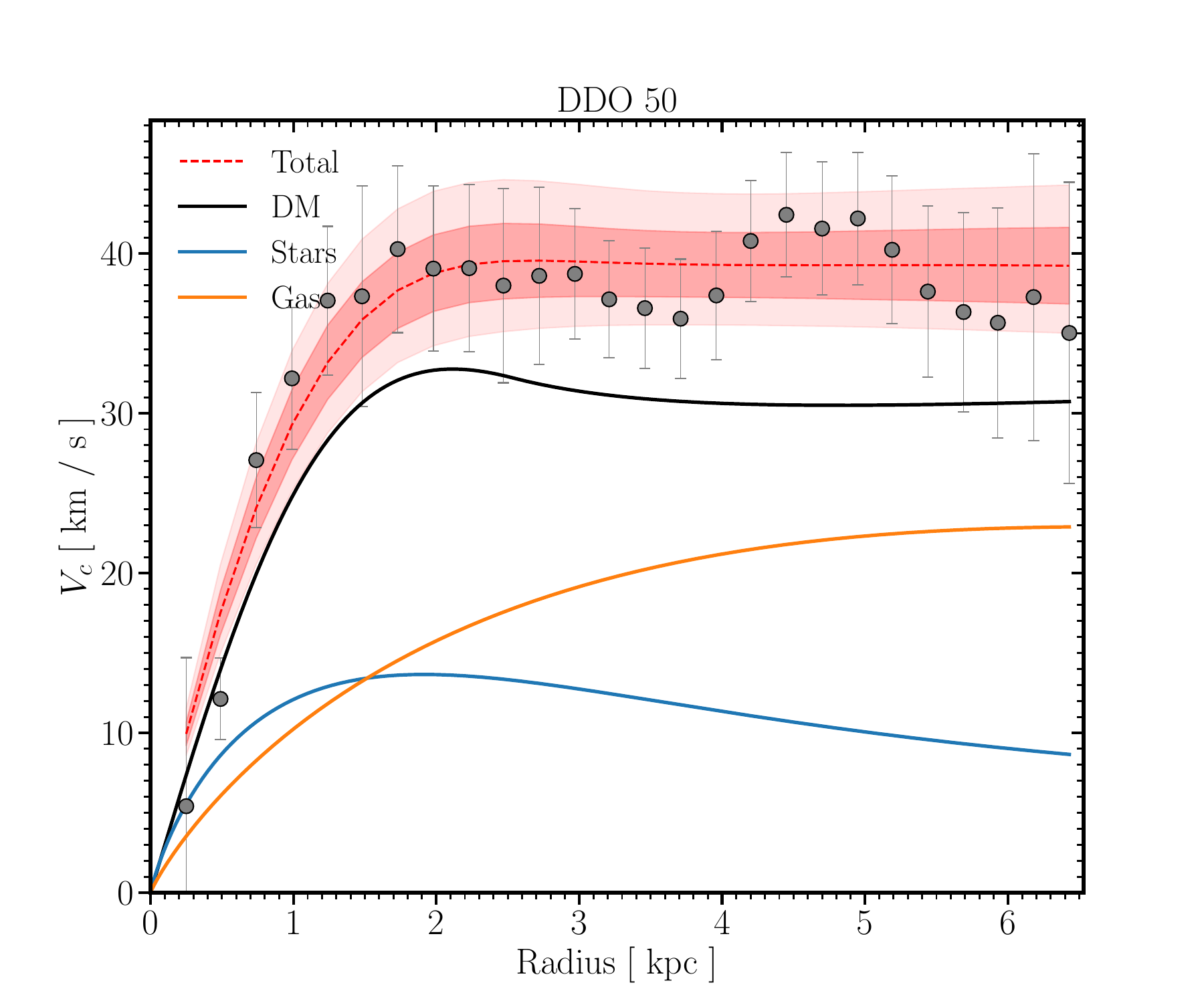} 
&\includegraphics[width=0.34\textwidth]{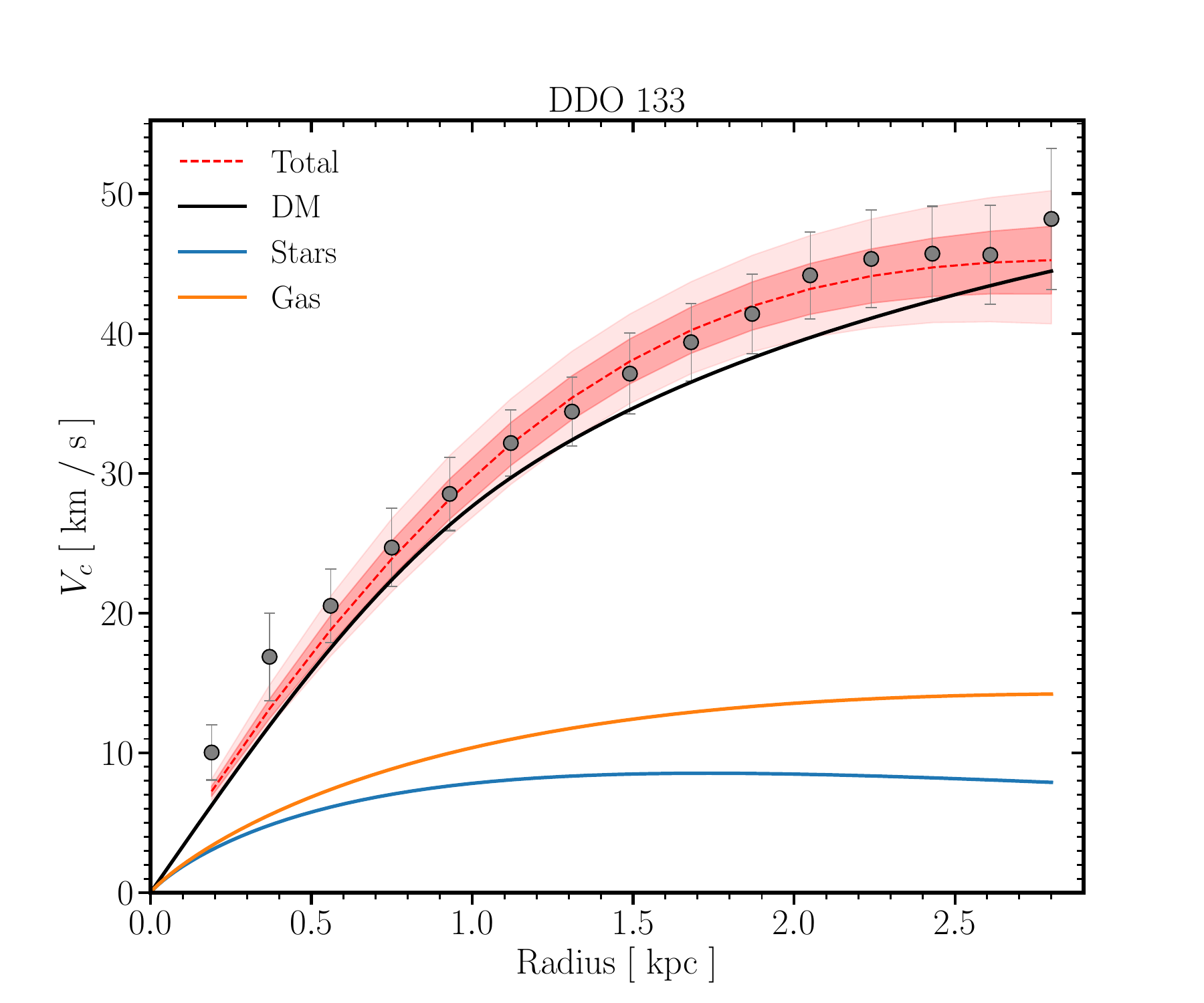}
\end{tabular}
\caption{Rotation curves in FDM for the inclination rogues. We represent the same information as in Fig.~\ref{fig:rcs_total}.
\label{fig:rcs_irogues}}
\end{figure*}

\begin{figure*}[t]
    \centering
    \includegraphics[scale=0.28]{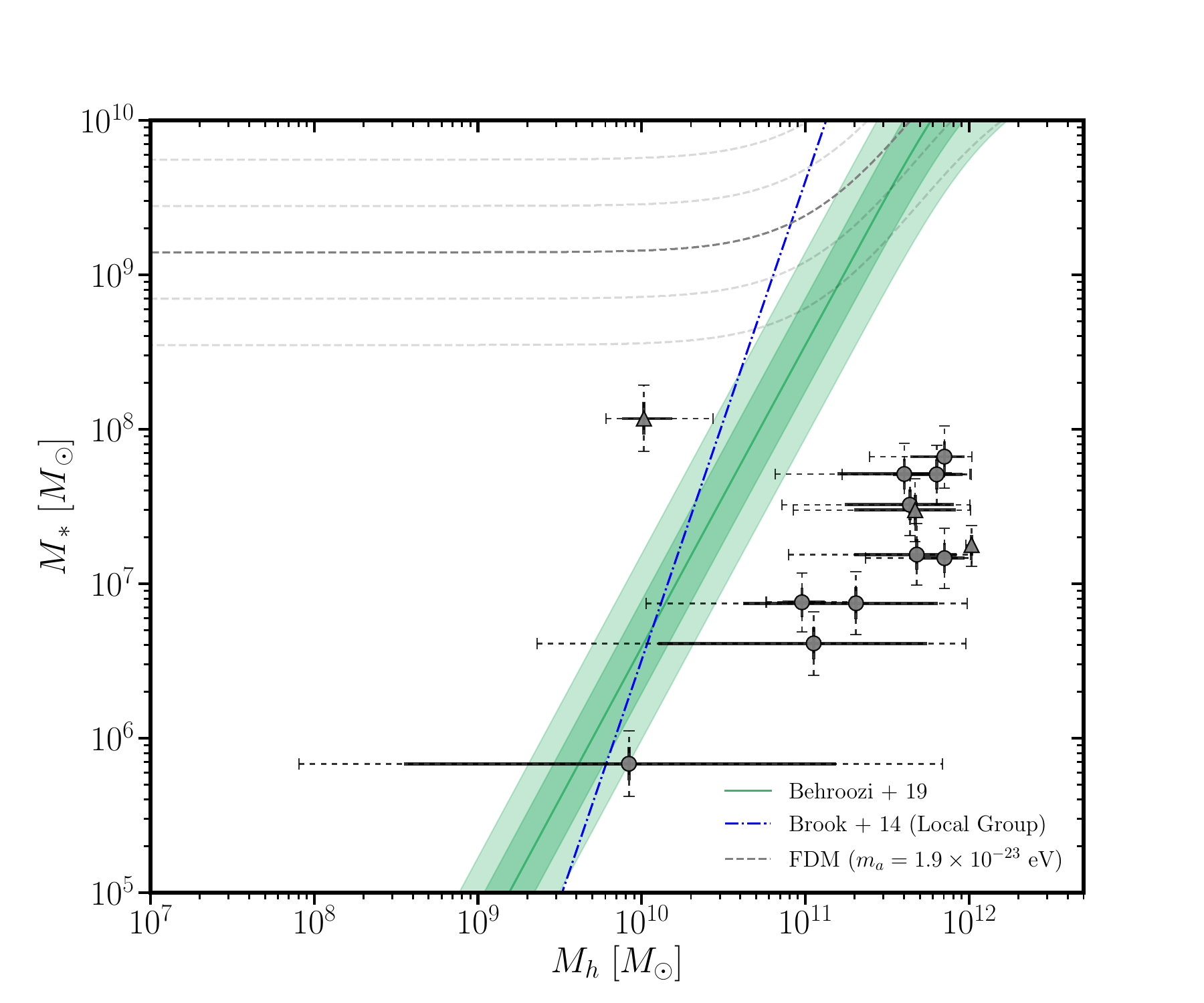}
    \includegraphics[scale=0.28]{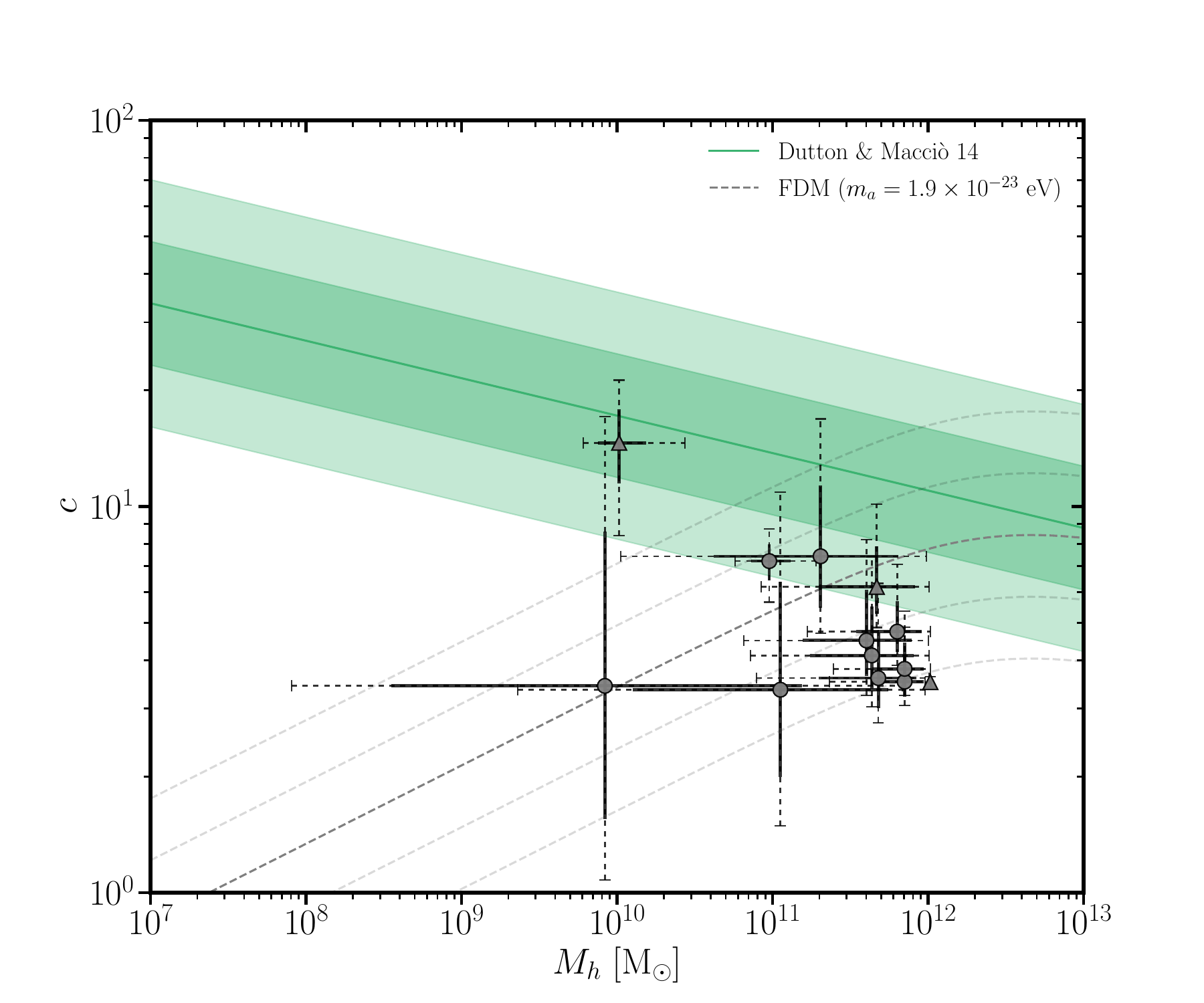}
    \caption{Median values of posterior distributions for the NFW model compared with $M_{*}-M_{h}$ (\emph{top}) and $c-M_{h}$  (\emph{bottom}) relations. Uncertainties in the data
points describe the 68$\%$ (solid) and 95$\%$ (dashed) CL ranges. The green bands describe $\pm1$ and 2 $\sigma$ scatter with $ \sigma= 0.3$ dex for the top panel and $\sigma=0.16$ dex for the bottom panel. Results for FDM are included for comparison.}
    \label{fig:relationsNFW}
\end{figure*}

\begin{table*}[t!]
\caption{Median and 68\% CL of the  posterior distributions of the fitted parameters $\MSt$, $M_h$, and $c$ in the NFW + baryons model for the galaxies from the LTs catalog, including some of the rogues and NGC~6822.}
\centering
\setlength{\tabcolsep}{0.7em}
\renewcommand{\arraystretch}{1.5}
\setlength{\arrayrulewidth}{.30mm}
\centering
\begin{tabular}{lcccc|c}\hline \hline
\multirow{2}{*}{Galaxy} & $\MSt$ & $M_{h}$ & \multirow{2}{*}{$c$} & \multirow{2}{*}{$\chi_{\nu}^{2}$}&\multirow{2}{*}{$\chi_{\nu,{\rm FDM}}^{2}$}\\
   &[10$^7$ $\MS$]&[$10^{11}$ $\MS$]&&  \\ \hline
 NGC~2366 & $6.7^{+1.7}_{-1.4}$ & $7.1^{+2.3}_{-2.7}$ & $3.80^{+0.64}_{-0.35}$ &  $3.26$ &1.29\\
 DDO~168  & $5.1^{+1.3}_{-1.0}$ & $6.3^{+2.8}_{-2.9}$ & $4.74^{+0.95}_{-0.53}$ & $2.24$  &0.56\\
 DDO~52  & $5.1^{+1.3}_{-1.1}$ & $4.0^{+3.8}_{-2.5}$ & $4.5^{+1.6}_{-0.9}$ & $0.42$  &0.13\\
 DDO~87  & $3.25^{+0.83}_{-0.67}$ & $4.3^{+3.7}_{-2.6}$ & $4.1^{+1.4}_{-0.7}$ & $0.39$  &0.27\\
 DDO~126  & $1.54^{+0.40}_{-0.31}$ & $4.8^{+3.6}_{-2.8}$ & $3.6^{+1.1}_{-0.6}$ & $2.08$ &0.49\\
 WLM  & $1.46^{+0.37}_{-0.30}$ & $7.1^{+2.3}_{-2.7}$ & $3.5^{+0.56}_{-0.30}$ & $1.55$ &0.68\\
 DDO~154  & $0.76^{+0.19}_{-0.15}$ & $0.95^{+0.35}_{-0.23}$ & $7.2^{+0.77}_{-0.78}$ & $1.98$ &0.94\\
 UGC~8508 & $0.75^{+0.20}_{-0.16}$ & $2.0^{+4.4}_{-1.6}$ & $7.4^{+3.9}_{-2.0}$ & $0.43$ &0.083\\
 CvnIdwA  & $0.41^{+0.11}_{-0.09}$ & $1.1^{+4.4}_{-1.0}$ & $3.4^{+3.0}_{-1.4}$ & $0.32$ &0.43\\
 DDO~210         & $0.068^{+0.019}_{-0.015}$ & $0.1^{+1.5}_{-0.1}$ & $3.4^{+5.2}_{-1.9}$ & $0.97$ &0.71\\
 & & -\emph{rogues}- &  &  \\
 DDO~50  & $11.7^{+3.3}_{-2.5}$ & $0.103^{+0.050}_{-0.028}$ & $15.0^{+3.3}_{-3.1}$ & $1.03$ &0.53\\
 DDO~133 & $3.0^{+0.78}_{-0.62}$ & $4.7^{+3.6}_{-2.7}$ & $6.2^{+1.7}_{-0.9}$ & $0.41$ &0.33\\
 NGC~6822$^{*}$ & $1.78^{+0.29}_{-0.26}$ & $10.3^{+0.15}_{-0.26}$ & $3.51^{+0.06}_{-0.05}$ & $32.5$&1.84\\
   \hline
   \hline
\end{tabular}
\tablefoot{ The last two columns include the reduced chi-square of the maximum posteriors for the NFW fit compared to the ones obtained for the FDM fits shown in Table~\ref{tab:ResultsLTs}.}
\label{tab:NFWResultsLTs}
\end{table*}

\subsection{Inclination rogues}
\label{app:info_fits:irogues}

We describe in detail the method used to fit the mass models to the galaxies with an inclination of $i<40^{\circ}$ (inclination rogues). The inclination $i$ is defined as the angle between
the line-of-sight and the vector normal to the plane of the galaxy~\citep{2016MNRAS.462.3628R,2017MNRAS.467.2019R}. Hence, a galaxy is edge-on when $i=90^{\circ}$ and face-on when $i=0^{\circ}$.  To account for this, we use the same method to fit the rotation curves (Sect. \ref{sec:fits}) but adding now the inclination as a new parameter. The dependence on  $i$ of the observed rotation velocity $V_{R}$ (and its uncertainty) is:
\begin{align}
\label{eq:icorrection}
V_{R}'=V_{R}\frac{\sin(i_{0})}{\sin({i})},\;\;\;\; \text{ and }\;\;\;\; \delta V_{R}'=\delta V_{R}\frac{\sin(i_{0})}{\sin({i})},
\end{align}
where $i_{0}$ is a ``reference'' angle.
Therefore, for the regime of $\emph{i}< 40^{\circ}$ (nearly face-on galaxies), small errors on $i$ can have a strong impact on the reconstruction of the observed rotational velocity. The $i_{0}$ values are taken (with uncertainties) from the original extraction of the rotation curves reported in~\cite{2017MNRAS.467.2019R}. To compute the circular velocity, we add in quadrature the gravitational components and the asymmetric drift correction. Correspondingly, the final velocity and errors are obtained from the transformed velocities and the errors with the inclusion of  the asymmetric drift correction $V_a$ as:
\begin{align}
\label{eq:icorr2}
V_{c}' = \sqrt{V_{R}^{'2} + V_{a}^{2}}
\;\;\;\; \text{ and }\;\;\;\;
\delta V_{c}' = \frac{\sqrt{V_{R}'^{2}\delta V_{R}'^{2} + V_{a}^{2}\delta V_{a}^{2}}}{V_{c}'}.
\end{align}
For the fits to the total velocity of Eq.~\eqref{eq:modelpredict} with the mass model in Eq.~\eqref{eq:densityFDM}, we assume a normal distribution prior centered around $i_{0}$ and error determined by its uncertainty. The remaining parameters follow the standard priors  of Table~\ref{tab:priors}, and the fitting algorithm, based on MCMC, uses the same attributes as those described in Sect.~\ref{sec:FitStrategy}. 

An additional correction for the log-likelihood is necessary since now the errors in Eqs.~\eqref{eq:icorrection} and \eqref{eq:icorr2} are $i$-dependent:
\begin{align}
\label{eq:loglike2}
    \ln \mathcal{L}(\vec\theta_{\rm FDM},\MSt, i)=\ln \mathcal{L}'(\vec\theta_{\rm FDM},\MSt, i)-\frac{1}{2}\sum_{j}\ln(\delta V_{c, j}'^{2}),
\end{align}
where $\mathcal{L}'(\vec\theta_{\rm FDM},\MSt,i)$ is the log-likelihood function in Eq.~\eqref{eq:loglike} changing the data points for velocities and their errors using the transformations of Eqs.~\eqref{eq:icorrection} and \eqref{eq:icorr2}. 

The inclination rogues analyzed in our work are DDO 50 and DDO 133, which have stellar masses in the selected range of $\MSt\lesssim 10^{8}$ $\MS$. In the case of FDM with variable mass, the parameters are reported in the last rows of Table~\ref{tab:ResultsLTs}. We show the corresponding rotation curves compared to the data in Fig.~\ref{fig:rcs_irogues}.  We also obtained the inclination angles $i=(40.7^{+2.3}_{-2.4})^{\circ}$ (DDO 50) and $i=(38.1^{+1.6}_{-1.7})^{\circ}$ (DDO 133),
which agree within $<5\%$ with the central values of $i_{0}$ used in the priors, \text{namely}, $37.4^{\circ}$ and $36.9^{\circ}$ for DDO 50 and DDO 133, respectively. 

\subsection{Fits with a NFW profile}
\label{app:info_fits:NFW}

We present here the results of the fits to the rotation curves of the LTs catalog using the NFW profile in Eq.~\eqref{eq:densityNFW}. We follow the same fit procedure as described in Sect.~\ref{sec:FitStrategy}, where the theoretical circular velocity in Eq.~\eqref{eq:modelpredict} depends now on $\vec\theta_{\rm NFW}=(c,M_{h})$ in the DM model instead of $\vec\theta_{\rm FDM}$. The priors for the halo parameters and $\MSt$ are equal to those described in Table~\ref{tab:priors}. As the NFW profile only has three parameters (two for the DM contribution and one for the baryionic model), the MCMCs are faster than in FDM for the convergence tests.  In Table~\ref{tab:NFWResultsLTs}, we summarize the results for the core LTs galaxies plus, the two inclination rogues (DDO~50 and DDO~133) and NGC~6822. We confirm that the NFW profile fails in providing a good fit to most of the LTs galaxies, in contrast to the fits to a ``cored'' profile such as the FDM model. 

Moreover, the values of the obtained halo parameters are inconsistent with the predictions of $\Lambda$CDM. In particular, in Fig.~\ref{fig:relationsNFW} we test the population of DM halos and the $\MSt-M_{h}$ relation via abundance matching~\citep{2019MNRAS.488.3143B} and the $c-M_{h}$ relation~\citep{2014MNRAS.441.3359D}. We find that most of the galaxies in the LTs sample are beyond the 2$\sigma$ bands in the case of abundance matching while the data cluster in regions below to the Dutton $\&$ Macci\`o relation for the $c-M_{h}$ plane. 

\subsection{Fitting linear scaling correlations}
\label{app:info_fits:correlations}

Here, we summarize the approach we use to quantify the linear correlations observed in our (variable mass) fits throughout Figs.~\ref{fig:AxionMass} and \ref{fig:solitonScaling}, namely, in the $(\log_{10} m_a,\: \log_{10} \MSt),$ 
$(\log_{10} \rho_c,\: \log_{10} r_c)$ and $(\log_{10} r_c,\: \log_{10} M_c)$ planes. To do this, we follow the process described in Sects. 7 and 8 of \cite{2010arXiv1008.4686H}.

Taking the case of the axion - stellar mass correlation as an example, we are interested in a fit of the form
\begin{equation}
\log_{10} m_a = \beta \log_{10} \MSt + \alpha^\prime
\label{eq:linear}
,\end{equation}
to our data from the posterior distributions of such parameters (obtained from our previous fits using MCMCs), taking the median and mean 68$\%$ CL errors as the data to be fitted. The idea is, then, to formulate a likelihood function that we can implement with MCMCs in much the same way as before and derive the posterior distributions of $\alpha^\prime$ and $\beta,$ with $\beta$ being the  key variable to assess the degree of correlation as quantified by the slope.

While the idea is simple, we need to account for the fact that we are dealing with errors in two dimensions. To this end,  we compute residues as the orthogonal distance between the line and each of the observed central values and qw compute the corresponding projections of the errors, under the assumption that these are normal and independent. To outline the specific way these are computed, it is more efficient to parametrize by angle $\theta,$ where $\beta \equiv \tan \theta,$ and the orthogonal distance between the line and the origin
$b_{\perp} \equiv \alpha^\prime \cos \theta,$ which are the ones we use internally. 

The resulting orthogonal residual for a given point $(x_i,\: y_i)$ of median observed values in the $(\log_{10} m_a,\: \log_{10} \MSt)$ plane  is given by:
\begin{equation}
    \Delta_i = y_i \cos \theta - x_i \sin \theta - b_{\perp},
    \label{eq:delta_i}
\end{equation}
while the resulting projected error $\Sigma_i$ is given by
\begin{equation}
    \Sigma_i^2 = \delta x_i^2 \sin^2 \theta + \delta y_i^2 \cos^2 \theta,
    \label{eq:sigma_i}    
\end{equation}
where $ \delta x_i,  \delta y_i$ denote the mean $68 \%$ CL errors from the median.

We also introduce an additional parameter $v$ to account for potential systematic underestimation of errors. The final log-likelihood function is then given by: 
\begin{equation}
   \ln \mathcal{L} (\theta, b_{\perp}, v) = -\frac{1}{2} \sum_i \frac{\Delta_i^2 }{\Sigma_i^2 + v} -\frac{1}{2} \sum_i \ln \Big( \Sigma_i^2 + v \Big).
\end{equation}
For the purposes of implementing this likelihood function into the MCMC fitting routine, we set the flat priors $-\pi/2 < \theta < \pi/2,$ $-100 \le b_{\perp} \le 100,$ and $0 \le v \le 100,$ which amply covers the set of expected values. Following the discussion in Section~\ref{sec:fits}, we run the 48-walker chains for a suitably long period that allows for reliable sampling of the distribution, which in this case amounts to 100~000 steps with a 50~000-step burn-in period. This is significantly more than necessary for convergence purposes, but allowed us to probe more extremal values of the distribution that are useful to establish the statistical significance of the observed correlations.

During the course of the fits, the values obtained for $v$ were quite small, indicating that the effect of such correction is not very significant. The results of these fits can be found in the \texttt{Variable Mass/Relations/Scaling Correlations} folder of the \href{https://github.com/acastillodm/FuzzyDM}{\texttt{FuzzyDM}\faGithub} GitHub repository.

\section{Mass functions and abundance matching}
\label{app:abundance_hmf}

\begin{figure}[t]
    \centering   \includegraphics[scale=0.27]{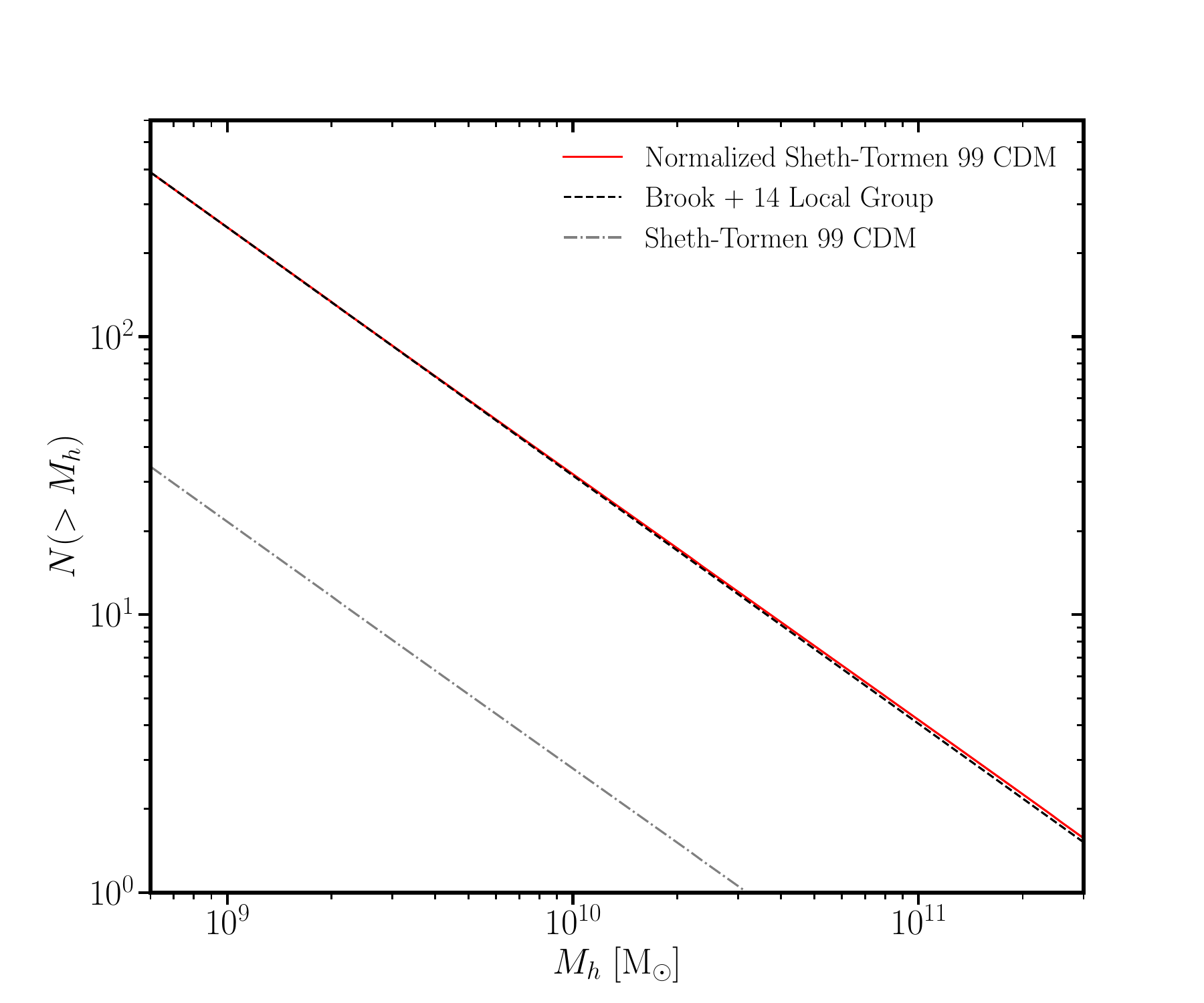}
    \caption{Comparison of accumulated abundances between the fitting function found in \cite{2014ApJ...784L..14B} (black, dashed) from CDM simulations in the LGV and the one obtained from the Sheth-Tormen HMF for CDM \citep{1999MNRAS.308..119S} normalized to match the same total abundance as the previous one (solid, red). We also include the original Sheth-Tormen abundance without the normalization.
    \label{fig:Sheth99-Brook14}}
\end{figure}

Here we provide more details on the procedure we use for computing the abundance matching results and the associated mass function constraints from Fig.~\ref{fig:halo}.
Given a mass function $\frac{d n}{ d\ln M}$ we can define the total number of halos within a given mass range $M_1 < M < M_2$ and volume $V$ as 
\begin{equation}
N = V \int_{\ln M_1}^{ \ln M_2} d\ln M \frac{d n}{ d \ln M}.
\label{eq:number}
\end{equation}
For our purposes, we define the LGV consisting of a sphere of radius $1.8$ Mpc. While we are not aware of FDM simulations being performed precisely in this environment and scale, we 
estimate the HMF of FDM by applying the suppression factor in Eq.~\eqref{eq:HMF-FDM} to the CDM mass function derived from \cite{2014ApJ...784L..14B}. While in this paper the mass function (as defined above) was not explicitly given, the cumulative mass function for the abundance of halos and subhalos above a given mass in the LGV was shown (Figure 1). We 
can accurately reproduce these results by introducing a normalization factor to the original Sheth-Tormen HMF for CDM \citep{1999MNRAS.308..119S} to ensure that the total abundances are the same. This is shown in Fig.~\ref{fig:Sheth99-Brook14}, where we observe excellent agreement. We emphasize
that a significant multiplicative factor is expected, since the LGV environment has a naturally higher abundance of galaxies than standard cosmological environments and we needs to account for the presence of subhalos.

With the CDM mass function, we can subsequently apply the suppression factor in Eq.~\eqref{eq:HMF-FDM} from~\cite{Schive:2015kza} to obtain the resulting FDM mass function that enables us to compute halo abundances via Eq.~\eqref{eq:number}. We note that while this suppression factor was originally obtained from simulations in higher redshift environments, the factor itself does not have a redshift dependence.

Regarding the abundance matching procedure, in addition to the discussion in Section~\ref{sec:res:halo}, we note that to obtain the effective cumulative SMF (as in Eq.~\eqref{eq:abundance_matching}, from a given $M$ to $\infty$), we computed the cumulative HMF using the Sheth-Tormen HMF for each $M_h$ and mapped it to the corresponding $\MSt$ via the abundance matching relation of \cite{2019MNRAS.488.3143B}.~\footnote{We did this as such data were not explicitly available in \cite{2019MNRAS.488.3143B} due to their use of a non-parametric algorithm.} We then subsequently performed the abundance matching procedure with the HMF of FDM (without the normalization factor discussed earlier in the appendix). The Sheth-Tormen HMF is computed via the \texttt{Colossus}  package in Python developed by~\cite{2018ApJS..239...35D}. 

\section{Comparison to the soliton-host halo relation}
\label{app:bar18}

As discussed in Sect.~\ref{sec:discussion}, \cite{Bar:2018acw} independently rederived  a core-halo relation  from~\cite{Schive:2014hza} and applied it to the analysis of rotation curves of the SPARC database. An important outcome of these studies is the prediction of a ``double-bump'' in the rotation curves, the first of which should become prominent in the inner part of the galaxies for $m_a\gtrsim10^{-22}$ eV. In our analysis we implement the core-halo relations spanned by Eq.~\eqref{eq:ferreira} from~\cite{Chan:2021bja}, which also includes  Eq.~\eqref{eq:Schive_2014}~\citep{Schive:2014hza}, and this structure is not  generically present in our theoretical rotation curves (see Fig.~\ref{fig:my_rc2}).
In order to understand this aspect, we first present  a brief summary of the derivation of the core-halo relation  in~\cite{Bar:2018acw}. 

\cite{Schive:2014dra} found a relation connecting the core mass of their FDM distributions and the halo specific energy $|E_h|/{M_h}$ (in natural units), 
\begin{equation}
    M_c \approx \Bigg(\frac{|E_h|}{M_h}\Bigg)^{\frac{1}{2}} \frac{m_p^2}{m_a}.
\label{eq:specific}    
\end{equation}
This can be equivalently expressed as~\citep{Bar:2018acw}
\begin{equation}
    \frac{E}{M} \bigg|_{\text{soliton}} \approx \frac{E}{M} \bigg|_{\text{halo}},
\end{equation}
due to an analytic property of the soliton where one can express the core mass in terms of the soliton specific energy. Assuming that the halo is dominated by a virialized NFW profile, its specific energy can be computed as:
\begin{equation}
\label{eq:SpE_halo}
  \frac{E}{M} \bigg|_{\text{halo}} \approx 
\frac{\pi}{M_h} \int_0^{r_{\text{vir}}} dx x^2 \rho_{\text{NFW}}(x)\phi_{\text{NFW}}(x),
\end{equation}
where $\phi_{\text{NFW}}(x)$ is the NFW potential. In addition,~\cite{Bar:2018acw} used a  
convention to define the virial radius as the point, where the average density equals 200 times the cosmological critical density, with the corresponding convention for the halo mass denoted by $M_{200}$.

Replacing  Eq.~\eqref{eq:SpE_halo} in Eq.~\eqref{eq:specific}, one obtains the core-halo relation reported by~\cite{Bar:2018acw}, 
\begin{equation}
    M_c \approx 5.7 \times 10^9 \bigg(\frac{m_a}{10^{-23} \text{ eV}} \bigg)^{-1} 
    \bigg(\frac{H(z)}{H_0}\bigg)^{\frac{1}{3}}
    \bigg(\frac{M_{200}}{10^{12} \, \MS} \bigg)^{\frac{1}{3}}
    f(c)\,\MS,
    \label{eq:bar_relation}
\end{equation}
where 
\begin{align}
\label{eq:fc}
f(c) = 0.54 \sqrt{\bigg(\frac{c}{1+c}\bigg) \frac{c - \ln(1+c)}{\big(\ln(1+c) - \frac{c}{1+c}\big)^2}},
\end{align}
is a function of $c$ that varies between $0.9 \lesssim f(c) \lesssim 1.1$ for 
$5 \le c \le 30$.

The core-halo relation in Eq.~\eqref{eq:bar_relation} is analogous to the one in Eq.~\eqref{eq:Schive_2014} but it is defined using a different convention for the halo mass. It is straightforward to transform the NFW conventions from a critical spherical overdensity definition of 200 to one of 100, which is close to the one used by \cite{Schive:2014hza, Chan:2021bja}. One uses a transformation of $(c_{200}, M_{200}) \to (c_{100}, M_{100})$ that parametrizes the same NFW density profile in Eq.~\eqref{eq:densityNFW}. 
Transforming $M_{200}$ to $M_h$ increases the given halo mass by $20-10\%$ for $c_{200}$ in the range $5-30$, yielding
\begin{equation}
    M_c \approx 5.5 \times 10^9 \bigg(\frac{m_a}{10^{-23} \text{ eV}} \bigg)^{-1}
    \bigg(\frac{M_h}{10^{12} \, \MS} \bigg)^{\frac{1}{3}}\,\MS,
    \label{eq:bar_relation2}
\end{equation}
where we have taken $z=0$ and $f(c)=1$ for simplicity. Therefore,  given a halo mass $M_h$, Eq.~\eqref{eq:bar_relation2} predicts a core mass that is $\approx80\%$ heavier than the one predicted by Eq.~\eqref{eq:Schive_2014}.  

The difference between the two relations is ultimately due to the way the specific energy of the halo is computed. In particular, as shown by \cite{Schive:2014hza}, one arrives at the original relation in Eq.~\eqref{eq:Schive_2014} by taking the halo specific energy to be that of a sphere of uniform density at the virial radius as 
\begin{equation}
    \frac{E}{M} \bigg|_{\text{halo}} \approx - \frac{3 G M_h}{10 r_{\text{vir}}}
\end{equation}
and applying it to the empirical relation in Eq.~\eqref{eq:specific}.~\footnote{In fact, this difference was also noticed by~\cite{Bar:2018acw} and discussed in its Appendix A. There, the difference in the computations of $E/M|_\text{halo}$ is absorbed into a redefinition of the halo mass which is not equivalent to the simple reparametrization of the NFW profile mentioned above.} This implies an important physical difference, since the potential (and implicitly the virialized energy) of a uniform spherical distribution is not equivalent to that of a self-gravitating NFW profile. It appears that such energy distribution, in combination with Eq.~\eqref{eq:specific}, is the median of the results of the simulations, as seen in Figs. 2 and 4 of~\cite{Schive:2014hza}, although  Eq.~\eqref{eq:bar_relation2} is compatible with an upward scatter of the data. 

\begin{figure}[t]
    \centering   \includegraphics[scale=0.27]{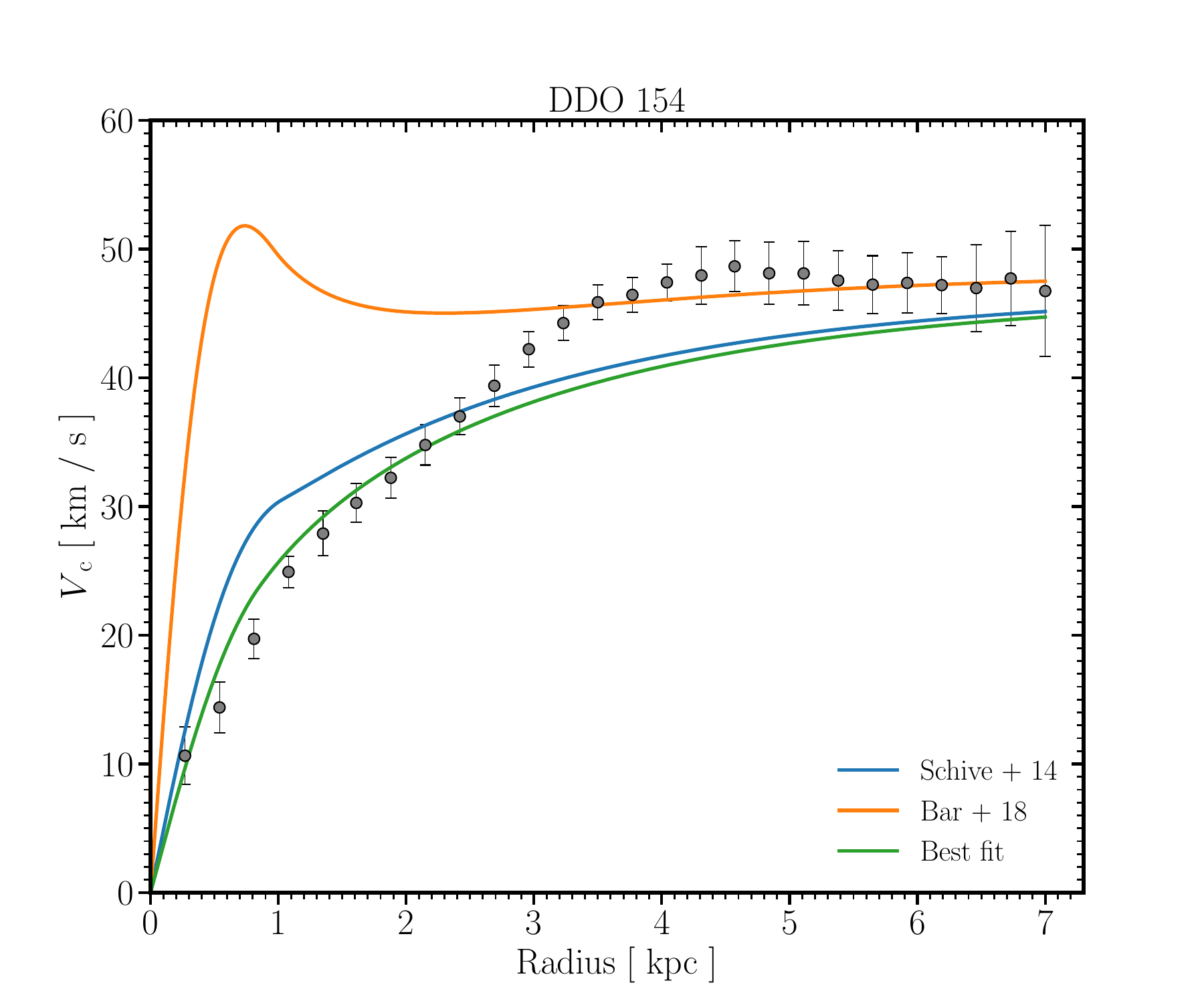}
    \caption{Comparison of the effect of imposing different core-halo relations in the rotation curve of DDO~154. Theoretical rotation curves are DM-only (i.e., excluding baryonic components) for visual clarity.
    \label{fig:ddo154_Bar}}
\end{figure}

This relation is also contained within the Eq.~\eqref{eq:ferreira} from \cite{Chan:2021bja} that is used in our analysis. However, this region of core-halo relations, leading to heavy solitons, is not favored by our fits for relatively heavy axion masses, $m_a\gtrsim10^{-22}$ eV. In particular, the rotation curve produced by a soliton has a maximum $v_{c,\rm max}$ at a radius $r_{c,\rm max}$ scaling with the core mass as $v_{c,\rm max}\propto M_c$ and $r_{c,\rm max}\propto1/M_c$~\citep{Bar:2018acw}. A heavier $M_c$  by a factor of $\sim2$ for a given halo mass $M_h$ will lead to a factor of $\sim 2$ higher rotation velocity peak at a distance $\sim1/2$ shorter from the center of the galaxy. This is illustrated in Fig.~\ref{fig:ddo154_Bar} where we show our maximum posterior fit to the LTs galaxy DDO~154 for $m_a=10^{-22}$ eV compared to the predictions that one would obtain imposing Eq.~\eqref{eq:Schive_2014} or~\eqref{eq:bar_relation2} and using the same $M_h$. The larger core mass implied by the latter suffices to produce the type of bump in the rotation curve that was discussed in~\cite{Bar:2018acw}. 
\end{appendix} 
\end{document}